\documentclass[a4paper,11pt]{article}
\pdfoutput=1 

\usepackage{jcappub} 

\usepackage[utf8]{inputenc}
\usepackage{graphicx}  
\usepackage{dcolumn}   
\usepackage{bm}  
\usepackage{multirow}
\usepackage{amssymb,amsmath,diagbox}
\usepackage{color}
\usepackage{appendix}
\usepackage{placeins} 
\newcommand\red[1]{#1}

\title{Reconstructing the Thermal Sunyaev Zeldovich Power Spectrum from {\it Planck} using the ABS Method}%

\author[a]{Zhaoxuan Zhang,}
\author[a,b,1]{Le Zhang\note{Corresponding author.},}
\author[c,d,e]{Pengjie Zhang}
\affiliation[a]{School of Physics and Astronomy, Sun Yat-sen University, Zhuhai 519082, People’s Republic of China}
\affiliation[b]{CSST Science Center for the Guangdong-Hongkong-Macau Greater Bay Area, SYSU, Zhuhai 519082, China}
\affiliation[c]{Shanghai Jiao Tong University, People’s Republic of China}%
\affiliation[d]{Tsung-Dao Lee Institute, Shanghai Jiao Tong University, Shanghai, 200240, China}
\affiliation[e]{Key Laboratory for Particle Astrophysics and Cosmology (MOE) / Shanghai Key Laboratory for Particle Physics and Cosmology, China}

\emailAdd{zhangzhx27@mail2.sysu.edu.cn, zhangle7@mail.sysu.edu.cn, zhangpj@sjtu.edu.cn}
\abstract{
This study presents a novel approach to reconstructing the thermal Sunyaev–Zeldovich (tSZ) power spectrum from {\it Planck} data using the Analytical Blind Separation (ABS) method. ABS enhances the recovery of weak signals by excluding low–signal-to-noise eigenmodes and introducing a shift parameter to stabilize the reconstruction. {\red 
Using the {\it Planck} PR3 full-mission data, ABS yields lower amplitudes of the $y$-map power spectrum at $\ell \gtrsim 300$ than the {\it Planck} 2015 MILCA and NILC band powers, suggesting reduced contamination from foreground residuals. After marginalizing over residual foregrounds, we estimate the tSZ power spectrum both with and without including the trispectrum contribution in the covariance. Excluding the trispectrum, the inferred tSZ amplitude is $\sim 9\%$ lower than the {\it Planck} 2015 best-fit value, yet consistent within $2\sigma$ \citep{2016A&A...594A..22P}. Including the trispectrum yields an amplitude $\sim 18\%$ higher than the ``Bolliet 2018'' best-fit value, corresponding to a $\sim 1.2\sigma$ deviation~\citep{2018MNRAS.477.4957B}. In both cases, the results are consistent with the ``Battaglia 2012'' model at the $1\sigma$ level~\citep{2012ApJ...758...74B}. These results indicate that ABS provides an independent and complementary measurement of the tSZ power spectrum.}
}

\begin{document}
\maketitle
\section{Introduction}\label{sec:intro}

The intracluster medium (ICM), consisting of hot gas in galaxy groups and clusters, serves as a secondary source of anisotropies in the cosmic microwave background (CMB). As CMB photons traverse a cluster, they are likely to undergo inverse Compton scattering by the hot electrons in the ICM, resulting in a slight energy gain for the photons. This interaction causes a small intensity or temperature decrement at radio wavelengths and a corresponding increment at millimeter wavelengths, an effect referred to as the thermal Sunyaev-Zel’dovich effect (hereafter, tSZ;~\cite{1972CoASP...4..173S}, see~\cite{1999PhR...310...97B} for a review).

The tSZ signal observed in the sky is highly sensitive to key cosmological parameters that govern the growth of galaxy clusters, providing an important and independent means of measuring cosmological parameters like $\sigma_8$, $\Omega_m$, and $H_0$~\citep{2002ARA&A..40..643C}. Additionally, it serves as a valuable tool for testing models of dark energy evolution~\citep{2018MNRAS.477.4957B}. Moreover, SZ observables have significant potential to probe extended cosmologies, including those related to primordial non-Gaussianity, massive neutrinos~\citep{2020MNRAS.497.1332B}. Furthermore, the evolution of the number of galaxy clusters, or ``cluster counts''~\citep{2003PhRvD..68h3506B,2018A&A...614A..13S}, as a function of redshift has long been recognized as a highly sensitive probe of cosmology.

In recent decades, extensive research has been conducted on the statistics of the tSZ signal, covering both analytical developments (e.g.,~\cite{2000PhRvD..62j3506C,2002ApJ...577..555Z,2003ApJ...598...49Z,2005ApJ...632....1C,2009ApJ...702..368S} and numerical simulations (e.g.,~\cite{2010ApJ...725.1452S,2011ApJ...727...94T,2012ApJ...758...74B,2017MNRAS.469..394H}). The angular power spectrum of the tSZ effect is influenced by the amplitude of matter fluctuations~\citep{1999ApJ...526L...1K,2002MNRAS.336.1256K}. However, this is complicated by the tSZ signal's sensitivity to the astrophysical processes governing the thermal state of the ICM, as its magnitude is directly related to the pressure of the hot gas. At high multipoles ($\ell \gtrsim 1000$), the power spectrum is impacted by the detailed pressure profiles within halos, whereas at lower multipoles, this dependence is significantly reduced~\citep{2014MNRAS.440.3645M}. Consequently, the tSZ power spectrum at $\ell \lesssim 1000$ is considered a valuable probe of cosmology.

It is therefore unsurprising that a significant number of tSZ surveys have been conducted. Recent experiments have yielded extensive catalogs of cluster SZ observations, including those from the Atacama Cosmology Telescope (ACT;~\cite{2013JCAP...07..008H}), the South Pole Telescope (SPT,~\cite{2013ApJ...763..127R,2015ApJS..216...27B}), and the {\it Planck}  satellite~\citep{2016A&A...594A..27P}. Furthermore, the {\it Planck}  Collaboration~\citep{2014A&A...571A..21P} produced the inaugural all-sky Compton-$y$ map and its corresponding SZ power spectrum. From these datasets, various studies have conducted analyses aimed at estimating cosmological parameters~\citep{2014JCAP...02..030H,2016A&A...594A..22P,2016A&A...594A..24P,2017A&A...604A..71H,2019ApJ...878...55B,2018MNRAS.477.4957B,2018A&A...614A..13S,2024ApJS..270...16I}.
 
The majority of these analyses have yielded a best-fitting cosmology with a matter clustering amplitude, $\sigma_8$, that is notably lower than the value derived from the {\it Planck}  2015 primary CMB data~\citep{2016A&A...594A..13P} by approximately 1--2$\sigma$. This discrepancy has prompted a significant research effort to identify the cause of the discrepancy (e.g.,~\cite{2016MNRAS.456.2361B,2016MNRAS.463.1797D,2017MNRAS.469..394H}).

The tSZ effect signal in the {\it Planck} multi-frequency maps is relatively weak compared to CMB and other foreground emissions. Unlike the diffuse CMB, the tSZ effect from galaxy clusters is spatially confined, producing a distinctly non-Gaussian signal. The regular component-separation methods designed for CMB analysis~\citep{2014A&A...571A..12P} are not well-suited for extracting the tSZ signal. To achieve separation, specialized component-separation algorithms were applied to extract the tSZ signal from {\it Planck}  frequency maps. There algorithms exploit both the spatial localization of astrophysical components and their distinct spectral characteristics. 

Specifically, the {\it Planck}  Collaboration~\citep{2014A&A...571A..21P,2016A&A...594A..22P} employed the MILCA (Modified Internal Linear Combination Algorithm;~\cite{2013A&A...558A.118H}) and NILC (Needlet Independent Linear Combination;~\cite{2011MNRAS.410.2481R}) methods. Both methods are based on the Internal Linear Combination (ILC) technique, which seeks a linear combination of input maps that minimizes the variance of the reconstructed map while maintaining unit gain for the target component here, the SZ effect, whose frequency dependence is well characterized. These algorithms have undergone extensive testing on simulated {\it Planck}  data. More recently, \cite{2022MNRAS.509..300T} enhanced the all-sky tSZ map reconstruction using the 100 to 857 GHz channel maps from {\it Planck} Data Release 4 (PR4)~\citep{2020A&A...643A..42P}, reducing noise and systematic effects, and producing a $y$-map with about 7\% lower noise. Recently, using the refined PR4 data,~\cite{2023MNRAS.526.5682C,2024PhRvD.109b3528M} demonstrate a slightly lower $y$-map power spectrum.

As known, residual foreground contamination--primarily from thermal dust emission at large angular scales and from the cosmic infrared background and extragalactic infrared and radio point sources at smaller angular scales--can lead to biased estimates of $y$-maps. This raises an important question: {\it are there optimized foreground removal methods that could yield more accurate estimates of the tSZ signal and effectively reduce the impact of foreground residuals?}

Recently, a new and computationally efficient method, ABS, has been proposed by~\citep{10.1093/mnras/stz091} for the blind separation of the CMB from foregrounds. Unlike ILC, ABS employs a distinct methodology that involves the exclusion of low signal-to-noise ratio (S/N) eigenmodes, enabling the robust and efficient recovery of weak signals. ABS analytically solves the CMB band power spectrum based on measured cross-band power, bypassing multi-parameter fitting. The method has been validated with simulated {\it Planck}  temperature maps~\citep{2018ApJS..239...36Y} and has successfully recovered $E$- and $B$-mode power spectra from simulated CMB polarization observations~\citep{2021A&A...650A..65S,2022JCAP...10..063G,2024arXiv240201233Z}. In this study, we reanalyze the {\it Planck} PR3 data, utilizing the ABS method to estimate the angular power spectrum of the tSZ signal from the frequency maps. We then compare our results with the power spectrum amplitudes reported in the literature. 

The paper is organized as follows: Sect.~\ref{sect:2} provides a brief overview of the {\it Planck} data and the simulations employed to reconstruct and validate the tSZ power spectrum. Sect.~\ref{sect:abs} introduces the ABS method used for signal separation. In Sect.~\ref{sect:5}, we present the results from reanalyzing the full-mission {\it Planck} PR3 data using the ABS method. Additionally, the residual foreground components are further removed through a likelihood analysis. We then compare the amplitude of the marginalized tSZ power spectrum with the best-fitting models from the literature. Finally, we summarize our findings in Sect.~\ref{sect:con}. 

\section{Simulations and Data}\label{sect:2}
\subsection{Compton $y$ parameter}
The Compton $y$ parameter is proportional to the line-of-sight integral of electron pressure, $P_{\rm e}=n_{\rm e} k_{\rm B} T_{\rm e}$, where $n_{\rm e}$ is the physical electron number density, $k_{\rm B}$ is the Boltzmann constant, and $T_{\rm e}$ is the electron temperature. In a given angular direction, $\hat{\boldsymbol{n}}$, the Compton $y$ parameter can be expressed as follows~\cite{1972CoASP...4..173S}:
\begin{eqnarray}
y (\vec{n}) = \int n_{\rm e} \frac{k_{\rm B} T_{\rm e}}{m_{\rm e} c^{2} } \sigma_{\rm T} \  {\rm d}s\,,
\end{eqnarray}
where d$s$ is the distance along the line of sight, $\sigma_{\rm T}$ is the Thomson scattering cross-section. The temperature shift in CMB due to the tSZ effect at a frequency $\nu$ is expressed as
\begin{equation}\label{eq:tsz}
\frac{\Delta T}{T_{\rm CMB}} = g(\nu)y\,,
\end{equation}
where, neglecting relativistic corrections~\citep{2016A&A...596A..61H,2019MNRAS.483.3459R,2023MNRAS.519.2138A}, the frequency-dependent factor in the thermodynamic temperature unit is given by $g(\nu) = x \coth(x/2) - 4 $, with $ x = h \nu/(k_{\rm B} T_{\rm CMB})$, and $T_{\rm CMB} = 2.726 \pm 0.001$ K. The tSZ effect results in a negative temperature shift at frequencies below 217 GHz and a positive shift at higher frequencies.

\subsection{tSZ angular power spectrum}
We based our analysis of the tSZ angular power spectrum on the methodologies outlined by~\cite{2014A&A...571A..21P,2016A&A...594A..22P}.

To reconstruct the tSZ angular power spectrum, we utilized both the auto power spectrum of the full dataset and the cross-angular power spectrum between the first (F) and last (L) halves of the data (denoted as F/L). The advantage of using the cross power spectrum is that it mitigates the bias introduced by noise and  potential systematic errors in the auto power spectrum. However, this approach introduces an increased level of statistical uncertainty in comparison to the auto one.

The cross-power spectrum was computed using \texttt{NaMaster}~\citep{2019MNRAS.484.4127A}, which employs the pseudo-$C_\ell$ framework to account for beam convolution, pixelization, and mode-coupling effects induced by the mask. We adopted the same multipole binning scheme as used in~\cite{2016A&A...594A..22P}. The tSZ reconstruction is influenced by considerable thermal dust emissions from our Galaxy, along with emissions from infrared and radio point sources. To reduce this contamination, we use the mask specifically designed for tSZ separation. The mask was taken from~\cite{2023MNRAS.526.5682C} (hereafter referred to as the \texttt{tSZ mask}), resulting in a fraction of the sky available for analysis being reduced to $f_{\rm sky}\approx0.55$. We also compared the tSZ power spectrum using the mask from the Planck 2015 analysis, but found no significant difference in the results. Therefore, to maximize the effective sky area, we chose the mask that removes fewer point sources. Additionally, uncertainties in the spectrum were derived directly from simulations in this study.

To minimize spurious signals from sharp edges and reduce mode coupling between different $\ell$-bins, we applied the \texttt{C2} scheme with an apodization scale of  $0.3^\circ$ to the \texttt{tSZ mask} before computing the power spectra. The \texttt{C2} function, denoted as $f$, is implemented in \texttt{NaMaster} and tapers the mask using a specified apodization scale, $\theta^*$, through
\begin{equation}
f=\left\{\begin{array}{cc}
\frac{1}{2}[1-\cos (\pi x)] & x<1\,, \\
1 & {\rm otherwise}\,.
\end{array}\right.
\end{equation}
Here $x=\sqrt{(1-\cos \theta) /\left(1-\cos \theta_{*}\right)}$ and $\theta$ represents the angular separation of a pixel from the nearest masked pixel.

\begin{figure}[htbp]
        \centering
        \includegraphics[width=0.45\textwidth]{ 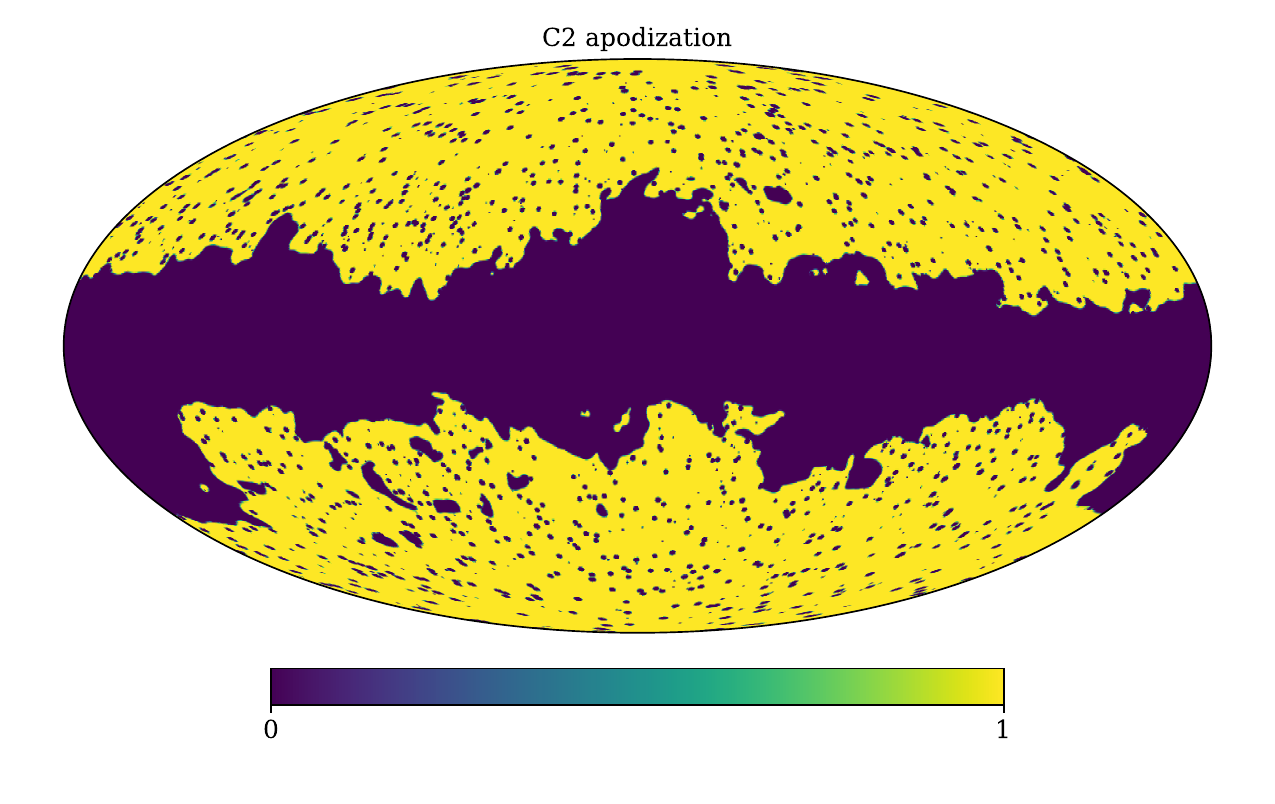}
        \caption{\texttt{tSZ mask} used in our analysis, combining the Planck Galactic mask and the point-source mask, leaving approximately $f_{\rm sky} \approx 55\%$ of the sky available for the tSZ reconstruction. For power spectrum computations, the mask was apodized using the \texttt{C2} apodization scheme in \texttt{NaMaster}, with a $0.3^\circ$ transition length.}
        \label{fig:mask}
    \end{figure}


\subsection{Simulations}

To validate our approach for separating the tSZ signal, we performed our investigations using sky simulations created with the {\red Python Sky Model} across all {\it Planck}  HFI channels. For these simulations, we utilized the latest version, \texttt{PySM3}~\citep{Thorne_2017,Zonca_2021}, which includes key sky components at microwave and millimeter frequencies,  such as the CMB signal, the tSZ effect, various contributions from galactic interstellar medium (ISM) emissions (including thermal and spinning dust, synchrotron, and free-free emissions), as well as emissions from point sources (both radio and infrared) and CO. {\red In particular, the extragalactic components in \texttt{PySM3} are derived from the \texttt{WebSky} simulations~\citep{2020JCAP...10..012S}}. The tSZ signal was constructed from the Compton-$y$ parameter map of WebSky\footnote{https://portal.nersc.gov/project/cmb/pysm-data/websky/0.4/}, applying the frequency dependency parameter for each channel while ignoring relativistic corrections. The CMB and noise maps from Full Focal Plane (FFP10) simulations are available at the {\it Planck}  Legacy Archive (PLA)\footnote{https://pla.esac.esa.int}, which cover both full and half mission datasets, respectively. 

The {\it Planck}  circular Gaussian beams and mask map were also taken into account to ensure that the {\it Planck}  simulation maps are as realistic as possible. The simulations were initially produced at $N_{\text {side }}=2048$, and we subsequently downgraded them to $N_{\text {side }}=1024$ for our analysis. This resolution is sufficient for accurately calculating the power spectrum down to scales of interest, up to $\ell \sim 1000$. In addition, due to the finite size of the \texttt{HEALPix} pixels, we accurately computed the angular power spectrum by incorporating corrections from the $\ell$-space window function. 

\begin{figure*}[htbp]
        \centering
        \includegraphics[width=0.95\textwidth]{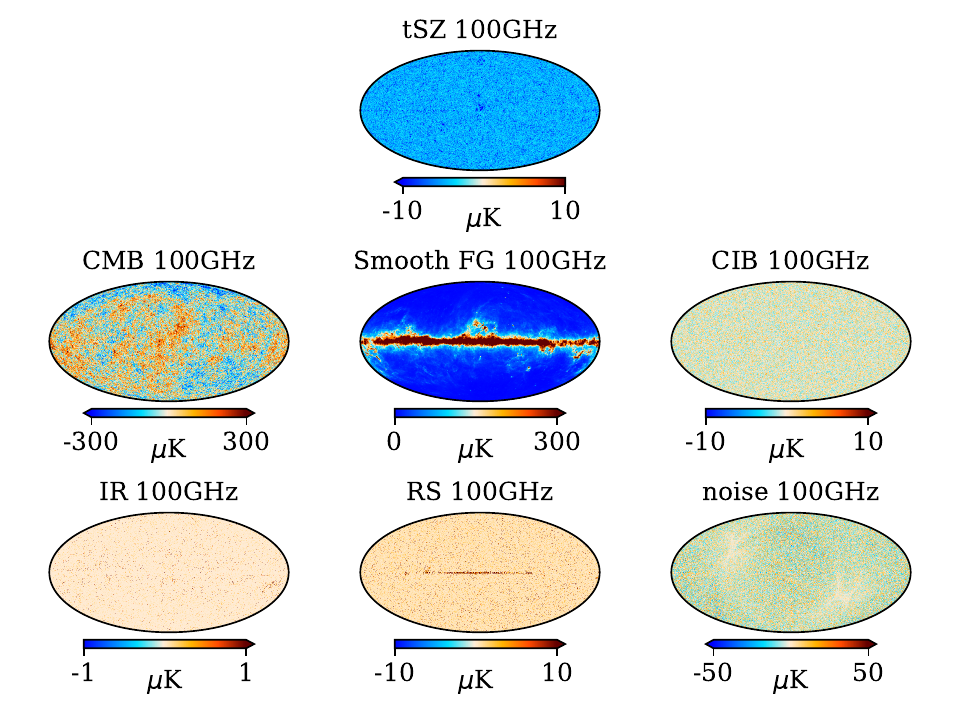}
        \caption{Comparison of simulated {\it Planck} maps ($N_{\rm side}=1024$) for the tSZ, CMB, foregrounds, and noise at 100 GHz, in units of $\mu{\rm K}$. The foreground components are generated using {\red \texttt{PySM3}}, whereas the CMB, instrumental noise, and  radio sources (RS) and {\red infrared point sources (IR)} are adopted from the PLA simulated data. The CMB and noise maps are taken from PLA realizations, with the noise corresponding to one {\it Planck} PR3 simulation. Note that the maximum and minimum values in the color bars are not the actual data limits but are adjusted to clearly highlight the features.}
        \label{fig:sim_100}
    \end{figure*}

{\red The simulated tSZ and foreground components, together with instrumental noise, of the {\it Planck} maps at 100 GHz are shown in Fig.~\ref{fig:sim_100}, where the CIB map traces the diffuse extragalactic infrared emission from unresolved distant galaxies, whereas the IR map represents extragalactic infrared point sources.}

\subsection{The {\it Planck}  data}

We utilized the publicly available total intensity dataset from the Low-Frequency Instrument (LFI; $<$100 GHz) and the High-Frequency Instrument (HFI; $\geq$100 GHz) collected over the full {\it Planck} mission. This dataset includes the nine frequency channel maps, spanning from 30 to 857 GHz, at their native resolution~\citep{2016A&A...594A..22P}. The complete {\it Planck}  datasets can be accessed through the {\it Planck}  Legacy Archive (PLA). To focus on the reconstruction of the tSZ signal, we also utilized publicly available mask map designed to discard regions of the sky strongly affected by point sources and galactic emissions. The FFP10 noise simulation maps at each frequency are also employed to estimate the noise variance, which is integral to our ABS signal separation method, for both full missions and half missions, respectively. The mask map and the noise realizations can be found at PLA. 
We employed {\it Planck}  circular Gaussian beams with Full Width at Half Maximum (FWHM) values derived from~\cite{2016A&A...594A...7P}, together with tSZ transmission values from~\cite{2016A&A...594A..22P,2014A&A...571A...9P} for the spectral bandpasses.  The characteristics of the {\it Planck}  maps are summarized in Tab.~\ref{tab:units}, presenting the conversion factors for the Compton parameter to CMB temperature for each frequency channel, based on integration over the instrumental bandpasses.

We chose to use the {\it Planck} PR3 dataset because it provides an improved estimate of the dipolar component due to the enhanced calibration. This refinement allows for a more accurate assessment of the signal at large angular scales. Furthermore, the inter-frequency calibration remains consistent with that of PR2, ensuring a precision level better than 1\%.

\begin{table}[!h]
         \caption{Conversion factors for the tSZ Compton parameter $y$ to CMB temperature units, along with the Full Width at Half Maximum (FWHM) of the {\it Planck}  channel beams.}
         \centering
         \begin{tabular}{crc}
          \hline\hline
            Frequency &  $ g (\nu)T_{\rm CMB}$ & FWHM\cr
            \omit\hfil$[$GHz$]$\hfil& $[{\rm K_{CMB}}]$ & [arcmin]\cr
            \hline
            30 & $-5.3364$ & 32.29\cr
            44 & $-5.1782$ & 27.00\cr
            70 & $-4.7662$ & 13.21\cr
            100& $-4.03121$& 9.66\cr
            143& $-2.78564$& 7.27\cr
            217& $ 0.18763$& 5.01\cr
            353& $ 6.20518$& 4.86\cr
            545& $14.45559$& 4.84\cr
            857& $26.33521$& 4.63\cr\hline\hline
         \end{tabular}
         \label{tab:units}
      \end{table}

\begin{figure*}[htbp]
        \centering
        \includegraphics[width=0.6\textwidth]{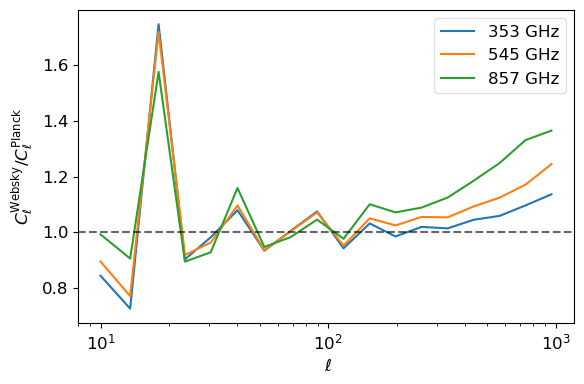}
        \caption{{\red Ratio of the CIB power spectra derived from the WebSky simulation and the {\it Planck} GNILC CIB maps, $C_\ell^{\mathrm{Websky}}/C_\ell^{\mathrm{Planck}}$, for the three available frequencies. The ratios are generally close to unity across most multipoles, indicating good agreement between the simulation and the {\it Planck} measurements.}}
        \label{fig:ratio}
    \end{figure*}

{\red To quantify the level of agreement between the simulation and observations more clearly, we compute the ratio of the CIB power spectra derived from the WebSky simulation and the {\it Planck} GNILC CIB maps. This ratio provides a straightforward way to highlight relative differences in amplitude across multipoles.}

{\red As seen in Fig.~\ref{fig:ratio}, at the lowest multipole bins, the ratios show relatively large fluctuations. For $\ell \gtrsim 50$, the ratios are typically within $\sim10\%$ of unity, indicating good agreement between the two datasets. Toward smaller angular scales ($\ell \gtrsim 300$), the ratio gradually increases, reaching $\sim1.2$--$1.35$, suggesting that the WebSky simulation predicts slightly higher small-scale power than the {\it Planck} CIB maps under our fiducial masking scheme. Overall, the agreement across the full multipole range is good, supporting the consistency of the WebSky CIB model with the Planck measurements.
}

\section{Reconstruction method}\label{sect:abs}
The reconstruction methods described in~\cite{2014A&A...571A..21P,2016A&A...594A..22P} employ two algorithms: MILCA~\citep{2013A&A...558A.118H} and NILC~\citep{2011MNRAS.410.2481R}. Both algorithms are based on the well-established ILC approach, which seeks to determine a linear combination of the input maps that minimizes the variance of the final reconstructed map while guaranteeing a unit gain for the target component. 

In contrast to MILCA and NILC, this study focuses on employing a novel source separation method--ABS--to estimate the tSZ power spectrum from the {\it Planck}  data. ABS utilizes a distinct foreground cleaning approach compared to the aforementioned algorithms and may provide an advantage for extracting weak signals. Below, we outline the ABS method in detail.

The ABS method~\citep{2018ApJS..239...36Y,10.1093/mnras/stz091,2021A&A...650A..65S} offers a blind and analytical approach to the source separation problem. It utilizes the measured cross‑bandpower between different frequency bands to analytically derive the bandpower of the desired signal, provided that the frequency dependence of the signal is precisely known. {\red ABS operates directly in the angular power spectrum domain and returns the signal power spectrum to be extracted.} 

Specifically, for estimating the tSZ bandpower--the power spectrum of the $y$ parameter as defined in Eq.~\ref{eq:tsz}--the measurement equation for the cross bandpower of temperature maps (in thermodynamic temperature units) at the $i$- and $j$-th frequency channels, denoted as $\mathcal{D}_{ij}^{\rm obs}(\ell)$, in the multipole bin $\ell$, can be expressed as follows:
\begin{eqnarray} \label{eq:abs}
        \mathcal{D}^{\rm obs}_{i j}(\ell) = 
        f_i f_j \mathcal{D}^{yy}(\ell) + \mathcal{D}^{\rm cmb}(\ell) +  \mathcal{D}^{\rm fore}_{i j}(\ell) + \delta \mathcal{D}^{\rm noise}_{i j}(\ell)\,.
    \end{eqnarray}
Here, $\mathcal{D}^{\rm fore}_{ij}$ denotes the cross bandpower matrix of the foreground, while $\mathcal{D}^{\rm cmb}$ represents the CMB cross bandpower, which remains constant across frequencies due to the properties of blackbody radiation. $\mathcal{D}^{yy}_{i j}(\ell)$  is the cross-power spectrum of tSZ Compton $y$-maps and $f_i$ is the frequency dependency parameter for the $i$-th frequency channel, which can be precisely computed. For the {\it Planck}  experiment, this parameter for each channel is shown in the second column of Tab.~\ref{tab:units}. 

The measured cross power spectrum is inevitably affected by instrumental noise, $\delta \mathcal{D}^{\rm noise}_{ij}$, which represents the fluctuations of the instrumental noise in the measurements. Note that the ensemble average of this noise has been implicitly subtracted beforehand. Moreover, we assume that the instrumental noise follows an uncorrelated Gaussian distribution with a mean of zero and root mean square (rms) levels of noise $\sigma^{\rm noise}_i$ for the $i$-th frequency channel. The
residual noise hence has the following properties:
    \begin{eqnarray}\label{eq:noise}
    &&\left<\delta \mathcal{D}_{i j}^{\rm noise}\right> =0\,,\nonumber \\
    &&\left<(\delta \mathcal{D}_{ij}^{\rm noise})^2\right> = \frac{1}{2} \sigma_{i}^{\rm noise} \sigma_{j}^{\rm noise} (1+\delta_{i j})\,.
    \end{eqnarray}
{\red Note that, because $\delta \mathcal{D}_{i j}^{\rm noise }$ is defined as the fluctuation of the noise band-power about its ensemble mean, its expectation is zero even for $i=j$. However, the variance of $\delta \mathcal{D}^{\rm noise }$ is nonzero and still contributes to the covariance. In addition, we have suppressed the $\ell$-dependence in Eq.~\ref{eq:noise} for simplicity; thus, in what follows, all terms related to band powers are understood to be 
$\ell$-dependent. }

Accounting for instrumental noise, ~\cite{10.1093/mnras/stz091} demonstrates that the signal power spectrum can generally be computed analytically using the following formula:
\begin{equation}\label{eq:abs1}
\mathcal{\hat{D}}^{yy}(\ell) = \left( \sum^{\tilde{\lambda}_{\mu}\geq \lambda_{\rm cut}} \tilde{G}^2_{\mu}\tilde{\lambda}_{\mu}^{-1}\right)^{-1} - \mathcal{S}\,.
\end{equation}
Here, we have introduced new variables, defined by
\begin{eqnarray}\label{eq:noiseD}
\mathcal{\tilde{D}}^{\rm obs}_{ij}\equiv \frac{\mathcal{D}^{\rm obs}_{ij}}{\sqrt{\sigma_{\mathcal{D},i}^{\rm noise}\sigma_{\mathcal{D},j}^{\rm noise}}} + \tilde{f}_i\tilde{f}_j\mathcal{S}\,,\\
 {\rm with}\quad\tilde{f_i} \equiv \frac{f_i}{\sqrt{\sigma_{\mathcal{D},i}^{\rm noise}}}\,,~\tilde{{\red G}}_{\mu}\equiv {\bf \tilde{f}}\cdot {\bf \tilde{E}}^\mu\,.
\end{eqnarray}
Here, the variance in each element of the residual noise matrix is represented by$ \sigma_{ij}^{\rm noise} \equiv \big< \left( \delta \mathcal{D}_{ij}^{\rm noise} \right)^2 \big>$. Furthermore, ${\bf \tilde{E}}^\mu$ and $\tilde{\lambda}_\mu$ denote the $\mu$-th eigenvector and its corresponding eigenvalue of $\mathcal{\tilde{D}}^{\rm obs}_{ij}$, respectively. Instrumental noise can introduce nonphysical eigenmodes characterized by eigenvalues of $|\tilde{\lambda}_\mu| \lesssim 1$ in $\mathcal{\tilde{D}}^{\rm obs}_{ij}$. The ABS method applies a threshold to the eigenvalues $\tilde{\lambda}_\mu$, retaining only the signal-dominated modes with $\tilde{\lambda}_{\rm cut} \gtrsim 1$ as recommended in \cite{2018ApJS..239...36Y}. {\red In this study, we set $\tilde{\lambda}_{\rm cut}=1$ for $\ell<250$ and adopt a slightly more aggressive mode-selection criterion, $\tilde{\lambda}_{\rm cut}=3$, for $\ell>250$. This choice is motivated by the observed drop-off and increased fluctuations in the recovered $yy$ power spectrum at intermediate multipoles for the real {\it Planck} data. We find that $\tilde{\lambda}_{\rm cut}=3$ significantly suppresses residual foreground and noise contamination, leading to a more stable recovered $C_\ell^{yy}$ in the range $250 \lesssim \ell \lesssim 500$, while having a negligible impact on the recovered signal at $\ell>500$ compared to $\tilde{\lambda}_{\rm cut}=1$.
}

In Eq.~\ref{eq:noiseD}, the parameter $\mathcal{S}$ shifts the amplitude of the input signal power spectrum. A significant positive shift is essential for stabilizing computations, ensuring that the signal modes within the eigenvector subspace are retained during the thresholding process. This ``shift'' strategy is particularly important in low signal-to-noise scenarios, where the shift value is typically chosen to be comparable to the signal level.  Specifically, for the tSZ reconstruction, we set $\mathcal{S} = 0.1$ for $10^{12}\ell(\ell+1)C^{yy}_\ell/(2\pi)$, observing no significant changes when increasing this value. Further discussion of ABS and how this relates to the ILC can be found in Appendix~\ref{sect:3}.   



\section{Power spectrum analysis for {\it Planck} PR3 Data}\label{sect:5}
\begin{figure}[htbp]
        \centering
        \includegraphics[width=0.48\textwidth]{ 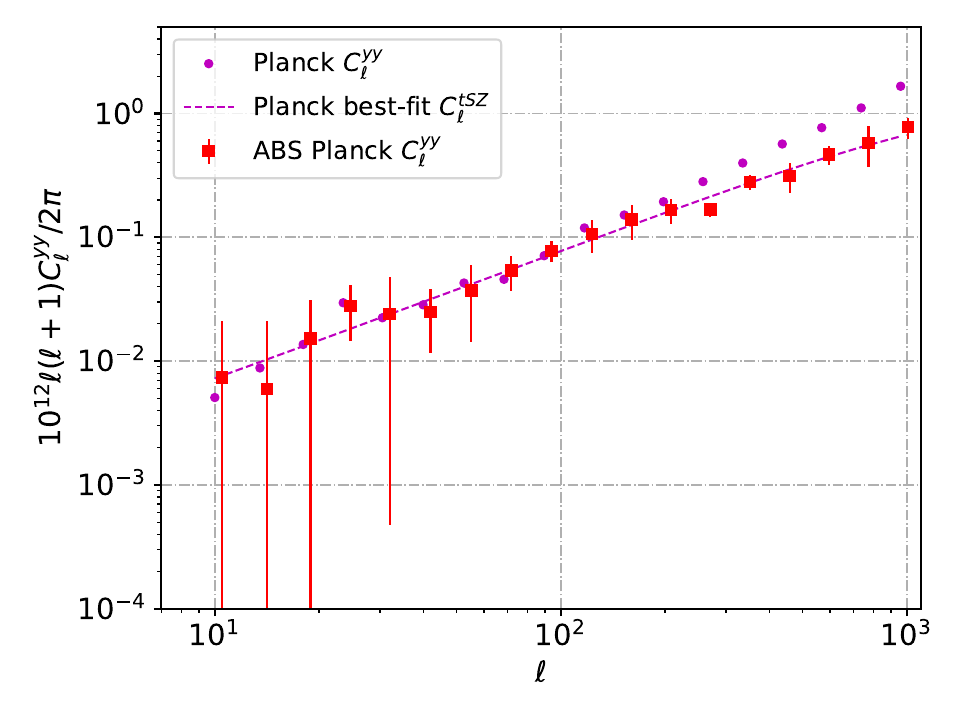}
        \caption{Comparison of the ABS-derived $y$-map power spectrum with {\it Planck} 2016 results. The ABS-derived $C_\ell^{yy}$ (red squares) is based on the PR3 full-mission data, where all nine frequency channels are used.  Also shown is the {\it Planck} F/L-estimated $C_\ell^{yy}$ (magenta dots). Note that we should directly compare the red squares and magenta dots, as they are derived from the foreground cleaning algorithms. Additionally, we show the {\it Planck} best-fit tSZ power spectrum (magenta dashed line), which is obtained through further residual foreground subtraction from the magenta dots by jointly fitting the signal and various modeled foreground residuals. When $\ell \lesssim 100$, the ABS results agree well with the {\it Planck} $C_\ell^{yy}$. However, at the higher multipoles, the ABS-derived amplitudes are lower than {\it Planck} $y$-map power spectrum, suggesting that the foreground contamination is substantially reduced by ABS. }
        \label{fig:planck_full}
    \end{figure} 

We now present the ABS results based on an analysis of the {\it Planck} full-mission data, utilizing all nine frequency channels (30–857 GHz). In Fig.~\ref{fig:planck_full}, the ABS-derived $y$-map power spectrum is compared with the {\it Planck} 2016 results, including the {\it Planck} F/L-estimated $C_\ell^{yy}$ (magenta dots) and the {\it Planck} best-fit tSZ power spectrum (magenta dashed), derived from a joint fit of the signal and various foreground residuals. All error bars represent the $2\sigma$ statistical uncertainties, estimated from simulations.



As shown, the ABS results are broadly consistent with the \textit{Planck} F/L estimate on large angular scales ($\ell \lesssim 100$). At higher multipoles, the ABS-derived $C_\ell^{yy}$ amplitudes are systematically lower than those obtained from the \textit{Planck} NILC--MILCA $y$-map, where residual foreground contamination is known to be more significant.

{\red At $\ell \gtrsim 250$, the ABS-reconstructed power spectrum departs from the NILC--MILCA estimates, yielding reduced power. This behavior is consistent with the expectation that ABS intrinsically suppresses residual foreground contributions, even without applying any additional post-processing or explicit foreground-residual subtraction. We emphasize that this comparison is intended to illustrate the relative level of foreground suppression achieved by ABS, rather than to suggest that any particular reference spectrum represents the true underlying $yy$ signal.}

{\red
At these high multipoles, instrumental noise also becomes increasingly important. Nevertheless, the ABS method remains effective in jointly mitigating noise amplification and suppressing foreground leakage. If any residual contamination remains in the ABS reconstruction, it would bias the recovered $C_\ell^{yy}$ amplitude high, implying that the true underlying tSZ power would be equal to or smaller than the ABS-derived values.}

{\red For the highest six multipole bins, the absolute deviations with respect to the {\it Planck} best-fit curve are 0.036, 0.016, 0.026, 0.037, 0.043, and 0.121, respectively, corresponding to absolute fractional differences of approximately 17.5\%, 5.9\%, 7.8\%, 8.5\%, 8.0\%, and 18.5\%. The reduced amplitudes recovered by ABS--obtained without any explicit subtraction of residual foregrounds--demonstrate that the algorithm achieves efficient intrinsic suppression of foreground contamination.}

{\red Subtracting the mean noise power spectrum from the auto-power spectrum may introduce a systematic bias if the noise level is imperfectly modeled. In modern CMB analyses, this issue is often mitigated by using cross-split power spectra, which remove the noise bias without requiring explicit subtraction. Applying the ABS pipeline to cross-split spectra will be explored in future work.
To assess the impact of possible noise misestimation in the present analysis, we varied the assumed noise power spectrum by $\pm10\%$ and propagated the change through the ABS pipeline. The recovered $C_\ell^{yy}$ varies approximately linearly with the noise level. In the signal-dominated regime ($\ell \lesssim 200$), a $10\%$ noise uncertainty leads to a comparable $\sim10\%$ variation in $C_\ell^{yy}$. At higher multipoles, where the measurement becomes increasingly noise-dominated, the sensitivity increases; for example, at $\ell\sim500$, a few-percent noise uncertainty can induce a $\sim10\%$ change in the recovered signal.}


The next step is to simultaneously fit both the signal and residuals, similar to the analysis in~\cite{2014A&A...571A..21P,2016A&A...594A..22P}, using the maximum likelihood estimation to provide a robust estimate of the tSZ spectrum.

\subsection{Maximum Likelihood Analysis}
In the following, we will perform a joint analysis to accurately estimate the tSZ power spectrum by jointly fitting the tSZ and foreground residual models.

Building upon the analysis of the marginalized band-powers of the {\it Planck} tSZ power spectrum~\citep{2014A&A...571A..21P,2016A&A...594A..22P}, we fit the measured $C^{yy,\rm obs}_\ell$ by considering both the tSZ component and three residual foreground components: the cosmic infrared background (CIB), radio sources (RS), infrared point sources (IR), and a correlated noise (CN) term. Therefore, the predicted total contribution to the measurement, $C_{\ell}^{yy,\rm pred}$, can be modeled as:
\begin{eqnarray} \label{eq:model} 
     C_{\ell}^{yy,\rm pred} = &&A_{\rm tSZ} \hat{C}_{\ell}^{\rm tSZ} + A_{\rm CIB} \hat{C}_{\ell}^{\rm CIB} + A_{\rm IR} 
     \hat{C}_{\ell}^{\rm IR} + A_{\rm RS} \hat{C}_{\ell}^{\rm RS} \nonumber\\
     &&+ A_{\rm CN} \hat{C}_{\ell}^{\rm CN}\,.
    \end{eqnarray}
For simplicity, we do not vary cosmological parameters such as $\Omega_m$ and $\sigma_8$ when generating the tSZ power spectrum. Instead, the band powers of $\hat{C}_{\ell}^{\rm tSZ}$ are fixed using an arbitrary template, and only the normalization factor, $A_{\rm tSZ}$, is varied during the fitting procedure. The templates for the various residual foreground components are represented by $\hat{C}_{\ell}^{\rm CIB}$, $\hat{C}_{\ell}^{\rm IR}$, $\hat{C}_{\ell}^{\rm RS}$, and $\hat{C}_{\ell}^{\rm CN}$, respectively. {\red Moreover, since the correlated noise term dominates over the other components at high multipoles, following~\citep{2018MNRAS.477.4957B}, we use the highest multipole data at $\ell=2742$ to determine $A_{\mathrm{CN}}$, which gives $A_{\rm CN} = 0.26$. We clarify that, beyond our baseline analysis limited to $\ell \le 1000$, we performed a dedicated power-spectrum estimation up to $\ell_{\rm max} \simeq 3\,N_{\mathrm{side}} \approx 3000$ in order to include modes up to $\ell = 2742$ required for the determination of $A_{\mathrm{CN}}$. The corresponding $\hat{C}_{\ell}^{\rm CN}$ term is derived from an empirical model, as described in~\citep{2018MNRAS.477.4957B}. } We verified that allowing $A_{\rm CN}$ to vary as a free parameter in the range of $[0,2]$ has a negligible impact on the estimation of the tSZ amplitude. 

Since our foreground cleaning algorithm differs from the one used by {\it Planck}, the residuals are not directly comparable to the templates provided by {\it Planck}. To estimate the angular power spectra of these residuals, we use the following procedure below. For a given foreground component (denoted as ``$X$'', where $ X\in\{\rm CIB, IR, RS\}$), we start from the simulated multi-frequency component maps. {\red For the IR component, we include both faint and bright point infrared sources, and for the RS component, we consider radio emissions from galaxy clusters as well as faint and bright radio point sources.} All maps are derived from publicly available {\it Planck} simulation data provided through the PLA.

We then compute the frequency–frequency cross-angular power spectra for all frequency pairs, denoted as $\mathbf{C}^X$. To determine the template for each residual foreground component, we employ an analytical estimation by introducing a small perturbation in the data. Based on the ABS estimator, by adding $\epsilon \mathbf{C}^{X}$ to the data, the first-order perturbation in the estimated band power spectrum due to foreground $X$ can serve as the template for $X$. This increase in the ABS-estimated power spectrum is denoted as $\epsilon\hat{C}_\ell^{X}$, and the template is thus given by (see Appendix~\ref{sect:abs-template} for details):  
\begin{equation}
\hat{C}^X_\ell = \frac{\tilde{\mathbf{f}}^T \tilde{\mathbf{C}}^{\dagger} \mathbf{C}^{X} \tilde{\mathbf{C}}^{\dagger} \tilde{\mathbf{f}}}{\left(\tilde{\mathbf{f}}^T \tilde{\mathbf{C}}^{\dagger} \tilde{\mathbf{f}}\right)^2}\,,
\end{equation}
where $\hat{C}^X_\ell$ serves as the constructed template model used in Eq.~\ref{eq:model}.


{\red Furthermore, we find the residual Galactic diffuse emission is sub-dominant at $\ell \gtrsim 200$, consistent with \citep{2023MNRAS.526.5682C}, and the CMB residual power is also sub-dominant, allowing these components to be neglected in the fitting. This is because the adopted Galactic mask efficiently suppresses large-scale foreground emission, and the parameter constraints are dominated by high-$\ell$ modes, where both the diffuse Galactic foreground and CMB residuals are intrinsically small and carry little statistical weight compared to the tSZ signal.} 

We consider an effective multipole range of $10\leq\ell_{\rm eff}\leq 959.5$, with the effective multipoles defined as the midpoints of the eighteen bins within this range, following the approach used in the {\it Planck} tSZ analysis. The parameter space is explored using the Markov Chain Monte Carlo (MCMC) method, from which we obtain the joint posterior probability distributions of the parameters. During the MCMC analysis, we apply uniform priors to each parameter. The chosen parameter ranges are sufficiently broad that altering the upper or lower bounds does not notably impact the resulting posterior distributions.

The likelihood is computed as follows:  
    \begin{equation} \label{eq:likelihood}
    -2\ln\mathcal{L}=\chi^{2}+\ln|\mathbf{M}|+{\rm const.}\,, 
    \end{equation}
where the $\chi^2$ value is defined as 
    \begin{equation} \label{eq:chi2}
        \chi^2= \left(\mathbf{C}^{\rm obs}-\mathbf{C}^{\rm pred}\right)^T \mathbf{M}^{-1}\left(\mathbf{C}^{\rm obs}-\mathbf{C}^{\rm pred}\right)\,,
    \end{equation}
and the vectors $\mathbf{C}^{\rm obs}$ and $\mathbf{C}^{\rm pred}$ collect all $C^{yy}_\ell$ values for the measured {\it Planck} data and the predicted band powers based on our model, respectively. The matrix $\mathbf{M}$ represents the covariance matrix of the ABS-recovered $C^{yy}_\ell$ values, estimated from the {\it Planck} simulations. This matrix remains fixed throughout the MCMC run. 

As discussed in~\cite{2018MNRAS.477.4957B}, the trispectrum can dominate over the Gaussian term from instrumental noise at low multipoles. However, this contribution is not included in the {\it Planck} tSZ analysis. Neglecting the trispectrum in the covariance matrix leads to an underestimation of the error bars, particularly for $\ell \lesssim 100$. To assess its impact, we exhaustively consider three cases for the tSZ amplitude fitting: 

1) the \textbf{ trispectrum-excluded } case, consistent with the {\it Planck} analysis, where the trispectrum contribution to the covariance is ignored; 

2) a \textbf{high-$\ell$} case, identical to the  trispectrum-excluded  case but with the additional exclusion of all data points with $\ell < 100$

3) a \textbf{trispectrum-included} case, in which the trispectrum is explicitly incorporated into the covariance matrix, in contrast to the first case.

In general, the elements of the covariance matrix are given by
\begin{equation}\label{eq:Mll}
M_{\ell\ell'} = \left(\sigma_{\ell}^{yy}\right)^2 \delta_{\ell\ell'} + \frac{\ell (\ell+1) \ell' (\ell'+1)}{4 \pi^2} \frac{T_{\ell\ell'}}{4 \pi f_{\mathrm{sky}}},
\end{equation}
where the Gaussian term $\sigma_{\ell}^{yy}$ represents the standard deviation of the measured band pwoers $C^{yy}_\ell$. This term accounts for contributions from sampling variance, Gaussian instrumental noise, and residual fluctuations from foreground removal. Each of these contributions is estimated from simulations. The bin-to-bin correlations are negligible, owing to the use of wide bin widths and the effective correction of mode coupling induced by sky masking, which is achieved through \texttt{NaMaster}. {\red In the first term of Eq.~\ref{eq:Mll}, we assume that the foreground-residual contribution to the covariance matrix $M_{\ell\ell'}$ is diagonal. We explicitly validate this assumption by constructing an ensemble of foreground realizations in which the amplitudes of individual foreground components are varied by 5\%. The resulting covariance matrix is found to be nearly diagonal, and including the off-diagonal terms leads to a negligible change in the inferred tSZ power spectrum compared to its statistical uncertainty.}

The trispectrum is computed following the approach outlined in \cite{2002MNRAS.336.1256K}. It is given by  
\begin{equation}  
T_{\ell \ell^{\prime}} = \int \mathrm{d} z \, \frac{d V}{d z d \Omega} \int \mathrm{d} \ln M \, \frac{d n}{d \ln M} \left|y_{\ell}(M, z)\right|^2 \left|y_{\ell^{\prime}}(M, z)\right|^2,  
\end{equation}  
where $y_\ell$ represents the two-dimensional Fourier transform of the electron pressure profile, and $dn/dM$ is the halo mass function. For the numerical computation of the trispectrum, we utilize the public code \texttt{class\_sz}\footnote{\url{https://github.com/CLASS-SZ}}, with the detailed parameters for the calculations provided {\red in~\cite{2018MNRAS.477.4957B,2023JCAP...03..039B,2023arXiv231018482B,2025arXiv250707346B}.  }

\begin{figure*}[htpb]
        \centering
\includegraphics[width=0.45\textwidth]{ 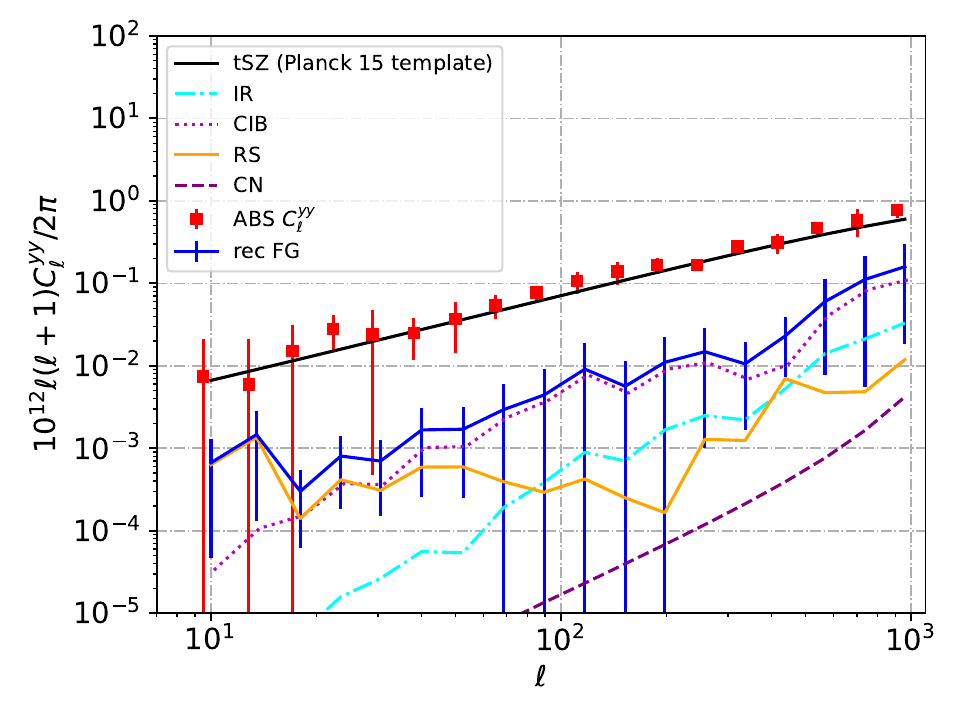}
\includegraphics[width=0.45\textwidth]{ 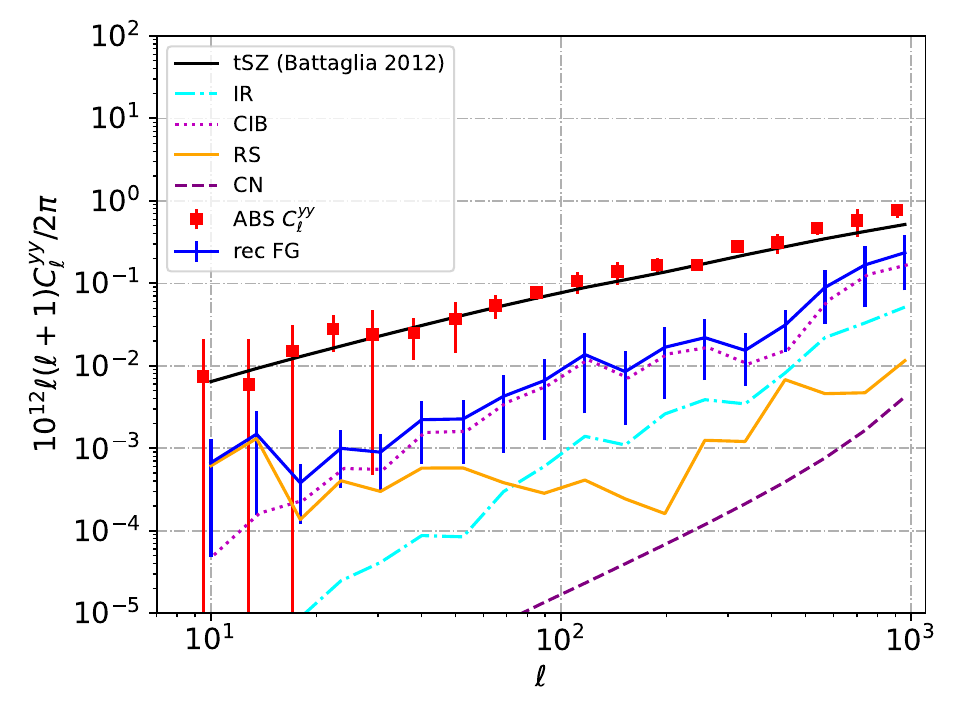}
        \caption{Best-fit tSZ power spectrum (solid black) and the power spectrum of the total residual components (labeled as ``rec FG''; solid blue), along with individual component power spectra, including the CIB, IR, RS, and CN components, for the  trispectrum-excluded  case. Same as in Fig.~\ref{fig:planck_full}, we show the measured $y$-map power spectrum from ABS (red squares), with error bars indicating the $2\sigma$ confidence level.  {\it Left}: the tSZ power spectrum template is chosen based on the best-fit {\it Planck} 2015 results from the joint model fitting, which is used for the further foreground subtraction.  {\it Right}: the fitting is based on the model from hydrodynamic simulations~\citep{2012ApJ...758...74B}. }
        \label{fig:tsz_all}
    \end{figure*} 
    
\begin{figure}[htpb]
        \centering
\includegraphics[width=0.45\textwidth]{ 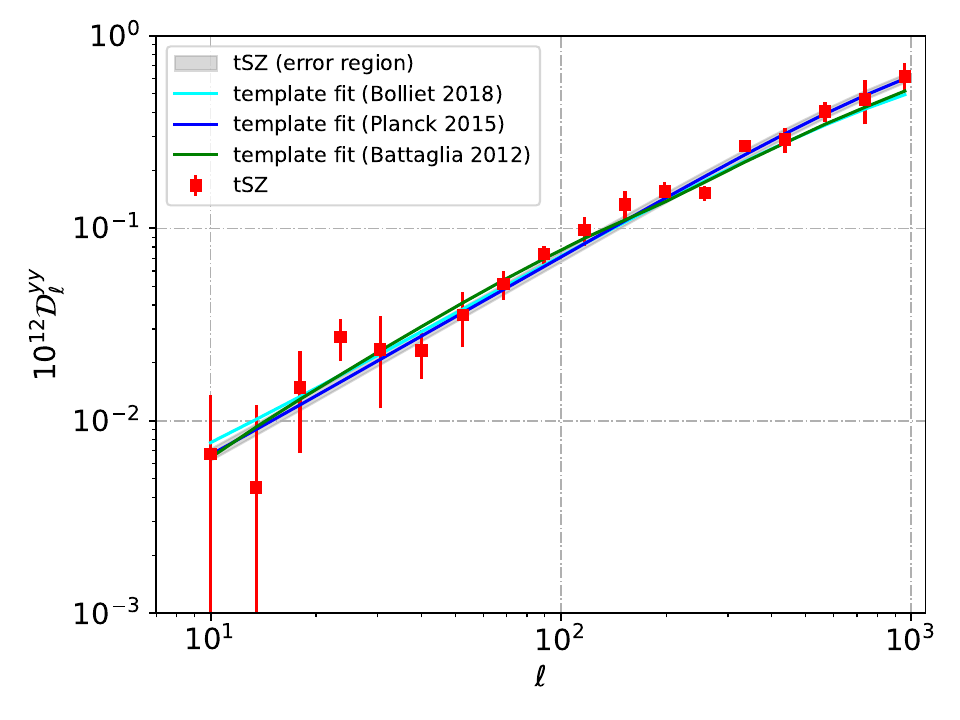}
\includegraphics[width=0.45\textwidth]{ 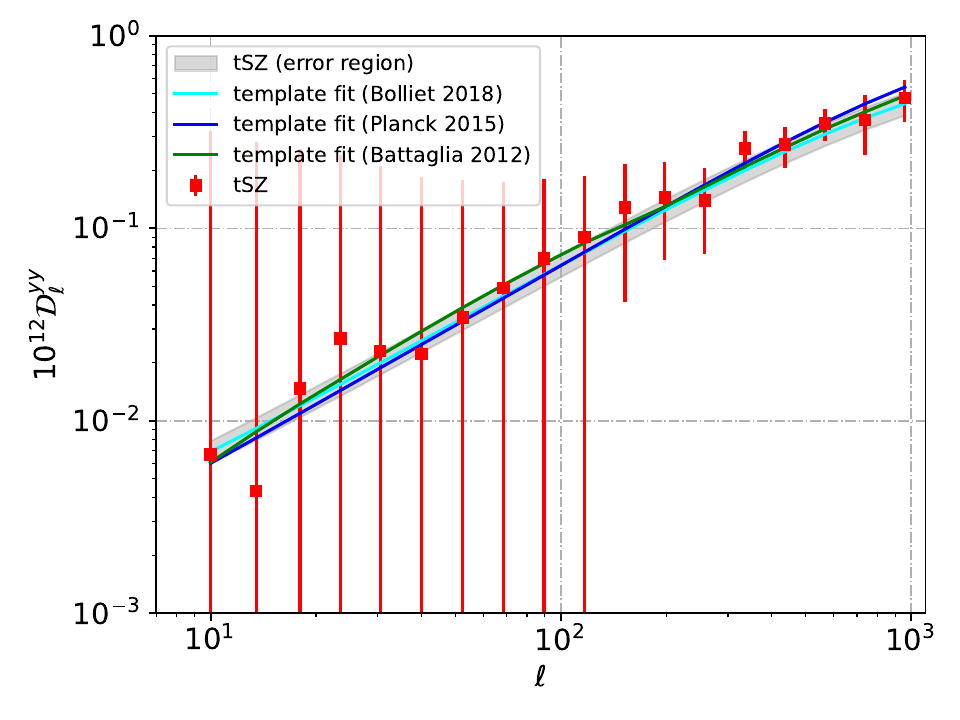}
        \caption{ Comparison of the tSZ power spectrum (red squares) between the  trispectrum-excluded  (upper) and trispectrum (lower) cases. The tSZ band powers are obtained by subtracting residual foreground contributions from the ABS-derived $y$-map power spectrum, using the best-fit foreground parameters (see Eq.~\ref{eq:model}). Error bars represent the 1$\sigma$ confidence interval, estimated from both the statistical uncertainties and the uncertainties induced by residual foregrounds. Rescaled tSZ power spectrum templates from the literature are also shown for comparison: cyan (``Bolliet 2018'';~\cite{2018MNRAS.477.4957B}), blue (``Planck 2015'';~\cite{2016A&A...594A..22P}), and green (``Battaglia 2012'';~\cite{2012ApJ...758...74B}). The scaling of each template is based on its respective best-fit $A_{\rm tSZ}$ value. The grey region indicates the allowed $\pm 1\sigma$ interval, estimated from the uncertainty in $A_{\rm tSZ}$.} 
        \label{fig:marg_tsz}
    \end{figure}

For the trispectrum-excluded case, in Fig.~\ref{fig:tsz_all}, we present our best-fitting results for the tSZ power spectrum, along with the individual foreground residual components and their sum, all shown with their associated {\red $2\sigma$} uncertainty levels. For comparison, two different tSZ templates are used to fit the overall amplitude $A_{\rm tSZ}$. The left panel shows the template from the {\it Planck} 2015 best-fit model, while the right panel displays the model derived from hydrodynamic simulations~\citep{2012ApJ...758...74B}. As seen, the CIB power spectra in both cases exhibit comparable amplitudes, and  essentially dominate the foreground residual contamination. The total contribution from the residuals (including the CN term) is represented by the blue solid line. In both cases, the amplitudes (denoted as ``rec FG'') and error bars are nearly identical, with changes in the mean amplitude and mean error bars being less than 5\%. This indicates that the results for the total residual foreground contribution are largely insensitive to the choice of the tSZ template. Moreover, in this study, we also consider the tSZ template based on the best-fit power spectrum obtained through a re-analysis of the {\it Planck} data~\citep{2018MNRAS.477.4957B}. The marginalized tSZ power spectrum, after being corrected for the residual foreground contributions, is shown in Fig.~\ref{fig:marg_tsz}.

         
\begin{figure*}[htpb]
        \centering
\includegraphics[width=0.45\textwidth]{ 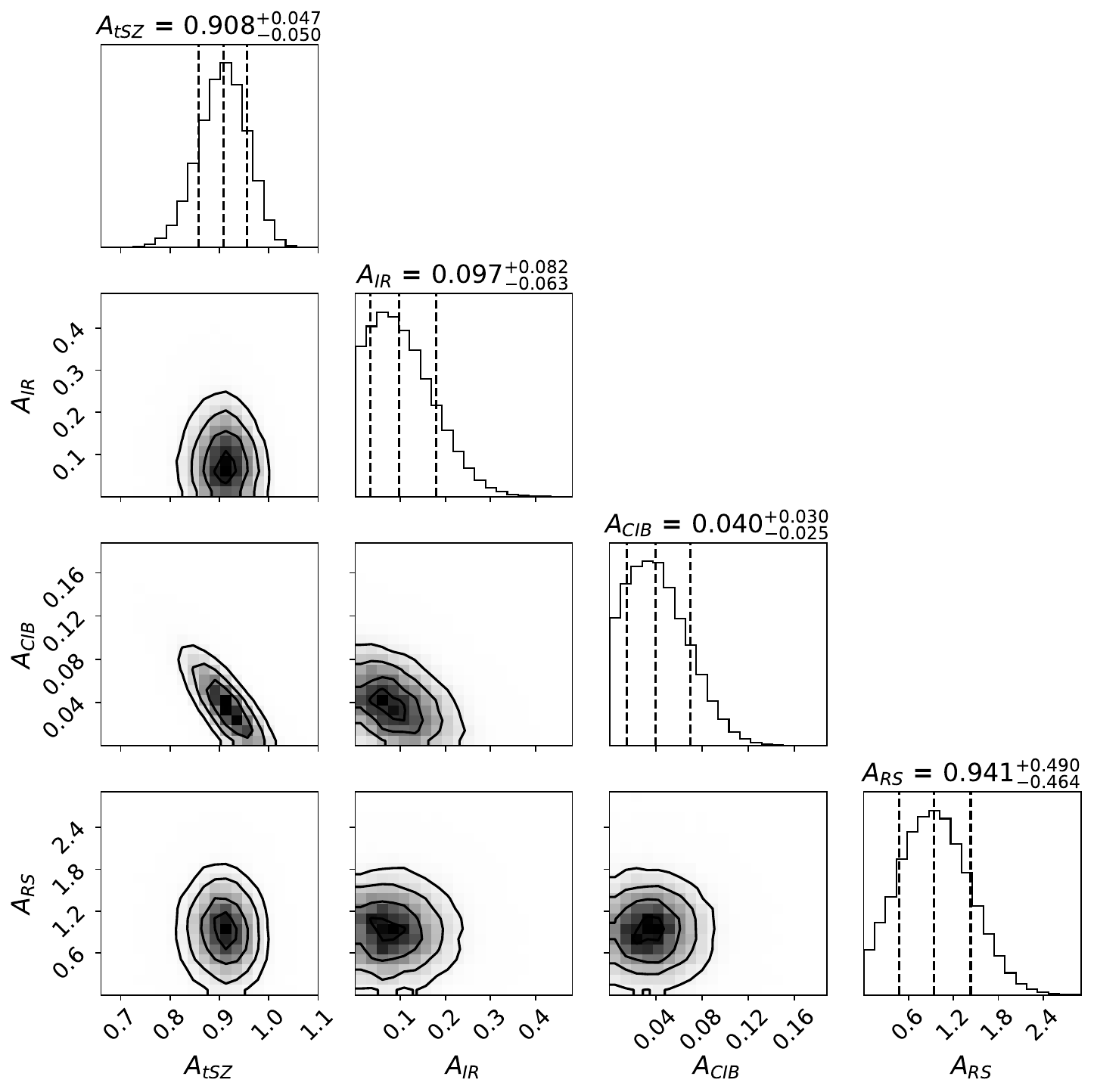}
\includegraphics[width=0.45\textwidth]{ 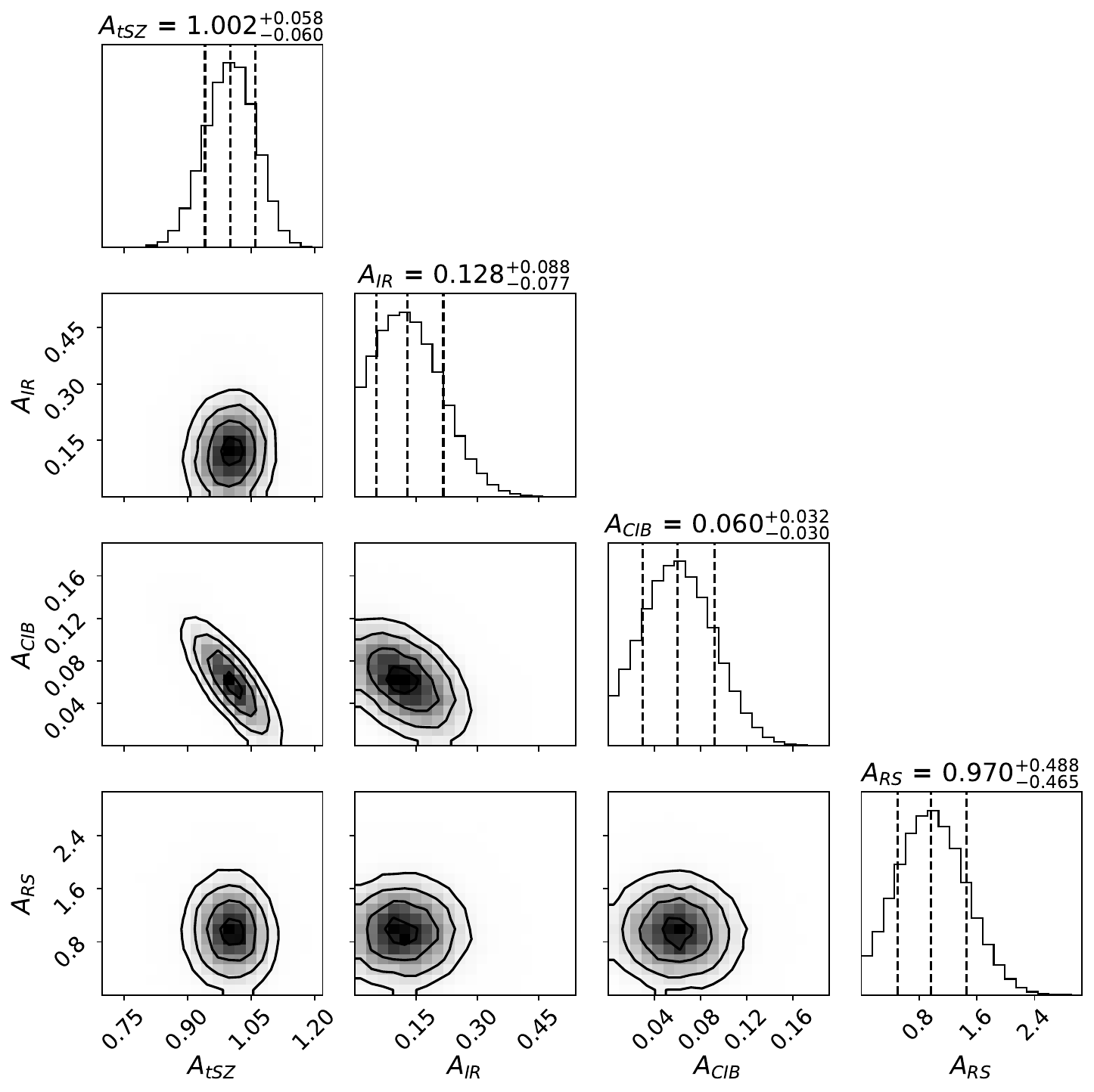}
        \caption{Marginalized (1D and 2D) joint posterior probability distributions of the tSZ amplitude and residual foreground components (CIB, IR, and RS), as defined in Eq.~\ref{eq:model}. The left panel displays results based on the {\it Planck} 15 best-fit tSZ power spectrum template, while the right panel corresponds to the model derived from hydrodynamic simulations, as shown in Fig.~\ref{fig:tsz_all}. The off-diagonal panels show the joint and marginalized constraint contours at $0.5, 1.0, 1.5, 2.0\sigma$ for the parameters, while the diagonal panels present the median and the $68\%$ credible intervals for the 1D marginalized distributions of each parameter.}
        \label{fig:tsz_contour}
    \end{figure*}

Furthermore, in Fig.~\ref{fig:tsz_contour}, the derived posterior distributions on our parameter sets are displayed, where the left and right panels showing the results for choosing the two different tSZ templates. {\red As shown, the median normalization amplitudes, $A_{\rm tSZ}=0.91$ and $1.0$ for the two cases, indicate that the marginalized amplitudes are respectively slightly lower than--yet fully consistent within $2\sigma$ uncertainties with--and well consistent with those of the corresponding tSZ templates.}. We also observe a strong correlation between the CIB and tSZ amplitudes, with the contributions from IR and RS being notably smaller than those of CIB. {\red In both cases, the mean total residual foreground contribution to $C^{yy}_\ell$, averaged over all $\ell$ bins, is about $0.03$.}

For the trispectrum-included case, the reconstructed band powers of the tSZ signal and various foreground residuals are presented in Appendix~\ref{app:tri}. As observed, the trispectrum's contribution reduces the weighting of low-$\ell$ data, causing the highest three $\ell$-bins to dominate the fitting process. This is due to increased uncertainty at lower multipoles and the trispectrum impacts on the signal.  

Fig.~\ref{fig:marg_tsz} shows the marginalized tSZ power spectrum after accounting for residual foreground contributions. The red data points represent the measured band powers, which were obtained after subtracting the sum of the foreground residuals. The associated 1$\sigma$ uncertainty is given by the sum of the statistical uncertainty and the uncertainties marginalized over foreground residuals. The gray region indicates the $1\sigma$ statistical uncertainty of the tSZ amplitude, derived from the posterior distribution of the parameter $A_{\rm tSZ}$.  For comparison, the rescaled tSZ power spectra, normalized by the best-fit amplitude parameter $A_{\rm tSZ}$, are also plotted: the solid cyan line represents ``Bolliet 2018''~\citep{2018MNRAS.477.4957B}, blue represents ``Planck 2015''~\citep{2016A&A...594A..22P}, and green represents ``Battaglia 2012''~\citep{2012ApJ...758...74B}, with each template corresponding to the best-fit power spectrum in the literature.

\FloatBarrier 
\begin{table*}[htbp]
\centering
\caption{Summary of the amplitude normalization parameters $A_{\rm tSZ}$ for the different cases, based on the three tSZ templates used for fitting. The table reports the marginalized posterior mean values (with standard deviations), the corresponding best-fit values, and {\red the minimum chi-square values, $\chi^2_{\rm min}$.} }
\label{tab:As_combined}
\resizebox{\textwidth}{!}{
\begin{tabular}{l|cc|cc|cc}
\hline
\multirow{2}{*}{Parameter} & \multicolumn{2}{c|}{\textbf{ trispectrum-excluded }} & \multicolumn{2}{c|}{\textbf{high-$\ell$}} & \multicolumn{2}{c}{\textbf{trispectrum-included}} \\
\cline{2-7}
 & Mean $\pm 1\sigma$ & Best fit & Mean $\pm 1\sigma$ & Best fit & Mean $\pm 1\sigma$ & Best fit \\
\hline
$A_{\rm tSZ}$ (``Planck 2015'') & $0.91 \pm 0.05$ & 0.92 ($\chi_{\rm min}^2 =19.2$) & $0.89 \pm 0.06$ & 0.90 ($\chi_{\rm min}^2 =9.6$) & $0.80 \pm 0.12$ & 0.83 ($\chi_{\rm min}^2 =10.5$) \\
$A_{\rm tSZ}$ (``Battaglia 2012'') & $1.0 \pm 0.06$ & 1.0 ($\chi_{\rm min}^2 =17.7$) & $0.99 \pm 0.07$ & 1.02 ($\chi_{\rm min}^2 =9.0$) & $0.92 \pm 0.14$ & 0.94 ($\chi_{\rm min}^2 =9.9$)  \\
$A_{\rm tSZ}$ (``Bolliet 2018'') & $1.34 \pm 0.08$ & 1.34 ($\chi_{\rm min}^2 =19.1$)  & $1.31 \pm 0.08$ & 1.34 ($\chi_{\rm min}^2 =10.2$)  & $1.18 \pm 0.16$ & 1.20 ($\chi_{\rm min}^2 =11.0$) \\
\hline
\end{tabular}
}
\end{table*}
{\red Tab.~\ref{tab:As_combined} summarizes the tSZ amplitude normalization parameters $A_{\rm tSZ}$ for different cases based on three fitting templates, reporting the marginalized posterior means (with standard deviations), best-fit values, and minimum chi-square values $\chi^2_{\rm min}$. With 18 bandpowers and 4 fitted parameters for the signal and foregrounds, the number of degrees of freedom is $N_{\rm dof}=14$, 5, and 14 for the three cases shown from left to right. For a standard $\chi^2$ distribution, the corresponding $2\sigma$ ranges are $[5.63,\,26.12]$ for $N_{\rm dof}=14$ and $[0.83,\,12.8]$ for $N_{\rm dof}=5$. The $\chi^2$ values obtained for the trispectrum-excluded case and for the high-$\ell$ analysis fall within these intervals, indicating statistically acceptable fits under the Gaussian covariance assumption. Moreover, when the tSZ trispectrum is included in the covariance, the resulting $\chi^2$ remains within the conventional $2\sigma$ range but is reduced relative to the Gaussian-covariance case.}

Specifically, for the  trispectrum-excluded  case, we obtain best-fit values of {\red $A_{\rm tSZ} = 0.92$ }for the ``Planck 2015'' template, {\red $A_{\rm tSZ} = 1.0$} for ``Battaglia 2012'', and {\red $A_{\rm tSZ} = 1.34$} for ``Bolliet 2018''. These results indicate that the amplitude of our reconstructed tSZ power spectrum is approximately 8\% lower than the ``Planck 2015'' template, about well consistent with the ``Battaglia 2012'' template, and roughly 34\% higher than the ``Bolliet 2018'' template. We also observe the minimum $\chi^2$ values in the fitting: {\red $\chi^2_{\rm min} = 19.2$, 17.7, and 19.1} for the ``Bolliet 2018'', ``Planck 2015'', and ``Battaglia 2012'' templates, respectively. Considering $N_{\rm dof} =14$.  Overall, our analysis favors the ``Battaglia 2012'' template. {\red The results are consistent with the ``Planck 2015'' template at the $2\sigma$ level, but show a deviation exceeding $2\sigma$ when compared with  ``Bolliet 2018'' template.}

For the high-$\ell$ case, we find slightly lower values of the tSZ amplitude $A_{\rm tSZ}$, ranging from {\red 0.89 to 1.31} across the three templates. {\red These results are largely consistent with those obtained in the trispectrum-excluded case}. In addition, the error bars are slightly larger than in the trispectrum-excluded case, as expected due to the reduced number of data points used in the fit.

{\red When the trispectrum contribution is included in the covariance, the best-fit results generally prefer lower tSZ amplitudes. The fitted amplitude is in good agreement with the ``Battaglia 2012'' template, remaining consistent within the $1\sigma$ level. Compared to the ``Planck 2015'' template, the recovered amplitude is approximately $20\%$ lower, but still consistent within $2\sigma$. In contrast, the best-fit amplitude is about $20\%$ higher than the prediction of the ``Bolliet 2018'' template, corresponding to a deviation at the $\sim1.2\sigma$ level.} Moreover, the uncertainties from the trispectrum dominate across the entire $\ell$ range, leading to error bars on $A_{\rm tSZ}$ that are approximately two times larger than those in the other two cases.


{\red Recently, small-scale constraints on the tSZ power spectrum have been reported by ground-based experiments. The SPT collaboration measures a tSZ amplitude of $10^{12} \ell(\ell+1)/(2\pi) C^{yy}_\ell = 3.74 \pm 0.54~\mu{\rm K}^2$ for the band centered at $\ell = 3000$ at 143 GHz~\cite{2021ApJ...908..199R}, while the ACT collaboration finds $3.3 \pm 0.4~\mu{\rm K}^2$ at 150~GHz~\cite{AtacamaCosmologyTelescope:2025blo,Beringue:2025bur}. These measurements correspond to Compton-$y$ band powers of $ 0.46 \pm 0.07$ for SPT and $0.49 \pm 0.06$ for ACT, respectively. 
In addition, \citep{Efstathiou:2025ckq} reconstructed the tSZ power spectrum by jointly combining the {\it Planck}, ACT, and SPT likelihoods, adopting a more conservative analysis strategy for multipoles $\ell \gtrsim 200$. Their reconstructed tSZ spectrum yields an overall higher amplitude than that inferred by~\cite{2018MNRAS.477.4957B}, but remains marginally consistent at the $2\sigma$ level. In the analysis including the trispectrum contribution, we find that the reconstructed tSZ amplitude is approximately 20\% higher than the best-fit spectrum reported by \citep{2018MNRAS.477.4957B}. However, the inferred amplitude remains lower than that reported by \citep{Efstathiou:2025ckq}. At $\ell \simeq 1000$, we measure $10^{12}\ell(\ell+1)/(2\pi)C^{yy}_\ell = 0.45 \pm 0.06$ by using the ``Bolliet 2018'' template. This value is lower to their reconstruction, but lies slightly beyond the $2\sigma$ consistency level.}


Furthermore,~\cite{2002MNRAS.336.1256K,2018MNRAS.477.4957B} demonstrate that the scaling of the tSZ power spectrum can be effectively approximated by 
\begin{equation}
C_{\ell}^{\rm tSZ} \propto \sigma_8^{8.1} \Omega_{m}^{3.2} B^{-3.2} h^{-1.7}\quad {\rm for} \quad \ell \lesssim 10^3\,,
\end{equation}
where $B$ represents the mass bias. The dependence on $\sigma_8$ and $\Omega_m$ are consistent with the findings of~\cite{2016A&A...594A..22P}. Therefore, the scaling can be approximately expressed as $C^{\rm tSZ}_{\ell} \propto S_8^8$, assuming all other parameters are fixed. 

{\red For the case without the trispectrum contribution, adopting the ``Planck 2015'' template, the best-fit normalization is $A_{\rm tSZ} = 0.91 \pm 0.05$, corresponding to a $\sim 9\%$ lower amplitude relative to the template. This implies a fractional shift in $S_8$ of $\Delta S_8 / S_8 \simeq (1/8) (\Delta A_{\rm tSZ}/A_{\rm tSZ}) \approx 1.1\%$, with a $1\sigma$ uncertainty of $\sim 0.7\%$. For the case including the trispectrum contribution, we find $A_{\rm tSZ} = 1.18 \pm 0.16$, corresponding to a $\sim 21\%$ higher amplitude relative to the ``Bolliet 2018'' template. This leads to a fractional increase in $S_8$ of $\Delta S_8 / S_8 \simeq 0.18 / 8 \approx 2.3\%$, with a $1\sigma$ uncertainty of $\sim 1.7\%$.}

{\red 
In conclusion, compared to the template predictions, the inferred $S_8$ shifts in opposite directions depending on whether the trispectrum contribution is included: without the trispectrum, $S_8$ is slightly lower, with a fractional change of $\sim 1.1\%$, while including the trispectrum increases $S_8$ modestly by $\sim 2.3\%$. In both cases, these shifts are small compared to the corresponding measurement uncertainties. We note that this should be considered a qualitative estimate, intended primarily to illustrate the overall trend.}

\section{Concluding Remarks}\label{sect:con}
In this study, we successfully applied ABS to reconstruct the tSZ effect power spectrum from the {\it Planck} PR3 data. The ABS method offers several advantages for extracting weak signals, particularly its ability to recover signals in low signal-to-noise ratio regimes while effectively reducing foreground contamination.

Initially, we demonstrated the effectiveness of ABS in reconstructing the tSZ power spectrum using simulated data. By varying the amplitudes of both foregrounds and signals, we present the reconstructed tSZ power spectra derived from different input maps. The results show that ABS achieves high reconstruction accuracy, even when the input foreground and signal amplitudes deviate significantly from the trispectrum-excluded values. These simulations underscore the robustness and reliability of ABS in recovering the faint tSZ signal.

Next, we applied ABS to the {\it Planck} PR3 full-mission data. The ABS results show good agreement with the {\it Planck} F/L estimate for $\ell \lesssim 300$. However, at the higher multipoles, the ABS-derived amplitudes are noticeably lower than the {\it Planck} $y$-map power spectrum.  Then, we conducted a joint analysis to more accurately estimate the tSZ power spectrum by fitting both the tSZ and residual foreground models simultaneously, further reducing foreground contamination. This approach accounted for the tSZ and the cosmic infrared background, radio sources, infrared point sources, and an additional correlated noise term. We considered three tSZ power spectrum templates ``Planck 2015'', ``Battaglia 2012'' and ``Bolliet 2018'', derived from the best-fit values reported in the literature and fitted the normalization amplitudes for each component.  The parameter space was explored using the MCMC method, enabling us to derive the joint posterior probability distributions of the parameters. 

{\red After marginalizing over residual foregrounds, we derive the tSZ power spectrum using covariance estimates that both exclude and include the trispectrum contribution. In the absence of the trispectrum, the recovered tSZ amplitude is approximately $9\%$ lower than the ``Planck 2015'' best-fit value, while remaining consistent within $2\sigma$. When the trispectrum is included, the inferred amplitude is about $18\%$ higher than the ``Bolliet 2018'' best-fit value, corresponding to a $\sim 1.2\sigma$ deviation. In both treatments, the results are consistent with the ``Battaglia 2012'' model at the $1\sigma$ level.} Furthermore, the inclusion of the trispectrum also doubles the error bars.

These results suggest that ABS can serve as a complementary alternative for tSZ power-spectrum measurements. Reanalyzing Planck data with this approach may improve our understanding of the tSZ effect. Future work will extend this method to $\ell>1000$ to better separate foregrounds from the tSZ signal and investigate its cosmological implications.

\appendix 

\section{Fitting results for the trispectrum case}\label{app:tri}

\begin{figure*}[htpb]
        \centering
\includegraphics[width=0.32\textwidth]{ 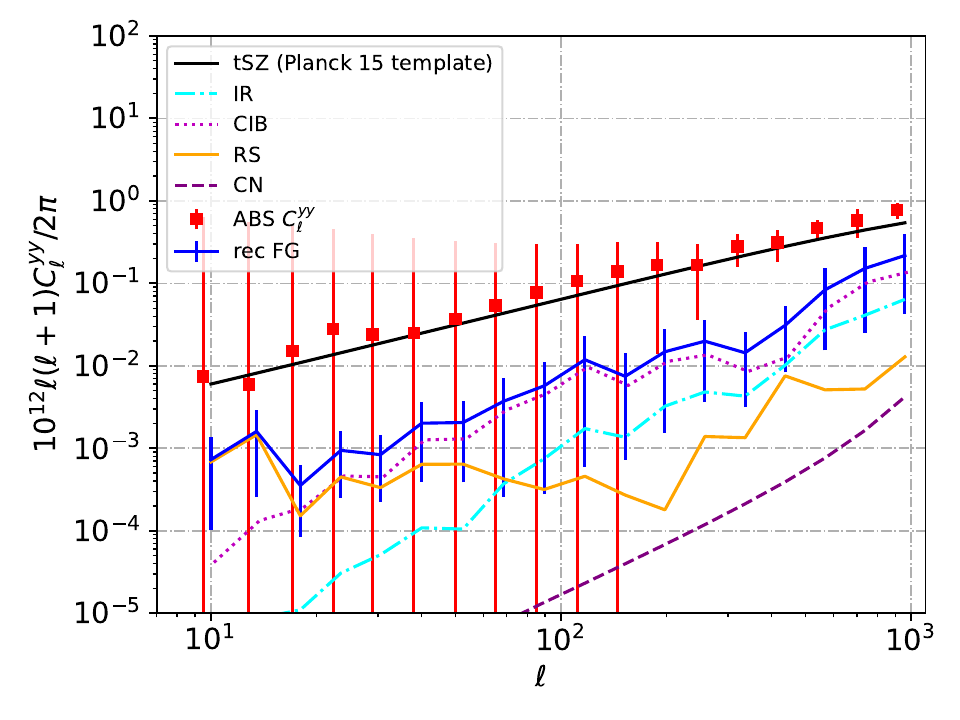}
\includegraphics[width=0.32\textwidth]{ 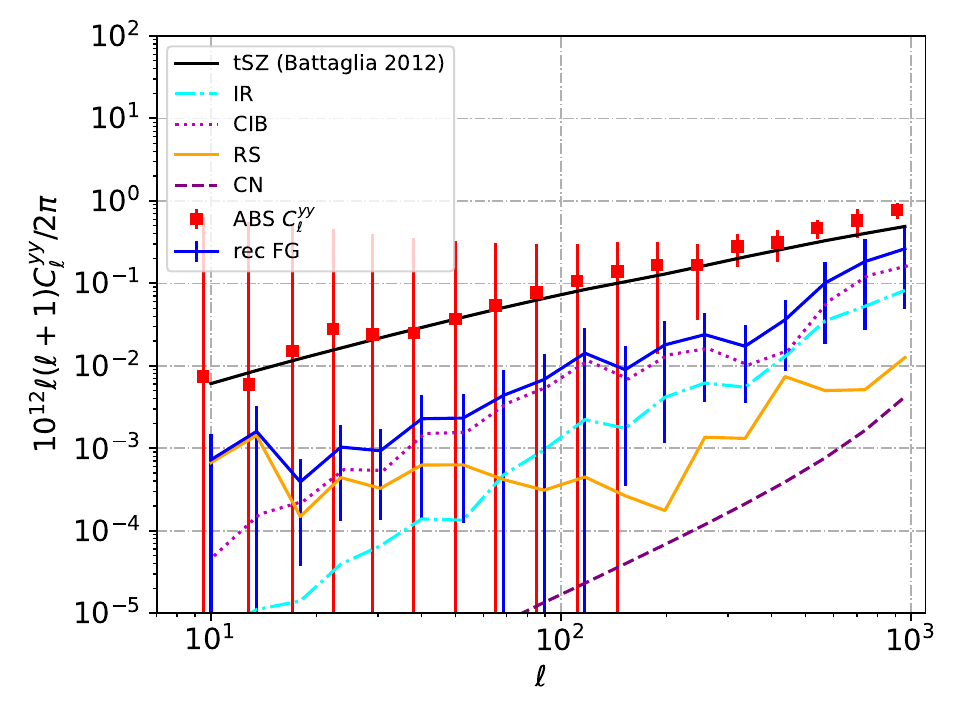}
\includegraphics[width=0.32\textwidth]
{ 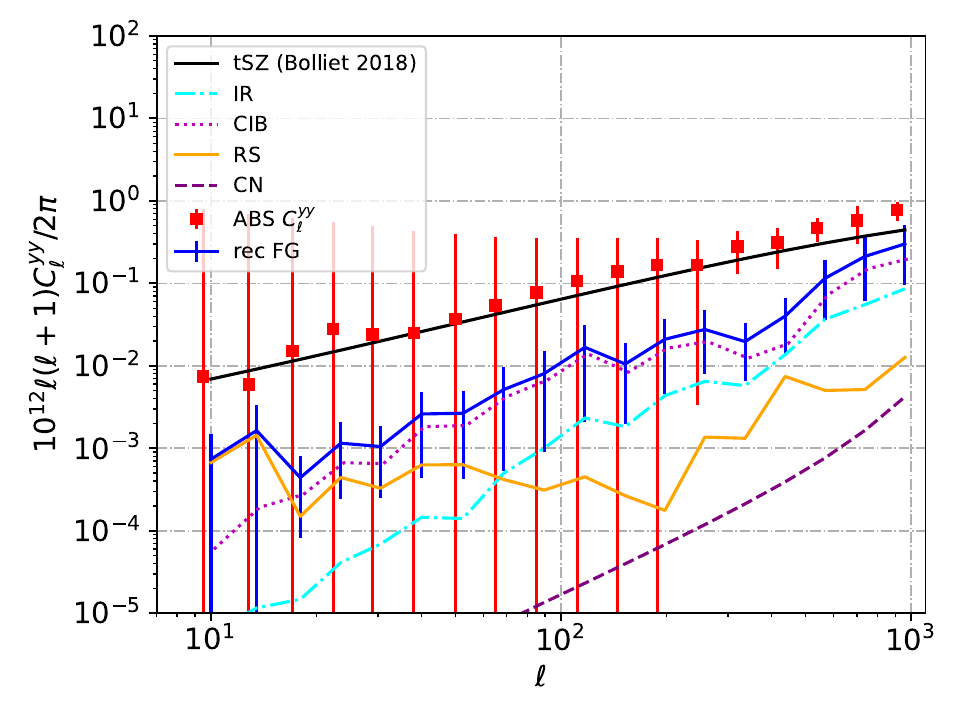}
        \caption{Same as in Fig.~\ref{fig:tsz_all}, but for the trispectrum-included case. As shown, the error bars are significantly larger compared to the trispectrum-excluded case, particularly for $\ell < 100$, due to the increased uncertainty from the trispectrum contribution.
        }
        \label{fig:tsz_all_tri}
    \end{figure*}

In the analysis presented in Fig.~\ref{fig:tsz_all_tri}, we now consider the case where the trispectrum contribution is included. As can be seen, the inclusion of the trispectrum introduces a substantial increase in the uncertainty, leading to significantly larger error bars compared to the trispectrum-excluded case, particularly at lower multipoles ($\ell < 100$).  It is also noteworthy that, when accounting for the trispectrum, the amplitude of the tSZ signal is found to be lower than that of the three templates under consideration.
The trispectrum's contribution suppresses the weighting of data points at low-$\ell$, leading to the highest three $\ell$-bins dominating the fit.

\section{tSZ Power Spectrum for Simulation}\label{app:2}

In this section, we assess the performance of ABS using a suite of simulations with varying input amplitudes of both the tSZ signal and astrophysical foregrounds. For each case, we compare the reconstructed $yy$ power spectra with the true input spectra, and quantify the reconstruction bias and residual foreground contamination. We show that ABS achieves a high level of reconstruction accuracy, even when the foreground and signal amplitudes differ substantially from the trispectrum-excluded reference values. {\red These tests indicate that the ABS reconstruction responds sensitively to variations in both the foreground level and the underlying tSZ signal, which is fully consistent with our expectations. By directly comparing the ABS-reconstructed $y$-map power spectrum with the input simulation, we obtain a noticeably large $\chi^2$, suggesting that residual foreground contamination is still present at a non-negligible level. This result does not imply a failure of the ABS method; rather, it highlights the necessity of a subsequent post–foreground-subtraction step, i.e., by marginalizing over foreground residual templates, in order to further mitigate residual contamination.}

\subsection{Power Spectrum Reconstruction}

\begin{figure}[htbp]
        \centering
        \includegraphics[width=0.49\textwidth]{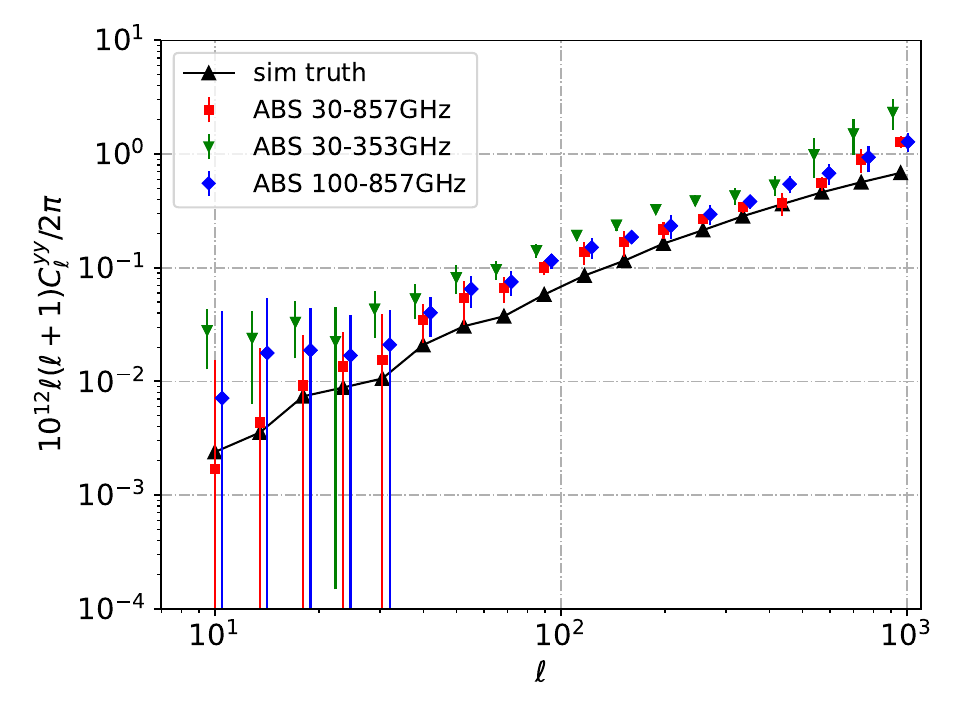}
        \caption{Comparison of the $y$-map angular power spectra for three ABS reconstructions is shown, together with the simulation truth (black solid). The reconstructions are evaluated using three different frequency coverages, as indicated, to demonstrate the effectiveness of ABS. The ABS reconstructions are based on the simulated {\it Planck} full-mission maps, with  \texttt{tSZ mask} applied. Error bars represent the 2$\sigma$ statistical uncertainties. The binning scheme follows that of \citep{2016A&A...594A..22P}, with effective $\ell$ values ranging from $10$ to $959.5$. The $\ell$ values in each case slightly adjusted to enhance clarity and prevent overlaps in the display.   }
        \label{fig:simulation}
    \end{figure}

In Fig.~\ref{fig:simulation}, we present the reconstructed $y$-map angular power spectrum.  For comparison, we also show the {\it Planck} NILC-MILCA F/L result and the simulation truth. All error bars represent the 2$\sigma$ statistical uncertainties. 

{\red Following the {\it Planck} analysis~\citep{2016A&A...594A..13P}, we include contributions from diffuse Galactic foregrounds, clustered CIB, and radio and IR point sources. The normalization amplitudes of each of these components are allowed to vary. To propagate the global modeling uncertainties into the estimated $yy$ power spectrum, we perform 50 independent all-sky simulations for each foreground component, each with distinct \texttt{FFP10} noise and CMB realizations, and varying the foreground models to reflect a 5\% uncertainty in their amplitudes. The uncertainties from each component, together with the cosmic variance of the tSZ power spectrum, are then combined in quadrature to obtain the total error. The 50 realizations are sufficient to accurately estimate the mean and statistical uncertainty, as additional realizations do not lead to any changes in the results.}

To evaluate the sensitivity of the ABS reconstruction to frequency coverage, we consider three cases: (A) excluding the two highest frequency channels (857 GHz and 545 GHz);  (B) omitting the three lowest frequency channels (i.e., the LFI channels: 30, 44, and 70 GHz), and (C) using all nine {\it Planck} channels. To quantify the reconstruction accuracy, we define the Mean-Absolute Error (MAE) between the reconstruction and the truth across all $N_\ell$ bins as:
\begin{equation}
{\rm MAE} = \frac{1}{N_\ell} \sum_\ell |C^{\rm rec}_\ell - C^{\rm truth}_\ell|\,, \end{equation}
and introduce the $\chi^2$ statistic, which is calculated as:
\begin{equation} \chi^2 \equiv \sum_\ell \frac{\big(C^{\rm rec}_\ell - C^{\rm truth}_\ell\big)^2}{\sigma^2_\ell}\,,
\end{equation}
where the statistical uncertainty $\sigma_\ell$ for ABS is assumed to have no correlation between neighboring $\ell$-bins. 

As shown, for case A (green), when comparing to the simulation truth, using the 30–353 GHz frequency coverage leads to an overestimate of the tSZ power spectrum across all relevant scales ($\ell \in [9, 1085]$), with significant deviations from the truth beyond $2\sigma$. 
{\red We find  ${\rm MAE}=0.24$ for all bins ($N_\ell=18$), while the $\chi^2$ statistic is $712.3$.} The strong deviation indicates the presence of strong foreground residuals in the reconstructed power spectrum. Since the 545 and 857 GHz channels are dominated by dust and IR emissions, and the tSZ signal has a unique frequency dependence at these frequencies, neglecting these channels results in an inability to properly disentangle the tSZ from the foregrounds, particularly at $\ell > 100$, leading to the observed overestimate on $C^{yy}_\ell$. 

For case B (blue), when only considering the {\it Planck} HFI channels, as seen, the estimated high-$\ell$ power spectrum decreases significantly, narrowing the gap to the truth. 
This demonstrates that the 545 and 857 GHz channels provide valuable information for reducing high-$\ell$ foreground emissions. However, for $\ell < 30$, the overestimate in amplitude remains, suggesting the presence of residuals at these larger scales. Since synchrotron and other Galactic diffuse foregrounds dominate at low frequencies ($v \lesssim 100$ GHz), the exclusion of the LFI channels results in strong residuals at these scales. {\red As a result, we find ${\rm MAE}=0.10$ for all bins ($N_\ell=18$), while the $\chi^2$ statistic is $223.4$. }

For case C (red), all {\it Planck} channels are utilized, which is expected to provide the highest accuracy in source separation. The broader frequency coverage enhances the capability for foreground cleaning. Compared to cases A and B, the reconstructed $C^{yy}_\ell$ is noticeably closer to the truth. Quantitatively, {\red the mean deviation is ${\rm MAE} = 0.07$ across all bins, and the $\chi^2$ statistic is $189.5$.} This corresponds to an absolute deviation that is approximately {\red 0.17 and 0.03} smaller than case A and B, respectively. Furthermore, the statistical uncertainty for case C is the smallest among all cases, as the inclusion of more frequency channels helps stabilize the signal estimation and reduces noise fluctuations.

\subsection{Sensitivity to foreground modeling}

Furthermore, to quantify the robustness of the ABS reconstruction against uncertainties in the foreground modeling, we artificially rescale the total foreground temperature maps at each frequency by factors of 1.2 and 1.5, while keeping all other components fixed. This procedure effectively enhances the corresponding foreground power spectra by factors of 1.44 and 2.25, respectively.

\begin{figure}[htbp]
        \centering
        \includegraphics[width=0.45\textwidth]{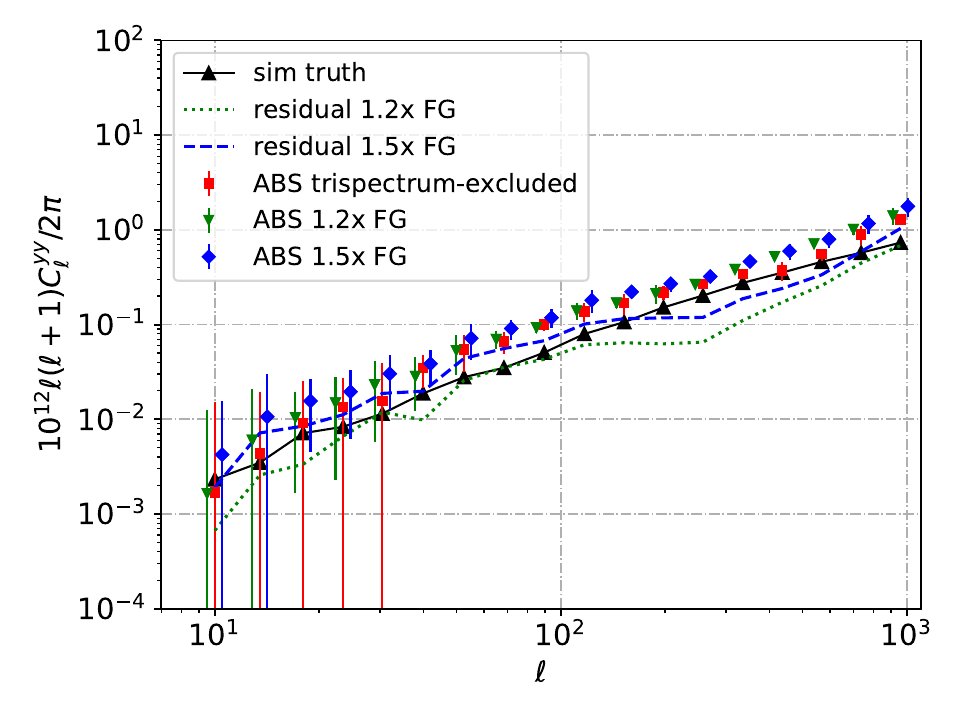}
        \caption{Same as in Fig.~\ref{fig:simulation}, but for comparing the reconstructed tSZ power spectra when the total foreground temperature map at each frequency is increased by factors of 1.2 (``1.2$\times$ FG''; green) and 1.5 (``1.5$\times$ FG''; blue). The ABS-derived power spectrum (red) is based on all {\it Planck} frequency channels. The absolute deviations between the simulation truth and the ``1.2$\times$ FG'' (green dotted) and ``1.5$\times$ FG'' (blue dashed) cases are also shown for comparison.  The error bars indicate the $2\sigma$ level for the measurements.}
        \label{fig:refg}
    \end{figure}

In Fig.~\ref{fig:refg}, as expected, we observe that a stronger foreground contamination leads to an overestimation of the tSZ signal. Compared to the  trispectrum-excluded  case, the foreground residuals increase progressively.  Given that the standard deviation of the total foreground amplitude (after masking) across all channels is about 0.3 K, the map-level foreground amplitude is enhanced by 20\% and 50\%, respectively, leading to an increase in the power spectrum by the square of these factors. 


Quantitatively, for the cases ``1.2$\times$ FG'' and ``1.5$\times$ FG'', the resulting deviations are MAE = 0.11 and 0.17, respectively, with corresponding $\chi^2$ values of {\red 218.7 and 796.1}. The significantly higher $\chi^2$ in the ``1.5$\times$ FG'' case indicates that the measured foreground residuals are substantial compared to the associated statistical uncertainties. Compared to the fiducial trispectrum-excluded case (MAE=0.07), the MAE values show only a slight increase. This result demonstrates that our algorithm can effectively suppresses foreground contamination and remains robust against increased foreground levels.

\subsection{Null test}

\begin{figure}[htbp]
        \centering
        \includegraphics[width=0.45\textwidth]{ 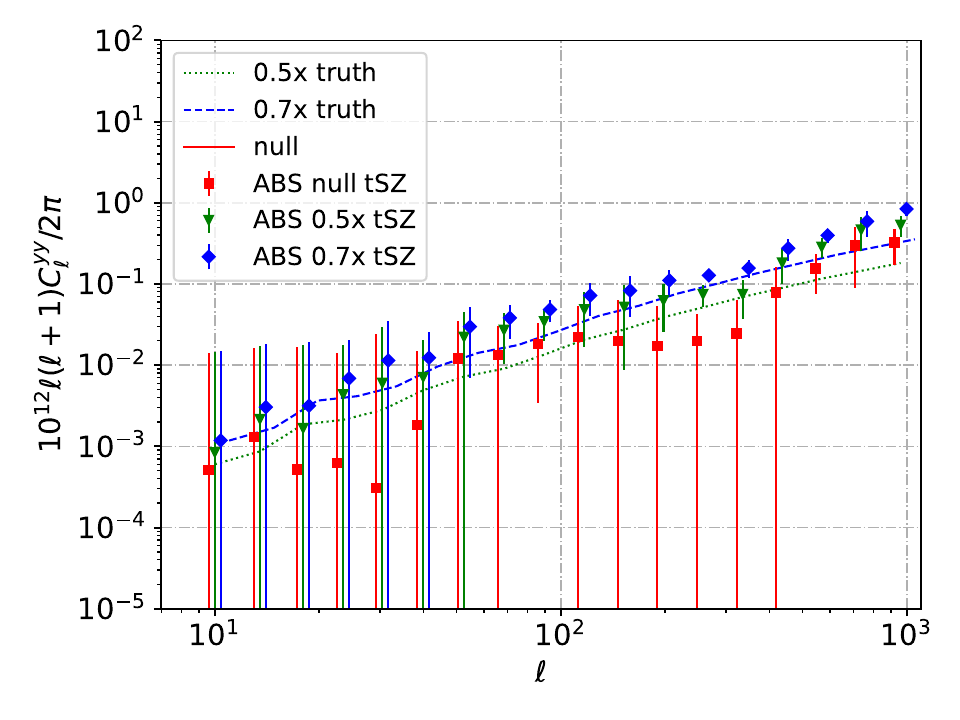}
        \caption{Comparison of the reconstructed tSZ power spectra when the tSZ temperature map at each frequency is reduced by factors of 0.5 (``0.5$\times$ tSZ''; green), 0.7 (``0.7$\times$ tSZ''; blue), and fully nulled (``null tSZ''; red). The simulated true values for ``0.5$\times$ truth'' (green dotted), ``0.7$\times$ truth'' (blue dashed),  are also shown for comparison. The error bars indicate the $2\sigma$ level for the measurements.}
        \label{fig:null}
    \end{figure}  
    
In Fig.~\ref{fig:null}, the null test results are presented to further validate the effectiveness of ABS in simulated {\it Planck} data. Three scenarios are considered, where the amplitude of the input tSZ $y$-map is progressively reduced by factors of 0.7, 0.5, and 0 (complete nulling). These cases are labeled as ``ABS 0.7$\times$ tSZ'' (blue), ``ABS 0.5$\times$ tSZ'' (green), and ``ABS null tSZ'' (red), respectively. On the power spectrum level, these reductions correspond to tSZ power spectrum amplitudes of 49\%, 25\%, and 0. The true power spectra for each case are also shown: the green dotted line corresponds to ``ABS 0.5$\times$ tSZ'', and the blue dashed line represents ``ABS 0.7$\times$ tSZ''.

Quantitatively, we compare the reconstructed values to the true values for each case. For case ``ABS 0.7$\times$ tSZ'', we find {\red ${\rm MAE} = 0.07$ and $\chi^2 =106.5$. Similarly, for the ``ABS 0.5$\times$ tSZ'' case, deviations become evident at high $\ell$, with ${\rm MAE} = 0.06$ and $\chi^2 =80.2$. For the ``ABS null tSZ'' case, a comparable trend is observed, yielding ${\rm MAE} = 0.05$ and $\chi^2 = 63.2$. As the input tSZ amplitude decreases, the reconstructed $yy$ power spectrum systematically shifts to lower amplitudes, while the relative impact of foreground residuals becomes more significant, particularly on small angular scales. This behavior indicates that residual foreground contamination contributes non-negligibly to the high-$\ell$ excess.}



\section{Key Concepts of ABS and the Connection with ILC}\label{sect:3}

Below, we describe the relationship between the ABS and ILC methods. The ILC estimator is a widely used technique for estimating {\red CMB power spectrum,  
$C_\ell$}, by minimizing the total variance of foregrounds and noise. The ILC estimator can be mathematically expressed as:

\begin{equation}\label{eq:ilc}
{\red C_\ell} = \frac{1}{\mathbf{f}^T [\mathbf{C}_\ell^{\rm obs}]^{-1}{\mathbf f}}\,,
\end{equation}
where $\mathbf{f}$ is the known frequency dependence of the signal $s$, and $\mathbf{C}_\ell^{\rm obs}$ is the observed covariance matrix of the data.

However, the ABS approach treats the noise in a different manner than ILC, by discarding the noise-dominated modes to achieve the signal estimate, rather than minimizing the total variance. To illustrate the connection between the ABS and ILC estimators, let us first de-weight the covariance matrix $\delta \mathbf{C}_\ell^{\rm obs}$ by normalizing it using the noise standard deviations in the ABS approach. Here, $\delta \mathbf{C}_\ell^{\rm obs} = \mathbf{C}_\ell^{\rm obs} - \mathbf{N}_\ell$, with the noise covariance $\mathbf{N}_\ell$ subtracted. In matrix form, the de-weighting yields:

\begin{equation}\label{eq:dc}
\tilde{\mathbf{C}}_\ell = \mathbf{\Sigma}^{-1/2}_\ell \delta \mathbf{C}_\ell^{\rm obs} \mathbf{\Sigma}_\ell^{-1/2}\,,
\end{equation}

where $\mathbf{\Sigma}_\ell$ is the diagonal matrix with diagonal elements $\Sigma_{ii}$ representing the sampling variance of the noise band power at frequency $i$, which is typically estimated from noise simulations.

By transforming $\mathbf{C}_\ell^{\rm obs}$ into $\tilde{\mathbf{C}}_\ell^{\rm obs}$ via Eq.~\ref{eq:dc}, the ILC estimator given in Eq.~\ref{eq:ilc} can be rewritten as:
\begin{equation}\label{eq:ilc1}
{\red C_\ell}  =  \frac{1}{ \tilde{\mathbf{f}}^T  \tilde{\mathbf{C}}_\ell^{-1}\tilde{\mathbf{f}}} \,,
\end{equation}
where we have introduced $\tilde{\mathbf{f}} = \mathbf{\Sigma}_\ell^{-1 / 2} \mathbf{f}$.

Applying the eigen-decomposition $\tilde{\mathbf{C}_\ell} = \tilde{\mathbf{E}} \tilde{\mathbf{\Lambda}} \tilde{\mathbf{E}}^T$, where $\tilde{\mathbf{E}}^{\mu}$ denotes the $\mu$th eigenvector, and $\tilde{\mathbf{\Lambda}}$ is the diagonal matrix of eigenvalues ( denoted by $\tilde{\lambda}_\mu$), we can express Eq.~\ref{eq:ilc1} as:

\begin{equation}\label{eq:ilc-2}
{\red C_\ell}  = \frac{1}{\tilde{\mathbf f}^T \left[\tilde{\mathbf{E}} \tilde{\mathbf{\Lambda}} \tilde{\mathbf{E}}^T\right]^{-1} \tilde{\mathbf f}} = \frac{1}{\tilde{\mathbf f}^T \left[\tilde{\mathbf{E}} \tilde{\mathbf{\Lambda}}^{-1} \tilde{\mathbf{E}}^T\right] \tilde{\mathbf f}}\,,
\end{equation}

Further introducing $\tilde{\mathbf{g}} = \tilde{\mathbf{E}}^T \tilde{\mathbf{f}}$, Eq.~\ref{eq:ilc-2} becomes:

\begin{equation}\label{eq:abss}
{\red C_\ell} = \frac{1}{\mathbf{g}^T \tilde{\mathbf{\Lambda}}^{-1} \tilde{\mathbf{g}}} \stackrel{\rm ILC\rightarrow ABS}{\approx} 
 \frac{1}{\mathbf{g}^T \tilde{\mathbf{\Lambda}}^{\dagger} \tilde{\mathbf{g}}} =\left(\sum_{\mu}^{\tilde{\lambda}_\mu \geq \lambda_{\rm cut}} \tilde{g}_\mu^2 \tilde{\lambda}_\mu^{-1}\right)^{-1}\,,
\end{equation}
where $\dagger$ denotes the pseudo-inverse, which maps all nonzero eigenvalues satisfying $\tilde{\lambda}_\mu \geq \lambda_{\rm cut}$ to their reciprocals, while setting the remaining eigenvalues to zero.

Physically, the eigenvalues $\tilde{\lambda}_{\mu}$ now represent the signal-to-noise ratio (S/N) of each mode. In the ABS approach, we apply a cutoff $\lambda_{\rm cut} \sim 1$, discarding all modes with $\tilde{\lambda}_{\mu} < \lambda_{\rm cut}$, thereby improving the robustness of the signal estimation by filtering out noise-dominated modes. The 
approximation in the last term corresponds to the ABS solution, suggesting that ABS exhibits certain similarities with ILC. However, ABS applies a noise-deweighting procedure to the data, which fundamentally differs from ILC. Additionally, the ABS approach incorporates a positive shift parameter, $\mathcal{S}$, to enhance computational stability, particularly when the signal is significantly weaker than the noise. This shift operation is another distinctive feature of ABS, further distinguishing it from ILC. Notably, for clarity, Eq.~\ref{eq:abss} does not include the shift parameter. 

{\red While both ABS and cross-ILC use the frequency–frequency cross bandpower matrices, they treat noise differently. The cross-ILC used in ~\citep{2014JCAP...02..030H} minimizes the cross-bandpower variance, assuming vanishing mean noise bandpower in cross correlation, whereas ABS analytically solves for the signal bandpower by discarding noise-dominated modes through eigenvalue selection. Moreover, the shift parameter in ABS further enhances its signal extraction capability, a feature absent in ILC. As a result, ABS yields a $yy$ power spectrum that closely matches the Planck tSZ measurement (Fig.~\ref{fig:planck_full}) even without any additional residual subtraction,  demonstrating its effectiveness in extracting such a weak signal.}

\subsection{ If  $M+1 \leq N_f$ fails...}\label{sect:rank}
Now, let us consider the case where the assumption $M+1 \leq N_f$ fails, where $M$ denotes the rank of the cross-band power matrix of the foreground, and $N_f$ is the number of frequency channels. This situation is more realistic when the frequency coverage is poor. Our tests indicate that violating this assumption leads to an overestimation of the CMB signal. Intuitively, this occurs because foreground power leaks into the signal estimate.  To illustrate this, consider a simple case of ILC in the absence of noise, assuming the foreground matrix $\mathbf{C}_F$ has full rank. Using the Sherman–Morrison formula, the ILC estimator can be rewritten as  
\begin{equation}
\frac{1}{\mathbf{f}^T\left({\red C_\ell}\cdot \mathbf{f} \mathbf{f}^T+\mathbf{C}_F\right)^{-1} \mathbf{f}}={\red C_\ell}+ \frac{1}{\mathbf{f}^T \mathbf{C}_F^{-1} \mathbf{f}}\,.
\end{equation}

If $\mathbf{C}_F$ is a covariance matrix, it will be positive semi-definite, meaning that the term $\mathbf{f}^T \mathbf{C}_F^{-1} \mathbf{f}$ is strictly positive, provided that $ \mathbf{f} $ is not entirely in the null space of $\mathbf{C}_F$. However, since $\mathbf{C}_F$ is constructed from observational data, representing only one realization of the underlying true covariance, it may not be strictly positive definite. In practice, however, our simulations show that even if negative eigenvalues appear, the eigenvalues of $\mathbf{C}_F^{-1}$ are typically 2-3 orders of magnitude smaller than the largest positive eigenvalue. Therefore, there is almost no risk of introducing a negative bias from $\mathbf{f}^T \mathbf{C}_F^{-1} \mathbf{f}$. Thus, an overestimation of $ C_\ell $ arises from the assumption of full-rank foreground covariance.

In fact, the ABS method can be interpreted as an analogy to this scenario. To simplify, let us consider the common case in which the signal is significantly smaller than the foreground, as is typically true for the tSZ effect and primary B-mode signals. Under this assumption, we approximate $\tilde{\mathbf{C}}_\ell^{-1} \approx \tilde{\mathbf{C}}_F^{-1}$. Consequently, the ABS solution, as given in Eq.~\ref{eq:abss}, corresponds to the pure residual foreground band powers and simplifies to  $C_\ell = \left(\sum_{\mu}^{\tilde{\lambda}_\mu \geq \lambda_{\rm cut}} \tilde{g}_\mu^2 \tilde{\lambda}_\mu^{-1} \right)^{-1}$. Since we choose $\tilde{\lambda}_\mu > 1$, this expression is always positive-definite. A more detailed discussion is beyond the scope of this paper.

\section{Construction of foreground templates}\label{sect:abs-template}
We now present the analytical construction of templates for each residual foreground. To achieve this, we introduce a small perturbation in the data and derive an analytical estimation. Specifically, by adding $\epsilon \mathbf{C}_\ell^{X}$ to the data, the first-order perturbation in the estimated band power spectrum due to the added foreground $X$ serves as its template.  

Referring to Eqs.~\ref{eq:ilc1} and~\ref{eq:abss}, the ABS solution (without the shift parameter) can be rewritten as  
$C_\ell = 1/(\tilde{\mathbf{f}}^T \tilde{\mathbf{C}}_\ell^{\dagger} \tilde{\mathbf{f}})$. 
After incorporating $\mathbf{C}_\ell^{X}$, the modified ABS solution becomes  
\begin{equation}
C'_\ell=\frac{1}{\tilde{\mathbf{f}}^T \big[\tilde{\mathbf{C}}_\ell+\epsilon\tilde{\mathbf{C}}_\ell^X\big]^{\dagger}\tilde{\mathbf{f}}}\,,
\end{equation}
where the same noise deweighting operation has been applied (recall Eq.~\ref{eq:dc}), i.e.,  
$\tilde{\mathbf{C}}_\ell^X = \mathbf{\Sigma}^{-1/2} \mathbf{C}_\ell^{X} \mathbf{\Sigma}^{-1/2}$. Note that when constructing the template, we do not include the shift parameter in the calculations, as it only contributes to the second-order term. We thus assume that the shift effect in the templates is negligible.

In general, consider the perturbed matrix \(\mathbf{A} + \epsilon \mathbf{B}\). Using the Woodbury matrix identity, its inverse can be approximated to first order as  
\begin{equation}
    (\mathbf{A} + \epsilon \mathbf{B})^{-1} \approx \mathbf{A}^{-1} - \epsilon \mathbf{A}^{-1} \mathbf{B} \mathbf{A}^{-1} + \mathcal{O}(\epsilon^2).
\end{equation}  
Applying this identity, we obtain  
\begin{equation}
    C'_\ell = C'_\ell + \delta C_\ell = \frac{1}{\tilde{\mathbf{f}}^T (\tilde{\mathbf{C}}_\ell + \epsilon \tilde{\mathbf{C}}_\ell^X)^{-1} \tilde{\mathbf{f}}} 
    \approx \frac{1}{\tilde{\mathbf{f}}^T \tilde{\mathbf{C}}_\ell^{-1} \tilde{\mathbf{f}}} + 
    \epsilon \frac{\tilde{\mathbf{f}}^T \tilde{\mathbf{C}}_\ell^{-1} \tilde{\mathbf{C}}_\ell^X \tilde{\mathbf{C}}_\ell^{-1} \tilde{\mathbf{f}}}{\big(\tilde{\mathbf{f}}^T \tilde{\mathbf{C}}_\ell^{-1} \tilde{\mathbf{f}}\big)^2},
\end{equation}
which leads to the template for residual $X$:  
\begin{equation}\label{eq:abs_res}
    \hat{C}^X_\ell = \frac{\tilde{\mathbf{f}}^T \tilde{\mathbf{C}}_\ell^{\dagger} \mathbf{C}_\ell^{X} \tilde{\mathbf{C}}_\ell^{\dagger} \tilde{\mathbf{f}}}{\big(\tilde{\mathbf{f}}^T \tilde{\mathbf{C}}_\ell^{\dagger} \tilde{\mathbf{f}}\big)^2}.
\end{equation}

\subsection{Linear perturbative estimate of ILC foreground residuals}
\label{app:ilc_residual}
Here we presents a linear perturbative derivation of the foreground residual power spectrum in the ILC framework. The goal is to clarify the validity of estimating individual foreground residuals using fixed ILC weights and to demonstrate explicitly that this procedure is mathematically equivalent, at leading order, to the residual estimator used in the ABS formalism.

We consider multi-frequency data
\begin{equation}
\mathbf{x} = s\,\mathbf{f} + \mathbf{g} + \mathbf{n},
\end{equation}
where $s$ is the target signal with known frequency dependence $\mathbf{f}$, 
$\mathbf{g}$ denotes foreground emission, and $\mathbf{n}$ is instrumental noise.
The covariance matrix is
\begin{equation}
\mathbf{C} \equiv \langle \mathbf{x}\mathbf{x}^T \rangle
= C_s\,\mathbf{f}\mathbf{f}^T + \mathbf{C}_g + \mathbf{N}.
\end{equation}

The ILC estimator is defined as
\begin{equation}
\hat{s} = \mathbf{w}^T \mathbf{x}, \qquad
\mathbf{w} =
\frac{\mathbf{C}^{-1}\mathbf{f}}
{\mathbf{f}^T \mathbf{C}^{-1}\mathbf{f}},
\end{equation}
which satisfies $\mathbf{w}^T\mathbf{f}=1$.

We treat the covariance of a given foreground component $X$ as a small
perturbation,
\begin{equation}
\mathbf{C} = \mathbf{C}_0 + \epsilon\,\mathbf{C}_X, \qquad
\mathbf{C}_0 \equiv C_s\,\mathbf{f}\mathbf{f}^T + \mathbf{N},
\end{equation}
with $\epsilon \ll 1$.
To first order,
\begin{equation}
\mathbf{C}^{-1}
= \mathbf{C}_0^{-1}
- \epsilon\,\mathbf{C}_0^{-1}\mathbf{C}_X\mathbf{C}_0^{-1}
+ \mathcal{O}(\epsilon^2).
\end{equation}

Defining $D_0 \equiv \mathbf{f}^T\mathbf{C}_0^{-1}\mathbf{f}$,
the ILC weights expand as
\begin{equation}
\mathbf{w} = \mathbf{w}_0 + \epsilon\,\delta\mathbf{w}, \qquad
\mathbf{w}_0 = \frac{\mathbf{C}_0^{-1}\mathbf{f}}{D_0},
\end{equation}
with
\begin{equation}
\delta\mathbf{w}
= -\frac{\mathbf{C}_0^{-1}\mathbf{C}_X\mathbf{C}_0^{-1}\mathbf{f}}{D_0}
+ \frac{\mathbf{C}_0^{-1}\mathbf{f}}{D_0^2}
\left(\mathbf{f}^T\mathbf{C}_0^{-1}\mathbf{C}_X\mathbf{C}_0^{-1}\mathbf{f}\right).
\end{equation}

The residual foreground contribution is
\begin{equation}
r_X \equiv \mathbf{w}^T \mathbf{g}_X ,
\end{equation}
with residual power spectrum
\begin{equation}
C_\ell^{\rm res, X}
= \langle r_X^2\rangle
= \mathbf{w}^T \mathbf{C}_X \mathbf{w}.
\end{equation}

Substituting the perturbative expansion of the weights yields
\begin{equation}\label{eq:ilc_fg}
C_\ell^{\rm res, X}
= \mathbf{w}_0^T \mathbf{C}_X \mathbf{w}_0
+ \mathcal{O}(\epsilon^2),
\end{equation}
or explicitly,
\begin{equation}\label{eq:ilc_res}
C_\ell^{\rm res, X}
=
\frac{
\mathbf{f}^T \mathbf{C}_0^{-1}
\mathbf{C}_X
\mathbf{C}_0^{-1}\mathbf{f}
}{
\left(\mathbf{f}^T \mathbf{C}_0^{-1}\mathbf{f}\right)^2
}
+ \mathcal{O}(\epsilon^2)\,.
\end{equation}

The linear-order expansion of Eq.~\ref{eq:ilc_fg} under the covariance perturbation formalism justifies our derivation of the residual foreground power spectrum and is fully consistent with existing treatments in the literature. In particular, the mathematical structure of Eq.~\ref{eq:ilc_res} is explicitly equivalent to the residual estimator in Eq.~\ref{eq:abs_res} within the ABS framework. This correspondence further justifies the common practice of estimating individual foreground residuals by fixing the ILC weights derived from the data covariance and projecting them onto simulated foreground-only maps.

\acknowledgments
This study is supported by the National SKA Program of China (2025SKA0160100), the National Science Foundation of China (12473097), the National Key R\&D Program of China (2020YFC2201600), Guangdong Basic and Applied Basic Research Foundation (2024A1515012309), the China Manned Space Project with No. CMS-CSST-2021 (A02, A03, B01). Some of the results presented in this paper were derived using the \texttt{HEALPix} package~\citep{Healpix2005} and the \texttt{emcee} Python package~\citep{2013PASP..125..306F}. Based on observations obtained with {\it Planck}  (\href{http://www.esa.int/Planck}{http://www.esa.int/Planck}), an ESA science mission with instruments and contributions directly funded by ESA Member States, NASA, and Canada.

\bibliography{sample631}

@article{Beringue:2025bur,
    author = "Beringue, Benjamin and others",
    title = "{The Atacama Cosmology Telescope: DR6 power spectrum foreground model and validation}",
    eprint = "2506.06274",
    archivePrefix = "arXiv",
    primaryClass = "astro-ph.CO",
    doi = "10.1088/1475-7516/2025/10/082",
    journal = "JCAP",
    volume = "10",
    pages = "082",
    year = "2025"
}

@article{AtacamaCosmologyTelescope:2025blo,
    author = "Louis, Thibaut and others",
    collaboration = "Atacama Cosmology Telescope",
    title = "{The Atacama Cosmology Telescope: DR6 power spectra, likelihoods and {\ensuremath{\Lambda}}CDM parameters}",
    eprint = "2503.14452",
    archivePrefix = "arXiv",
    primaryClass = "astro-ph.CO",
    reportNumber = "FERMILAB-PUB-25-0071-PPD",
    doi = "10.1088/1475-7516/2025/11/062",
    journal = "JCAP",
    volume = "11",
    pages = "062",
    year = "2025"
}

@article{Efstathiou:2025ckq,
    author = "Efstathiou, George and McCarthy, Fiona",
    title = "{The Power Spectrum of the Thermal Sunyaev-Zeldovich Effect}",
    eprint = "2502.10232",
    archivePrefix = "arXiv",
    primaryClass = "astro-ph.CO",
    month = "2",
    year = "2025"
}

@ARTICLE{2014JCAP...02..030H,
       author = {{Hill}, J. Colin and {Spergel}, David N.},
        title = "{Detection of thermal SZ-CMB lensing cross-correlation in Planck nominal mission data}",
      journal = {\jcap},
         year = 2014,
        month = feb,
       volume = {2014},
       number = {2},
          eid = {030},
        pages = {030},
          doi = {10.1088/1475-7516/2014/02/030},
archivePrefix = {arXiv},
       eprint = {1312.4525},
 primaryClass = {astro-ph.CO},
       adsurl = {https://ui.adsabs.harvard.edu/abs/2014JCAP...02..030H}
}

@ARTICLE{2025arXiv250707346B,
       author = {{Bolliet}, Boris and {Kusiak}, Aleksandra and {McCarthy}, Fiona and {Sabyr}, Alina and {Surrao}, Kristen and {Chluba}, Jens and {Embil Villagra}, Carmen and {Ferraro}, Simone and {Hadzhiyska}, Boryana and {Han}, Dongwon and {Hill}, J. Colin and {Mac{\'\i}as-P{\'e}rez}, Juan Francisco and {Maniyar}, Abhishek and {Mehta}, Yogesh and {Pandey}, Shivam and {Schaan}, Emmanuel and {Sherwin}, Blake and {Spurio Mancini}, Alessio and {Zubeldia}, {\'I}{\~n}igo},
        title = "{CLASS\_SZ II: Notes and Examples of Fast and Accurate Calculations of Halo Model, Large Scale Structure and Cosmic Microwave Background Observables}",
      journal = {arXiv e-prints},
         year = 2025,
        month = jul,
          eid = {arXiv:2507.07346},
        pages = {arXiv:2507.07346},
          doi = {10.48550/arXiv.2507.07346},
archivePrefix = {arXiv},
       eprint = {2507.07346},
 primaryClass = {astro-ph.CO},
       adsurl = {https://ui.adsabs.harvard.edu/abs/2025arXiv250707346B}
}

@ARTICLE{2023arXiv231018482B,
       author = {{Bolliet}, B. and {Kusiak}, A. and {McCarthy}, F. and {Sabyr}, A. and {Surrao}, K. and {Hill}, J.~C. and {Chluba}, J. and {Ferraro}, S. and {Hadzhiyska}, B. and {Han}, D. and {Mac{\'\i}as-P{\'e}rez}, J.~F. and {Madhavacheril}, M. and {Maniyar}, A. and {Mehta}, Y. and {Pandey}, S. and {Schaan}, E. and {Sherwin}, B. and {Spurio Mancini}, A. and {Zubeldia}, {\'I}.},
        title = "{class\_sz I: Overview}",
      journal = {arXiv e-prints},
         year = 2023,
        month = oct,
          eid = {arXiv:2310.18482},
        pages = {arXiv:2310.18482},
          doi = {10.48550/arXiv.2310.18482},
archivePrefix = {arXiv},
       eprint = {2310.18482},
 primaryClass = {astro-ph.IM},
       adsurl = {https://ui.adsabs.harvard.edu/abs/2023arXiv231018482B}
}

@ARTICLE{2023JCAP...03..039B,
       author = {{Bolliet}, Boris and {Colin Hill}, J. and {Ferraro}, Simone and {Kusiak}, Aleksandra and {Krolewski}, Alex},
        title = "{Projected-field kinetic Sunyaev-Zel'dovich Cross-correlations: halo model and forecasts}",
      journal = {\jcap},
         year = 2023,
        month = mar,
       volume = {2023},
       number = {3},
          eid = {039},
        pages = {039},
          doi = {10.1088/1475-7516/2023/03/039},
archivePrefix = {arXiv},
       eprint = {2208.07847},
 primaryClass = {astro-ph.CO},
       adsurl = {https://ui.adsabs.harvard.edu/abs/2023JCAP...03..039B}
}

@ARTICLE{2020JCAP...10..012S,
       author = {{Stein}, George and {Alvarez}, Marcelo A. and {Bond}, J. Richard and {van Engelen}, Alexander and {Battaglia}, Nicholas},
        title = "{The Websky extragalactic CMB simulations}",
      journal = {\jcap},
         year = 2020,
        month = oct,
       volume = {2020},
       number = {10},
          eid = {012},
        pages = {012},
          doi = {10.1088/1475-7516/2020/10/012},
archivePrefix = {arXiv},
       eprint = {2001.08787},
 primaryClass = {astro-ph.CO},
       adsurl = {https://ui.adsabs.harvard.edu/abs/2020JCAP...10..012S}
}

@ARTICLE{2013PASP..125..306F,
       author = {{Foreman-Mackey}, Daniel and {Hogg}, David W. and {Lang}, Dustin and {Goodman}, Jonathan},
        title = "{emcee: The MCMC Hammer}",
      journal = {\pasp},
         year = 2013,
        month = mar,
       volume = {125},
       number = {925},
        pages = {306},
          doi = {10.1086/670067},
archivePrefix = {arXiv},
       eprint = {1202.3665},
 primaryClass = {astro-ph.IM},
       adsurl = {https://ui.adsabs.harvard.edu/abs/2013PASP..125..306F}
}

@ARTICLE{2024ApJS..270...16I,
       author = {{Ibitoye}, Ayodeji and {Dai}, Wei-Ming and {Ma}, Yin-Zhe and {Vielva}, Patricio and {Tramonte}, Denis and {Abebe}, Amare and {Beesham}, Aroonkumar and {Chen}, Xuelei},
        title = "{Cross Correlation between the Thermal Sunyaev-Zeldovich Effect and the Integrated Sachs-Wolfe Effect}",
      journal = {\apjs},
         year = 2024,
        month = jan,
       volume = {270},
       number = {1},
          eid = {16},
        pages = {16},
          doi = {10.3847/1538-4365/ad08c5},
archivePrefix = {arXiv},
       eprint = {2310.18478},
 primaryClass = {astro-ph.CO},
       adsurl = {https://ui.adsabs.harvard.edu/abs/2024ApJS..270...16I}
}

@ARTICLE{2023MNRAS.526.5682C,
       author = {{Chandran}, Jyothis and {Remazeilles}, Mathieu and {Barreiro}, R.~B.},
        title = "{An improved Compton parameter map of thermal Sunyaev-Zeldovich effect from Planck PR4 data}",
      journal = {\mnras},
         year = 2023,
        month = dec,
       volume = {526},
       number = {4},
        pages = {5682-5698},
          doi = {10.1093/mnras/stad3156},
archivePrefix = {arXiv},
       eprint = {2305.10193},
 primaryClass = {astro-ph.CO},
       adsurl = {https://ui.adsabs.harvard.edu/abs/2023MNRAS.526.5682C}
}

@article{Zonca_2021,
  doi = {10.21105/joss.03783},
  url = {https://doi.org/10.21105/joss.03783},
  year = {2021},
  publisher = {The Open Journal},
  volume = {6},
  number = {67},
  pages = {3783},
  author = {Andrea Zonca and Ben Thorne and Nicoletta Krachmalnicoff and Julian Borrill},
  title = {{The Python Sky Model 3 software}},
  journal = {Journal of Open Source Software}
}

@article{Thorne_2017,
   title={{The Python Sky Model: software for simulating the Galactic microwave sky}},
   volume={469},
   ISSN={1365-2966},
   url={http://dx.doi.org/10.1093/mnras/stx949},
   DOI={10.1093/mnras/stx949},
   number={3},
   journal={Monthly Notices of the Royal Astronomical Society},
   publisher={Oxford University Press (OUP)},
   author={Thorne, B. and Dunkley, J. and Alonso, D. and Næss, S.},
   year={2017},
   month={May},
   pages={2821–2833}
}

@ARTICLE{2016A&A...594A...7P,
       author = {{Planck Collaboration} and {Adam}, R. and {Ade}, P.~A.~R. and {Aghanim}, N. and {Arnaud}, M. and {Ashdown}, M. and {Aumont}, J. and {Baccigalupi}, C. and {Banday}, A.~J. and {Barreiro}, R.~B. and {Bartolo}, N. and {Battaner}, E. and {Benabed}, K. and {Beno{\^\i}t}, A. and {Benoit-L{\'e}vy}, A. and {Bernard}, J. -P. and {Bersanelli}, M. and {Bertincourt}, B. and {Bielewicz}, P. and {Bock}, J.~J. and {Bonavera}, L. and {Bond}, J.~R. and {Borrill}, J. and {Bouchet}, F.~R. and {Boulanger}, F. and {Bucher}, M. and {Burigana}, C. and {Calabrese}, E. and {Cardoso}, J. -F. and {Catalano}, A. and {Challinor}, A. and {Chamballu}, A. and {Chary}, R. -R. and {Chiang}, H.~C. and {Christensen}, P.~R. and {Clements}, D.~L. and {Colombi}, S. and {Colombo}, L.~P.~L. and {Combet}, C. and {Couchot}, F. and {Coulais}, A. and {Crill}, B.~P. and {Curto}, A. and {Cuttaia}, F. and {Danese}, L. and {Davies}, R.~D. and {Davis}, R.~J. and {de Bernardis}, P. and {de Rosa}, A. and {de Zotti}, G. and {Delabrouille}, J. and {Delouis}, J. -M. and {D{\'e}sert}, F. -X. and {Diego}, J.~M. and {Dole}, H. and {Donzelli}, S. and {Dor{\'e}}, O. and {Douspis}, M. and {Ducout}, A. and {Dupac}, X. and {Efstathiou}, G. and {Elsner}, F. and {En{\ss}lin}, T.~A. and {Eriksen}, H.~K. and {Falgarone}, E. and {Fergusson}, J. and {Finelli}, F. and {Forni}, O. and {Frailis}, M. and {Fraisse}, A.~A. and {Franceschi}, E. and {Frejsel}, A. and {Galeotta}, S. and {Galli}, S. and {Ganga}, K. and {Ghosh}, T. and {Giard}, M. and {Giraud-H{\'e}raud}, Y. and {Gjerl{\o}w}, E. and {Gonz{\'a}lez-Nuevo}, J. and {G{\'o}rski}, K.~M. and {Gratton}, S. and {Gruppuso}, A. and {Gudmundsson}, J.~E. and {Hansen}, F.~K. and {Hanson}, D. and {Harrison}, D.~L. and {Henrot-Versill{\'e}}, S. and {Herranz}, D. and {Hildebrandt}, S.~R. and {Hivon}, E. and {Hobson}, M. and {Holmes}, W.~A. and {Hornstrup}, A. and {Hovest}, W. and {Huffenberger}, K.~M. and {Hurier}, G. and {Jaffe}, A.~H. and {Jaffe}, T.~R. and {Jones}, W.~C. and {Juvela}, M. and {Keih{\"a}nen}, E. and {Keskitalo}, R. and {Kisner}, T.~S. and {Kneissl}, R. and {Knoche}, J. and {Kunz}, M. and {Kurki-Suonio}, H. and {Lagache}, G. and {Lamarre}, J. -M. and {Lasenby}, A. and {Lattanzi}, M. and {Lawrence}, C.~R. and {Le Jeune}, M. and {Leahy}, J.~P. and {Lellouch}, E. and {Leonardi}, R. and {Lesgourgues}, J. and {Levrier}, F. and {Liguori}, M. and {Lilje}, P.~B. and {Linden-V{\o}rnle}, M. and {L{\'o}pez-Caniego}, M. and {Lubin}, P.~M. and {Mac{\'\i}as-P{\'e}rez}, J.~F. and {Maggio}, G. and {Maino}, D. and {Mandolesi}, N. and {Mangilli}, A. and {Maris}, M. and {Martin}, P.~G. and {Mart{\'\i}nez-Gonz{\'a}lez}, E. and {Masi}, S. and {Matarrese}, S. and {McGehee}, P. and {Melchiorri}, A. and {Mendes}, L. and {Mennella}, A. and {Migliaccio}, M. and {Mitra}, S. and {Miville-Desch{\^e}nes}, M. -A. and {Moneti}, A. and {Montier}, L. and {Moreno}, R. and {Morgante}, G. and {Mortlock}, D. and {Moss}, A. and {Mottet}, S. and {Munshi}, D. and {Murphy}, J.~A. and {Naselsky}, P. and {Nati}, F. and {Natoli}, P. and {Netterfield}, C.~B. and {N{\o}rgaard-Nielsen}, H.~U. and {Noviello}, F. and {Novikov}, D. and {Novikov}, I. and {Oxborrow}, C.~A. and {Paci}, F. and {Pagano}, L. and {Pajot}, F. and {Paoletti}, D. and {Pasian}, F. and {Patanchon}, G. and {Pearson}, T.~J. and {Perdereau}, O. and {Perotto}, L. and {Perrotta}, F. and {Pettorino}, V. and {Piacentini}, F. and {Piat}, M. and {Pierpaoli}, E. and {Pietrobon}, D. and {Plaszczynski}, S. and {Pointecouteau}, E. and {Polenta}, G. and {Pratt}, G.~W. and {Pr{\'e}zeau}, G. and {Prunet}, S. and {Puget}, J. -L. and {Rachen}, J.~P. and {Reinecke}, M. and {Remazeilles}, M. and {Renault}, C. and {Renzi}, A. and {Ristorcelli}, I. and {Rocha}, G. and {Rosset}, C. and {Rossetti}, M. and {Roudier}, G. and {Rowan-Robinson}, M. and {Rusholme}, B. and {Sandri}, M. and {Santos}, D. and {Sauv{\'e}}, A. and {Savelainen}, M. and {Savini}, G. and {Scott}, D. and {Seiffert}, M.~D. and {Shellard}, E.~P.~S. and {Spencer}, L.~D. and {Stolyarov}, V. and {Stompor}, R. and {Sudiwala}, R. and {Sutton}, D. and {Suur-Uski}, A. -S. and {Sygnet}, J. -F. and {Tauber}, J.~A. and {Terenzi}, L. and {Toffolatti}, L. and {Tomasi}, M. and {Tristram}, M. and {Tucci}, M. and {Tuovinen}, J. and {Valenziano}, L. and {Valiviita}, J. and {Van Tent}, B. and {Vibert}, L. and {Vielva}, P. and {Villa}, F. and {Wade}, L.~A. and {Wandelt}, B.~D. and {Watson}, R. and {Wehus}, I.~K. and {Yvon}, D. and {Zacchei}, A. and {Zonca}, A.},
        title = "{Planck 2015 results. VII. High Frequency Instrument data processing: Time-ordered information and beams}",
      journal = {\aap},
         year = 2016,
        month = sep,
       volume = {594},
          eid = {A7},
        pages = {A7},
          doi = {10.1051/0004-6361/201525844},
archivePrefix = {arXiv},
       eprint = {1502.01586},
 primaryClass = {astro-ph.IM},
       adsurl = {https://ui.adsabs.harvard.edu/abs/2016A&A...594A...7P}
}

@ARTICLE{2019MNRAS.484.4127A,
       author = {{Alonso}, David and {Sanchez}, Javier and {Slosar}, An{\v{z}}e and {LSST Dark Energy Science Collaboration}},
        title = "{A unified pseudo-C$_{{\ensuremath{\ell}}}$ framework}",
      journal = {\mnras},
         year = 2019,
        month = apr,
       volume = {484},
       number = {3},
        pages = {4127-4151},
          doi = {10.1093/mnras/stz093},
archivePrefix = {arXiv},
       eprint = {1809.09603},
 primaryClass = {astro-ph.CO},
       adsurl = {https://ui.adsabs.harvard.edu/abs/2019MNRAS.484.4127A}
}

@ARTICLE{2012ApJ...758...74B,
       author = {{Battaglia}, N. and {Bond}, J.~R. and {Pfrommer}, C. and {Sievers}, J.~L.},
        title = "{On the Cluster Physics of Sunyaev-Zel'dovich and X-Ray Surveys. I. The Influence of Feedback, Non-thermal Pressure, and Cluster Shapes on Y-M Scaling Relations}",
      journal = {\apj},
         year = 2012,
        month = oct,
       volume = {758},
       number = {2},
          eid = {74},
        pages = {74},
          doi = {10.1088/0004-637X/758/2/74},
archivePrefix = {arXiv},
       eprint = {1109.3709},
 primaryClass = {astro-ph.CO},
       adsurl = {https://ui.adsabs.harvard.edu/abs/2012ApJ...758...74B}
}

@ARTICLE{2011ApJ...727...94T,
       author = {{Trac}, Hy and {Bode}, Paul and {Ostriker}, Jeremiah P.},
        title = "{Templates for the Sunyaev-Zel'dovich Angular Power Spectrum}",
      journal = {\apj},
         year = 2011,
        month = feb,
       volume = {727},
       number = {2},
          eid = {94},
        pages = {94},
          doi = {10.1088/0004-637X/727/2/94},
archivePrefix = {arXiv},
       eprint = {1006.2828},
 primaryClass = {astro-ph.CO},
       adsurl = {https://ui.adsabs.harvard.edu/abs/2011ApJ...727...94T}
}

@ARTICLE{2010ApJ...725.1452S,
       author = {{Shaw}, Laurie D. and {Nagai}, Daisuke and {Bhattacharya}, Suman and {Lau}, Erwin T.},
        title = "{Impact of Cluster Physics on the Sunyaev-Zel'dovich Power Spectrum}",
      journal = {\apj},
         year = 2010,
        month = dec,
       volume = {725},
       number = {2},
        pages = {1452-1465},
          doi = {10.1088/0004-637X/725/2/1452},
archivePrefix = {arXiv},
       eprint = {1006.1945},
 primaryClass = {astro-ph.CO},
       adsurl = {https://ui.adsabs.harvard.edu/abs/2010ApJ...725.1452S}
}

@ARTICLE{2009ApJ...702..368S,
       author = {{Shaw}, Laurie D. and {Zahn}, Oliver and {Holder}, Gilbert P. and {Dor{\'e}}, Olivier},
        title = "{Sharpening the Precision of the Sunyaev-Zel'dovich Power Spectrum}",
      journal = {\apj},
         year = 2009,
        month = sep,
       volume = {702},
       number = {1},
        pages = {368-376},
          doi = {10.1088/0004-637X/702/1/368},
archivePrefix = {arXiv},
       eprint = {0903.5322},
 primaryClass = {astro-ph.CO},
       adsurl = {https://ui.adsabs.harvard.edu/abs/2009ApJ...702..368S}
}

@ARTICLE{2005ApJ...632....1C,
       author = {{Cohn}, J.~D. and {Kadota}, Kenji},
        title = "{Uncertainties in the Sunyaev-Zel'dovich-selected Cluster Angular Power Spectrum}",
      journal = {\apj},
         year = 2005,
        month = oct,
       volume = {632},
       number = {1},
        pages = {1-14},
          doi = {10.1086/432706},
archivePrefix = {arXiv},
       eprint = {astro-ph/0409657},
 primaryClass = {astro-ph},
       adsurl = {https://ui.adsabs.harvard.edu/abs/2005ApJ...632....1C}
}

@ARTICLE{2003ApJ...598...49Z,
       author = {{Zentner}, Andrew R. and {Bullock}, James S.},
        title = "{Halo Substructure and the Power Spectrum}",
      journal = {\apj},
         year = 2003,
        month = nov,
       volume = {598},
       number = {1},
        pages = {49-72},
          doi = {10.1086/378797},
archivePrefix = {arXiv},
       eprint = {astro-ph/0304292},
 primaryClass = {astro-ph},
       adsurl = {https://ui.adsabs.harvard.edu/abs/2003ApJ...598...49Z}
}

@ARTICLE{2002ApJ...577..555Z,
       author = {{Zhang}, Pengjie and {Pen}, Ue-Li and {Wang}, Benjamin},
        title = "{The Sunyaev-Zeldovich Effect: Simulations and Observations}",
      journal = {\apj},
         year = 2002,
        month = oct,
       volume = {577},
       number = {2},
        pages = {555-568},
          doi = {10.1086/342149},
archivePrefix = {arXiv},
       eprint = {astro-ph/0201375},
 primaryClass = {astro-ph},
       adsurl = {https://ui.adsabs.harvard.edu/abs/2002ApJ...577..555Z}
}

@ARTICLE{2000PhRvD..62j3506C,
       author = {{Cooray}, Asantha},
        title = "{Large scale pressure fluctuations and the Sunyaev-Zel'dovich effect}",
      journal = {\prd},
         year = 2000,
        month = nov,
       volume = {62},
       number = {10},
          eid = {103506},
        pages = {103506},
          doi = {10.1103/PhysRevD.62.103506},
archivePrefix = {arXiv},
       eprint = {astro-ph/0005287},
 primaryClass = {astro-ph},
       adsurl = {https://ui.adsabs.harvard.edu/abs/2000PhRvD..62j3506C}
}

@ARTICLE{2023MNRAS.519.2138A,
       author = {{Acharya}, Sandeep Kumar and {Chluba}, Jens},
        title = "{Importance of intracluster scattering and relativistic corrections from tSZ effect with cosmic infrared background}",
      journal = {\mnras},
         year = 2023,
        month = feb,
       volume = {519},
       number = {2},
        pages = {2138-2154},
          doi = {10.1093/mnras/stac3714},
archivePrefix = {arXiv},
       eprint = {2205.00857},
 primaryClass = {astro-ph.CO},
       adsurl = {https://ui.adsabs.harvard.edu/abs/2023MNRAS.519.2138A}
}

@ARTICLE{2016A&A...596A..61H,
       author = {{Hurier}, G.},
        title = "{High significance detection of the tSZ effect relativistic corrections}",
      journal = {\aap},
         year = 2016,
        month = dec,
       volume = {596},
          eid = {A61},
        pages = {A61},
          doi = {10.1051/0004-6361/201629726},
archivePrefix = {arXiv},
       eprint = {1701.09020},
 primaryClass = {astro-ph.CO},
       adsurl = {https://ui.adsabs.harvard.edu/abs/2016A&A...596A..61H}
}

@ARTICLE{2019MNRAS.483.3459R,
       author = {{Remazeilles}, Mathieu and {Bolliet}, Boris and {Rotti}, Aditya and {Chluba}, Jens},
        title = "{Can we neglect relativistic temperature corrections in the Planck thermal SZ analysis?}",
      journal = {\mnras},
         year = 2019,
        month = mar,
       volume = {483},
       number = {3},
        pages = {3459-3464},
          doi = {10.1093/mnras/sty3352},
archivePrefix = {arXiv},
       eprint = {1809.09666},
 primaryClass = {astro-ph.CO},
       adsurl = {https://ui.adsabs.harvard.edu/abs/2019MNRAS.483.3459R}
}

@ARTICLE{2021A&A...650A..65S,
       author = {{Santos}, Larissa and {Yao}, Jian and {Zhang}, Le and {Ghosh}, Shamik and {Zhang}, Pengjie and {Zhao}, Wen and {Villela}, Thyrso and {Chen}, Jiming and {Delabrouille}, Jacques},
        title = "{Testing the analytical blind separation method in simulated CMB polarization maps}",
      journal = {\aap},
         year = 2021,
        month = jun,
       volume = {650},
          eid = {A65},
        pages = {A65},
          doi = {10.1051/0004-6361/201936546},
       adsurl = {https://ui.adsabs.harvard.edu/abs/2021A&A...650A..65S}
}

@ARTICLE{2022JCAP...10..063G,
       author = {{Ghosh}, Shamik and {Liu}, Yang and {Zhang}, Le and {Li}, Siyu and {Zhang}, Junzhou and {Wang}, Jiaxin and {Dou}, Jiazheng and {Chen}, Jiming and {Delabrouille}, Jacques and {Remazeilles}, Mathieu and {Feng}, Chang and {Hu}, Bin and {Huang}, Zhi-Qi and {Liu}, Hao and {Santos}, Larissa and {Zhang}, Pengjie and {Zhang}, Zhaoxuan and {Zhao}, Wen and {Li}, Hong and {Zhang}, Xinmin},
        title = "{Performance forecasts for the primordial gravitational wave detection pipelines for AliCPT-1}",
      journal = {\jcap},
         year = 2022,
        month = oct,
       volume = {2022},
       number = {10},
          eid = {063},
        pages = {063},
          doi = {10.1088/1475-7516/2022/10/063},
archivePrefix = {arXiv},
       eprint = {2205.14804},
 primaryClass = {astro-ph.CO},
       adsurl = {https://ui.adsabs.harvard.edu/abs/2022JCAP...10..063G}
}

@ARTICLE{2024arXiv240201233Z,
       author = {{Zhang}, Junzhou and {Ghosh}, Shamik and {Dou}, Jiazheng and {Liu}, Yang and {Li}, Siyu and {Chen}, Jiming and {Wang}, Jiaxin and {Zhang}, Zhaoxuan and {Delabrouille}, Jacques and {Remazeilles}, Mathieu and {Feng}, Chang and {Hu}, Bin and {Liu}, Hao and {Santos}, Larissa and {Zhang}, Pengjie and {Zhao}, Wen and {Zhang}, Le and {Huang}, Zhi-Qi and {Li}, Hong and {Kuo}, Chao-Lin and {Zhang}, Xinmin},
        title = "{Forecast of foreground cleaning strategies for AliCPT-1}",
      journal = {arXiv e-prints},
         year = 2024,
        month = feb,
          eid = {arXiv:2402.01233},
        pages = {arXiv:2402.01233},
          doi = {10.48550/arXiv.2402.01233},
archivePrefix = {arXiv},
       eprint = {2402.01233},
 primaryClass = {astro-ph.CO},
       adsurl = {https://ui.adsabs.harvard.edu/abs/2024arXiv240201233Z}
}

@article{10.1093/mnras/stz091,
    author = {Zhang, Pengjie and Zhang, Jun and Zhang, Le},
    title = "{ABS: an analytical method of blind separation of CMB from foregrounds}",
    journal = {Monthly Notices of the Royal Astronomical Society},
    volume = {484},
    number = {2},
    pages = {1616-1626},
    year = {2019},
    month = {01},
    issn = {0035-8711},
    doi = {10.1093/mnras/stz091},
    url = {https://doi.org/10.1093/mnras/stz091},
    eprint = {https://academic.oup.com/mnras/article-pdf/484/2/1616/27583099/stz091.pdf},
}

@ARTICLE{Healpix2005,
       author = {{G{\'o}rski}, K.~M. and {Hivon}, E. and {Banday}, A.~J. and {Wandelt}, B.~D. and {Hansen}, F.~K. and {Reinecke}, M. and {Bartelmann}, M.},
        title = "{HEALPix: A Framework for High-Resolution Discretization and Fast Analysis of Data Distributed on the Sphere}",
      journal = {\apj},
         year = 2005,
        month = apr,
       volume = {622},
       number = {2},
        pages = {759-771},
          doi = {10.1086/427976},
archivePrefix = {arXiv},
       eprint = {astro-ph/0409513},
 primaryClass = {astro-ph},
       adsurl = {https://ui.adsabs.harvard.edu/abs/2005ApJ...622..759G}
}

@ARTICLE{2020A&A...643A..42P,
       author = {{Planck Collaboration} and {Akrami}, Y. and {Andersen}, K.~J. and {Ashdown}, M. and {Baccigalupi}, C. and {Ballardini}, M. and {Banday}, A.~J. and {Barreiro}, R.~B. and {Bartolo}, N. and {Basak}, S. and {Benabed}, K. and {Bernard}, J. -P. and {Bersanelli}, M. and {Bielewicz}, P. and {Bond}, J.~R. and {Borrill}, J. and {Burigana}, C. and {Butler}, R.~C. and {Calabrese}, E. and {Casaponsa}, B. and {Chiang}, H.~C. and {Colombo}, L.~P.~L. and {Combet}, C. and {Crill}, B.~P. and {Cuttaia}, F. and {de Bernardis}, P. and {de Rosa}, A. and {de Zotti}, G. and {Delabrouille}, J. and {Di Valentino}, E. and {Diego}, J.~M. and {Dor{\'e}}, O. and {Douspis}, M. and {Dupac}, X. and {Eriksen}, H.~K. and {Fernandez-Cobos}, R. and {Finelli}, F. and {Frailis}, M. and {Fraisse}, A.~A. and {Franceschi}, E. and {Frolov}, A. and {Galeotta}, S. and {Galli}, S. and {Ganga}, K. and {Gerbino}, M. and {Ghosh}, T. and {Gonz{\'a}lez-Nuevo}, J. and {G{\'o}rski}, K.~M. and {Gruppuso}, A. and {Gudmundsson}, J.~E. and {Handley}, W. and {Helou}, G. and {Herranz}, D. and {Hildebrandt}, S.~R. and {Hivon}, E. and {Huang}, Z. and {Jaffe}, A.~H. and {Jones}, W.~C. and {Keih{\"a}nen}, E. and {Keskitalo}, R. and {Kiiveri}, K. and {Kim}, J. and {Kisner}, T.~S. and {Krachmalnicoff}, N. and {Kunz}, M. and {Kurki-Suonio}, H. and {Lasenby}, A. and {Lattanzi}, M. and {Lawrence}, C.~R. and {Le Jeune}, M. and {Levrier}, F. and {Liguori}, M. and {Lilje}, P.~B. and {Lilley}, M. and {Lindholm}, V. and {L{\'o}pez-Caniego}, M. and {Lubin}, P.~M. and {Mac{\'\i}as-P{\'e}rez}, J.~F. and {Maino}, D. and {Mandolesi}, N. and {Marcos-Caballero}, A. and {Maris}, M. and {Martin}, P.~G. and {Mart{\'\i}nez-Gonz{\'a}lez}, E. and {Matarrese}, S. and {Mauri}, N. and {McEwen}, J.~D. and {Meinhold}, P.~R. and {Mennella}, A. and {Migliaccio}, M. and {Mitra}, S. and {Molinari}, D. and {Montier}, L. and {Morgante}, G. and {Moss}, A. and {Natoli}, P. and {Paoletti}, D. and {Partridge}, B. and {Patanchon}, G. and {Pearson}, D. and {Pearson}, T.~J. and {Perrotta}, F. and {Piacentini}, F. and {Polenta}, G. and {Rachen}, J.~P. and {Reinecke}, M. and {Remazeilles}, M. and {Renzi}, A. and {Rocha}, G. and {Rosset}, C. and {Roudier}, G. and {Rubi{\~n}o-Mart{\'\i}n}, J.~A. and {Ruiz-Granados}, B. and {Salvati}, L. and {Savelainen}, M. and {Scott}, D. and {Sirignano}, C. and {Sirri}, G. and {Spencer}, L.~D. and {Suur-Uski}, A. -S. and {Svalheim}, L.~T. and {Tauber}, J.~A. and {Tavagnacco}, D. and {Tenti}, M. and {Terenzi}, L. and {Thommesen}, H. and {Toffolatti}, L. and {Tomasi}, M. and {Tristram}, M. and {Trombetti}, T. and {Valiviita}, J. and {Van Tent}, B. and {Vielva}, P. and {Villa}, F. and {Vittorio}, N. and {Wandelt}, B.~D. and {Wehus}, I.~K. and {Zacchei}, A. and {Zonca}, A.},
        title = "{Planck intermediate results. LVII. Joint Planck LFI and HFI data processing}",
      journal = {\aap},
         year = 2020,
        month = nov,
       volume = {643},
          eid = {A42},
        pages = {A42},
          doi = {10.1051/0004-6361/202038073},
archivePrefix = {arXiv},
       eprint = {2007.04997},
 primaryClass = {astro-ph.CO},
       adsurl = {https://ui.adsabs.harvard.edu/abs/2020A&A...643A..42P}
}

@ARTICLE{2022MNRAS.509..300T,
       author = {{Tanimura}, Hideki and {Douspis}, Marian and {Aghanim}, Nabila and {Salvati}, Laura},
        title = "{Constraining cosmology with a new all-sky Compton parameter map from the Planck PR4 data}",
      journal = {\mnras},
         year = 2022,
        month = jan,
       volume = {509},
       number = {1},
        pages = {300-313},
          doi = {10.1093/mnras/stab2956},
archivePrefix = {arXiv},
       eprint = {2110.08880},
 primaryClass = {astro-ph.CO},
       adsurl = {https://ui.adsabs.harvard.edu/abs/2022MNRAS.509..300T}
}

@ARTICLE{2011MNRAS.410.2481R,
       author = {{Remazeilles}, Mathieu and {Delabrouille}, Jacques and {Cardoso}, Jean-Fran{\c{c}}ois},
        title = "{CMB and SZ effect separation with constrained Internal Linear Combinations}",
      journal = {\mnras},
         year = 2011,
        month = feb,
       volume = {410},
       number = {4},
        pages = {2481-2487},
          doi = {10.1111/j.1365-2966.2010.17624.x},
archivePrefix = {arXiv},
       eprint = {1006.5599},
 primaryClass = {astro-ph.CO},
       adsurl = {https://ui.adsabs.harvard.edu/abs/2011MNRAS.410.2481R}
}

@ARTICLE{2013A&A...558A.118H,
       author = {{Hurier}, G. and {Mac{\'\i}as-P{\'e}rez}, J.~F. and {Hildebrandt}, S.},
        title = "{MILCA, a modified internal linear combination algorithm to extract astrophysical emissions from multifrequency sky maps}",
      journal = {\aap},
         year = 2013,
        month = oct,
       volume = {558},
          eid = {A118},
        pages = {A118},
          doi = {10.1051/0004-6361/201321891},
archivePrefix = {arXiv},
       eprint = {1007.1149},
 primaryClass = {astro-ph.IM},
       adsurl = {https://ui.adsabs.harvard.edu/abs/2013A&A...558A.118H}
}

@ARTICLE{2014A&A...571A..12P,
       author = {{Planck Collaboration} and {Ade}, P.~A.~R. and {Aghanim}, N. and {Armitage-Caplan}, C. and {Arnaud}, M. and {Ashdown}, M. and {Atrio-Barandela}, F. and {Aumont}, J. and {Baccigalupi}, C. and {Banday}, A.~J. and {Barreiro}, R.~B. and {Bartlett}, J.~G. and {Battaner}, E. and {Benabed}, K. and {Beno{\^\i}t}, A. and {Benoit-L{\'e}vy}, A. and {Bernard}, J. -P. and {Bersanelli}, M. and {Bielewicz}, P. and {Bobin}, J. and {Bock}, J.~J. and {Bonaldi}, A. and {Bonavera}, L. and {Bond}, J.~R. and {Borrill}, J. and {Bouchet}, F.~R. and {Boulanger}, F. and {Bridges}, M. and {Bucher}, M. and {Burigana}, C. and {Butler}, R.~C. and {Cardoso}, J. -F. and {Castex}, G. and {Catalano}, A. and {Challinor}, A. and {Chamballu}, A. and {Chary}, R. -R. and {Chen}, X. and {Chiang}, H.~C. and {Chiang}, L. -Y. and {Christensen}, P.~R. and {Church}, S. and {Clements}, D.~L. and {Colombi}, S. and {Colombo}, L.~P.~L. and {Couchot}, F. and {Coulais}, A. and {Crill}, B.~P. and {Cruz}, M. and {Curto}, A. and {Cuttaia}, F. and {Danese}, L. and {Davies}, R.~D. and {Davis}, R.~J. and {de Bernardis}, P. and {de Rosa}, A. and {de Zotti}, G. and {Delabrouille}, J. and {Delouis}, J. -M. and {D{\'e}sert}, F. -X. and {Dickinson}, C. and {Diego}, J.~M. and {Dobler}, G. and {Dole}, H. and {Donzelli}, S. and {Dor{\'e}}, O. and {Douspis}, M. and {Dunkley}, J. and {Dupac}, X. and {Efstathiou}, G. and {En{\ss}lin}, T.~A. and {Eriksen}, H.~K. and {Falgarone}, E. and {Finelli}, F. and {Forni}, O. and {Frailis}, M. and {Fraisse}, A.~A. and {Franceschi}, E. and {Galeotta}, S. and {Ganga}, K. and {Giard}, M. and {Giardino}, G. and {Giraud-H{\'e}raud}, Y. and {Gonz{\'a}lez-Nuevo}, J. and {G{\'o}rski}, K.~M. and {Gratton}, S. and {Gregorio}, A. and {Gruppuso}, A. and {Hansen}, F.~K. and {Hanson}, D. and {Harrison}, D.~L. and {Helou}, G. and {Henrot-Versill{\'e}}, S. and {Hern{\'a}ndez-Monteagudo}, C. and {Herranz}, D. and {Hildebrandt}, S.~R. and {Hivon}, E. and {Hobson}, M. and {Holmes}, W.~A. and {Hornstrup}, A. and {Hovest}, W. and {Huey}, G. and {Huffenberger}, K.~M. and {Jaffe}, A.~H. and {Jaffe}, T.~R. and {Jewell}, J. and {Jones}, W.~C. and {Juvela}, M. and {Keih{\"a}nen}, E. and {Keskitalo}, R. and {Kisner}, T.~S. and {Kneissl}, R. and {Knoche}, J. and {Knox}, L. and {Kunz}, M. and {Kurki-Suonio}, H. and {Lagache}, G. and {L{\"a}hteenm{\"a}ki}, A. and {Lamarre}, J. -M. and {Lasenby}, A. and {Laureijs}, R.~J. and {Lawrence}, C.~R. and {Le Jeune}, M. and {Leach}, S. and {Leahy}, J.~P. and {Leonardi}, R. and {Lesgourgues}, J. and {Liguori}, M. and {Lilje}, P.~B. and {Linden-V{\o}rnle}, M. and {L{\'o}pez-Caniego}, M. and {Lubin}, P.~M. and {Mac{\'\i}as-P{\'e}rez}, J.~F. and {Maffei}, B. and {Maino}, D. and {Mandolesi}, N. and {Marcos-Caballero}, A. and {Maris}, M. and {Marshall}, D.~J. and {Martin}, P.~G. and {Mart{\'\i}nez-Gonz{\'a}lez}, E. and {Masi}, S. and {Massardi}, M. and {Matarrese}, S. and {Matthai}, F. and {Mazzotta}, P. and {Meinhold}, P.~R. and {Melchiorri}, A. and {Mendes}, L. and {Mennella}, A. and {Migliaccio}, M. and {Mikkelsen}, K. and {Mitra}, S. and {Miville-Desch{\^e}nes}, M. -A. and {Molinari}, D. and {Moneti}, A. and {Montier}, L. and {Morgante}, G. and {Mortlock}, D. and {Moss}, A. and {Munshi}, D. and {Murphy}, J.~A. and {Naselsky}, P. and {Nati}, F. and {Natoli}, P. and {Netterfield}, C.~B. and {N{\o}rgaard-Nielsen}, H.~U. and {Noviello}, F. and {Novikov}, D. and {Novikov}, I. and {O'Dwyer}, I.~J. and {Osborne}, S. and {Oxborrow}, C.~A. and {Paci}, F. and {Pagano}, L. and {Pajot}, F. and {Paladini}, R. and {Paoletti}, D. and {Partridge}, B. and {Pasian}, F. and {Patanchon}, G. and {Pearson}, T.~J. and {Perdereau}, O. and {Perotto}, L. and {Perrotta}, F. and {Pettorino}, V. and {Piacentini}, F. and {Piat}, M. and {Pierpaoli}, E. and {Pietrobon}, D. and {Plaszczynski}, S. and {Platania}, P. and {Pointecouteau}, E. and {Polenta}, G. and {Ponthieu}, N. and {Popa}, L. and {Poutanen}, T. and {Pratt}, G.~W. and {Pr{\'e}zeau}, G. and {Prunet}, S. and {Puget}, J. -L. and {Rachen}, J.~P. and {Reach}, W.~T. and {Rebolo}, R. and {Reinecke}, M. and {Remazeilles}, M. and {Renault}, C. and {Renzi}, A. and {Ricciardi}, S. and {Riller}, T. and {Ristorcelli}, I. and {Rocha}, G. and {Roman}, M. and {Rosset}, C. and {Roudier}, G. and {Rowan-Robinson}, M. and {Rubi{\~n}o-Mart{\'\i}n}, J.~A. and {Rusholme}, B. and {Salerno}, E. and {Sandri}, M. and {Santos}, D. and {Savini}, G. and {Schiavon}, F. and {Scott}, D. and {Seiffert}, M.~D. and {Shellard}, E.~P.~S. and {Spencer}, L.~D. and {Starck}, J. -L. and {Stompor}, R. and {Sudiwala}, R. and {Sunyaev}, R. and {Sureau}, F. and {Sutton}, D. and {Suur-Uski}, A. -S. and {Sygnet}, J. -F. and {Tauber}, J.~A. and {Tavagnacco}, D. and {Terenzi}, L. and {Toffolatti}, L. and {Tomasi}, M. and {Tristram}, M. and {Tucci}, M. and {Tuovinen}, J. and {T{\"u}rler}, M. and {Umana}, G. and {Valenziano}, L. and {Valiviita}, J. and {Van Tent}, B. and {Varis}, J. and {Viel}, M. and {Vielva}, P. and {Villa}, F. and {Vittorio}, N. and {Wade}, L.~A. and {Wandelt}, B.~D. and {Wehus}, I.~K. and {Wilkinson}, A. and {Xia}, J. -Q. and {Yvon}, D. and {Zacchei}, A. and {Zonca}, A.},
        title = "{Planck 2013 results. XII. Diffuse component separation}",
      journal = {\aap},
         year = 2014,
        month = nov,
       volume = {571},
          eid = {A12},
        pages = {A12},
          doi = {10.1051/0004-6361/201321580},
archivePrefix = {arXiv},
       eprint = {1303.5072},
 primaryClass = {astro-ph.CO},
       adsurl = {https://ui.adsabs.harvard.edu/abs/2014A&A...571A..12P}
}

@ARTICLE{2013ApJ...763..127R,
       author = {{Reichardt}, C.~L. and {Stalder}, B. and {Bleem}, L.~E. and {Montroy}, T.~E. and {Aird}, K.~A. and {Andersson}, K. and {Armstrong}, R. and {Ashby}, M.~L.~N. and {Bautz}, M. and {Bayliss}, M. and {Bazin}, G. and {Benson}, B.~A. and {Brodwin}, M. and {Carlstrom}, J.~E. and {Chang}, C.~L. and {Cho}, H.~M. and {Clocchiatti}, A. and {Crawford}, T.~M. and {Crites}, A.~T. and {de Haan}, T. and {Desai}, S. and {Dobbs}, M.~A. and {Dudley}, J.~P. and {Foley}, R.~J. and {Forman}, W.~R. and {George}, E.~M. and {Gladders}, M.~D. and {Gonzalez}, A.~H. and {Halverson}, N.~W. and {Harrington}, N.~L. and {High}, F.~W. and {Holder}, G.~P. and {Holzapfel}, W.~L. and {Hoover}, S. and {Hrubes}, J.~D. and {Jones}, C. and {Joy}, M. and {Keisler}, R. and {Knox}, L. and {Lee}, A.~T. and {Leitch}, E.~M. and {Liu}, J. and {Lueker}, M. and {Luong-Van}, D. and {Mantz}, A. and {Marrone}, D.~P. and {McDonald}, M. and {McMahon}, J.~J. and {Mehl}, J. and {Meyer}, S.~S. and {Mocanu}, L. and {Mohr}, J.~J. and {Murray}, S.~S. and {Natoli}, T. and {Padin}, S. and {Plagge}, T. and {Pryke}, C. and {Rest}, A. and {Ruel}, J. and {Ruhl}, J.~E. and {Saliwanchik}, B.~R. and {Saro}, A. and {Sayre}, J.~T. and {Schaffer}, K.~K. and {Shaw}, L. and {Shirokoff}, E. and {Song}, J. and {Spieler}, H.~G. and {Staniszewski}, Z. and {Stark}, A.~A. and {Story}, K. and {Stubbs}, C.~W. and {{\v{S}}uhada}, R. and {van Engelen}, A. and {Vanderlinde}, K. and {Vieira}, J.~D. and {Vikhlinin}, A. and {Williamson}, R. and {Zahn}, O. and {Zenteno}, A.},
        title = "{Galaxy Clusters Discovered via the Sunyaev-Zel'dovich Effect in the First 720 Square Degrees of the South Pole Telescope Survey}",
      journal = {\apj},
         year = 2013,
        month = feb,
       volume = {763},
       number = {2},
          eid = {127},
        pages = {127},
          doi = {10.1088/0004-637X/763/2/127},
archivePrefix = {arXiv},
       eprint = {1203.5775},
 primaryClass = {astro-ph.CO},
       adsurl = {https://ui.adsabs.harvard.edu/abs/2013ApJ...763..127R}
}

@ARTICLE{2017MNRAS.469..394H,
       author = {{Horowitz}, B. and {Seljak}, U.},
        title = "{Cosmological constraints from thermal Sunyaev-Zeldovich power spectrum revisited}",
      journal = {\mnras},
         year = 2017,
        month = jul,
       volume = {469},
       number = {1},
        pages = {394-400},
          doi = {10.1093/mnras/stx766},
archivePrefix = {arXiv},
       eprint = {1609.01850},
 primaryClass = {astro-ph.CO},
       adsurl = {https://ui.adsabs.harvard.edu/abs/2017MNRAS.469..394H}
}

@ARTICLE{2016MNRAS.463.1797D,
       author = {{Dolag}, K. and {Komatsu}, E. and {Sunyaev}, R.},
        title = "{SZ effects in the Magneticum Pathfinder simulation: comparison with the Planck, SPT, and ACT results}",
      journal = {\mnras},
         year = 2016,
        month = dec,
       volume = {463},
       number = {2},
        pages = {1797-1811},
          doi = {10.1093/mnras/stw2035},
archivePrefix = {arXiv},
       eprint = {1509.05134},
 primaryClass = {astro-ph.CO},
       adsurl = {https://ui.adsabs.harvard.edu/abs/2016MNRAS.463.1797D}
}

@ARTICLE{2016MNRAS.456.2361B,
       author = {{Bocquet}, Sebastian and {Saro}, Alex and {Dolag}, Klaus and {Mohr}, Joseph J.},
        title = "{Halo mass function: baryon impact, fitting formulae, and implications for cluster cosmology}",
      journal = {\mnras},
         year = 2016,
        month = mar,
       volume = {456},
       number = {3},
        pages = {2361-2373},
          doi = {10.1093/mnras/stv2657},
archivePrefix = {arXiv},
       eprint = {1502.07357},
 primaryClass = {astro-ph.CO},
       adsurl = {https://ui.adsabs.harvard.edu/abs/2016MNRAS.456.2361B}
}

@ARTICLE{2016A&A...594A..13P,
       author = {{Planck Collaboration} and {Ade}, P.~A.~R. and {Aghanim}, N. and {Arnaud}, M. and {Ashdown}, M. and {Aumont}, J. and {Baccigalupi}, C. and {Banday}, A.~J. and {Barreiro}, R.~B. and {Bartlett}, J.~G. and {Bartolo}, N. and {Battaner}, E. and {Battye}, R. and {Benabed}, K. and {Beno{\^\i}t}, A. and {Benoit-L{\'e}vy}, A. and {Bernard}, J. -P. and {Bersanelli}, M. and {Bielewicz}, P. and {Bock}, J.~J. and {Bonaldi}, A. and {Bonavera}, L. and {Bond}, J.~R. and {Borrill}, J. and {Bouchet}, F.~R. and {Boulanger}, F. and {Bucher}, M. and {Burigana}, C. and {Butler}, R.~C. and {Calabrese}, E. and {Cardoso}, J. -F. and {Catalano}, A. and {Challinor}, A. and {Chamballu}, A. and {Chary}, R. -R. and {Chiang}, H.~C. and {Chluba}, J. and {Christensen}, P.~R. and {Church}, S. and {Clements}, D.~L. and {Colombi}, S. and {Colombo}, L.~P.~L. and {Combet}, C. and {Coulais}, A. and {Crill}, B.~P. and {Curto}, A. and {Cuttaia}, F. and {Danese}, L. and {Davies}, R.~D. and {Davis}, R.~J. and {de Bernardis}, P. and {de Rosa}, A. and {de Zotti}, G. and {Delabrouille}, J. and {D{\'e}sert}, F. -X. and {Di Valentino}, E. and {Dickinson}, C. and {Diego}, J.~M. and {Dolag}, K. and {Dole}, H. and {Donzelli}, S. and {Dor{\'e}}, O. and {Douspis}, M. and {Ducout}, A. and {Dunkley}, J. and {Dupac}, X. and {Efstathiou}, G. and {Elsner}, F. and {En{\ss}lin}, T.~A. and {Eriksen}, H.~K. and {Farhang}, M. and {Fergusson}, J. and {Finelli}, F. and {Forni}, O. and {Frailis}, M. and {Fraisse}, A.~A. and {Franceschi}, E. and {Frejsel}, A. and {Galeotta}, S. and {Galli}, S. and {Ganga}, K. and {Gauthier}, C. and {Gerbino}, M. and {Ghosh}, T. and {Giard}, M. and {Giraud-H{\'e}raud}, Y. and {Giusarma}, E. and {Gjerl{\o}w}, E. and {Gonz{\'a}lez-Nuevo}, J. and {G{\'o}rski}, K.~M. and {Gratton}, S. and {Gregorio}, A. and {Gruppuso}, A. and {Gudmundsson}, J.~E. and {Hamann}, J. and {Hansen}, F.~K. and {Hanson}, D. and {Harrison}, D.~L. and {Helou}, G. and {Henrot-Versill{\'e}}, S. and {Hern{\'a}ndez-Monteagudo}, C. and {Herranz}, D. and {Hildebrandt}, S.~R. and {Hivon}, E. and {Hobson}, M. and {Holmes}, W.~A. and {Hornstrup}, A. and {Hovest}, W. and {Huang}, Z. and {Huffenberger}, K.~M. and {Hurier}, G. and {Jaffe}, A.~H. and {Jaffe}, T.~R. and {Jones}, W.~C. and {Juvela}, M. and {Keih{\"a}nen}, E. and {Keskitalo}, R. and {Kisner}, T.~S. and {Kneissl}, R. and {Knoche}, J. and {Knox}, L. and {Kunz}, M. and {Kurki-Suonio}, H. and {Lagache}, G. and {L{\"a}hteenm{\"a}ki}, A. and {Lamarre}, J. -M. and {Lasenby}, A. and {Lattanzi}, M. and {Lawrence}, C.~R. and {Leahy}, J.~P. and {Leonardi}, R. and {Lesgourgues}, J. and {Levrier}, F. and {Lewis}, A. and {Liguori}, M. and {Lilje}, P.~B. and {Linden-V{\o}rnle}, M. and {L{\'o}pez-Caniego}, M. and {Lubin}, P.~M. and {Mac{\'\i}as-P{\'e}rez}, J.~F. and {Maggio}, G. and {Maino}, D. and {Mandolesi}, N. and {Mangilli}, A. and {Marchini}, A. and {Maris}, M. and {Martin}, P.~G. and {Martinelli}, M. and {Mart{\'\i}nez-Gonz{\'a}lez}, E. and {Masi}, S. and {Matarrese}, S. and {McGehee}, P. and {Meinhold}, P.~R. and {Melchiorri}, A. and {Melin}, J. -B. and {Mendes}, L. and {Mennella}, A. and {Migliaccio}, M. and {Millea}, M. and {Mitra}, S. and {Miville-Desch{\^e}nes}, M. -A. and {Moneti}, A. and {Montier}, L. and {Morgante}, G. and {Mortlock}, D. and {Moss}, A. and {Munshi}, D. and {Murphy}, J.~A. and {Naselsky}, P. and {Nati}, F. and {Natoli}, P. and {Netterfield}, C.~B. and {N{\o}rgaard-Nielsen}, H.~U. and {Noviello}, F. and {Novikov}, D. and {Novikov}, I. and {Oxborrow}, C.~A. and {Paci}, F. and {Pagano}, L. and {Pajot}, F. and {Paladini}, R. and {Paoletti}, D. and {Partridge}, B. and {Pasian}, F. and {Patanchon}, G. and {Pearson}, T.~J. and {Perdereau}, O. and {Perotto}, L. and {Perrotta}, F. and {Pettorino}, V. and {Piacentini}, F. and {Piat}, M. and {Pierpaoli}, E. and {Pietrobon}, D. and {Plaszczynski}, S. and {Pointecouteau}, E. and {Polenta}, G. and {Popa}, L. and {Pratt}, G.~W. and {Pr{\'e}zeau}, G. and {Prunet}, S. and {Puget}, J. -L. and {Rachen}, J.~P. and {Reach}, W.~T. and {Rebolo}, R. and {Reinecke}, M. and {Remazeilles}, M. and {Renault}, C. and {Renzi}, A. and {Ristorcelli}, I. and {Rocha}, G. and {Rosset}, C. and {Rossetti}, M. and {Roudier}, G. and {Rouill{\'e} d'Orfeuil}, B. and {Rowan-Robinson}, M. and {Rubi{\~n}o-Mart{\'\i}n}, J.~A. and {Rusholme}, B. and {Said}, N. and {Salvatelli}, V. and {Salvati}, L. and {Sandri}, M. and {Santos}, D. and {Savelainen}, M. and {Savini}, G. and {Scott}, D. and {Seiffert}, M.~D. and {Serra}, P. and {Shellard}, E.~P.~S. and {Spencer}, L.~D. and {Spinelli}, M. and {Stolyarov}, V. and {Stompor}, R. and {Sudiwala}, R. and {Sunyaev}, R. and {Sutton}, D. and {Suur-Uski}, A. -S. and {Sygnet}, J. -F. and {Tauber}, J.~A. and {Terenzi}, L. and {Toffolatti}, L. and {Tomasi}, M. and {Tristram}, M. and {Trombetti}, T. and {Tucci}, M. and {Tuovinen}, J. and {T{\"u}rler}, M. and {Umana}, G. and {Valenziano}, L. and {Valiviita}, J. and {Van Tent}, F. and {Vielva}, P. and {Villa}, F. and {Wade}, L.~A. and {Wandelt}, B.~D. and {Wehus}, I.~K. and {White}, M. and {White}, S.~D.~M. and {Wilkinson}, A. and {Yvon}, D. and {Zacchei}, A. and {Zonca}, A.},
        title = "{Planck 2015 results. XIII. Cosmological parameters}",
      journal = {\aap},
         year = 2016,
        month = sep,
       volume = {594},
          eid = {A13},
        pages = {A13},
          doi = {10.1051/0004-6361/201525830},
archivePrefix = {arXiv},
       eprint = {1502.01589},
 primaryClass = {astro-ph.CO},
       adsurl = {https://ui.adsabs.harvard.edu/abs/2016A&A...594A..13P}
}

@ARTICLE{2019ApJ...878...55B,
       author = {{Bocquet}, S. and {Dietrich}, J.~P. and {Schrabback}, T. and {Bleem}, L.~E. and {Klein}, M. and {Allen}, S.~W. and {Applegate}, D.~E. and {Ashby}, M.~L.~N. and {Bautz}, M. and {Bayliss}, M. and {Benson}, B.~A. and {Brodwin}, M. and {Bulbul}, E. and {Canning}, R.~E.~A. and {Capasso}, R. and {Carlstrom}, J.~E. and {Chang}, C.~L. and {Chiu}, I. and {Cho}, H. -M. and {Clocchiatti}, A. and {Crawford}, T.~M. and {Crites}, A.~T. and {de Haan}, T. and {Desai}, S. and {Dobbs}, M.~A. and {Foley}, R.~J. and {Forman}, W.~R. and {Garmire}, G.~P. and {George}, E.~M. and {Gladders}, M.~D. and {Gonzalez}, A.~H. and {Grandis}, S. and {Gupta}, N. and {Halverson}, N.~W. and {Hlavacek-Larrondo}, J. and {Hoekstra}, H. and {Holder}, G.~P. and {Holzapfel}, W.~L. and {Hou}, Z. and {Hrubes}, J.~D. and {Huang}, N. and {Jones}, C. and {Khullar}, G. and {Knox}, L. and {Kraft}, R. and {Lee}, A.~T. and {von der Linden}, A. and {Luong-Van}, D. and {Mantz}, A. and {Marrone}, D.~P. and {McDonald}, M. and {McMahon}, J.~J. and {Meyer}, S.~S. and {Mocanu}, L.~M. and {Mohr}, J.~J. and {Morris}, R.~G. and {Padin}, S. and {Patil}, S. and {Pryke}, C. and {Rapetti}, D. and {Reichardt}, C.~L. and {Rest}, A. and {Ruhl}, J.~E. and {Saliwanchik}, B.~R. and {Saro}, A. and {Sayre}, J.~T. and {Schaffer}, K.~K. and {Shirokoff}, E. and {Stalder}, B. and {Stanford}, S.~A. and {Staniszewski}, Z. and {Stark}, A.~A. and {Story}, K.~T. and {Strazzullo}, V. and {Stubbs}, C.~W. and {Vanderlinde}, K. and {Vieira}, J.~D. and {Vikhlinin}, A. and {Williamson}, R. and {Zenteno}, A.},
        title = "{Cluster Cosmology Constraints from the 2500 deg$^{2}$ SPT-SZ Survey: Inclusion of Weak Gravitational Lensing Data from Magellan and the Hubble Space Telescope}",
      journal = {\apj},
         year = 2019,
        month = jun,
       volume = {878},
       number = {1},
          eid = {55},
        pages = {55},
          doi = {10.3847/1538-4357/ab1f10},
archivePrefix = {arXiv},
       eprint = {1812.01679},
 primaryClass = {astro-ph.CO},
       adsurl = {https://ui.adsabs.harvard.edu/abs/2019ApJ...878...55B}
}

@ARTICLE{2017A&A...604A..71H,
       author = {{Hurier}, G. and {Lacasa}, F.},
        title = "{Combined analysis of galaxy cluster number count, thermal Sunyaev-Zel'dovich power spectrum, and bispectrum}",
      journal = {\aap},
         year = 2017,
        month = aug,
       volume = {604},
          eid = {A71},
        pages = {A71},
          doi = {10.1051/0004-6361/201630041},
archivePrefix = {arXiv},
       eprint = {1701.09067},
 primaryClass = {astro-ph.CO},
       adsurl = {https://ui.adsabs.harvard.edu/abs/2017A&A...604A..71H}
}

@ARTICLE{2016A&A...594A..24P,
       author = {{Planck Collaboration} and {Ade}, P.~A.~R. and {Aghanim}, N. and {Arnaud}, M. and {Ashdown}, M. and {Aumont}, J. and {Baccigalupi}, C. and {Banday}, A.~J. and {Barreiro}, R.~B. and {Bartlett}, J.~G. and {Bartolo}, N. and {Battaner}, E. and {Battye}, R. and {Benabed}, K. and {Beno{\^\i}t}, A. and {Benoit-L{\'e}vy}, A. and {Bernard}, J. -P. and {Bersanelli}, M. and {Bielewicz}, P. and {Bock}, J.~J. and {Bonaldi}, A. and {Bonavera}, L. and {Bond}, J.~R. and {Borrill}, J. and {Bouchet}, F.~R. and {Bucher}, M. and {Burigana}, C. and {Butler}, R.~C. and {Calabrese}, E. and {Cardoso}, J. -F. and {Catalano}, A. and {Challinor}, A. and {Chamballu}, A. and {Chary}, R. -R. and {Chiang}, H.~C. and {Christensen}, P.~R. and {Church}, S. and {Clements}, D.~L. and {Colombi}, S. and {Colombo}, L.~P.~L. and {Combet}, C. and {Comis}, B. and {Couchot}, F. and {Coulais}, A. and {Crill}, B.~P. and {Curto}, A. and {Cuttaia}, F. and {Danese}, L. and {Davies}, R.~D. and {Davis}, R.~J. and {de Bernardis}, P. and {de Rosa}, A. and {de Zotti}, G. and {Delabrouille}, J. and {D{\'e}sert}, F. -X. and {Diego}, J.~M. and {Dolag}, K. and {Dole}, H. and {Donzelli}, S. and {Dor{\'e}}, O. and {Douspis}, M. and {Ducout}, A. and {Dupac}, X. and {Efstathiou}, G. and {Elsner}, F. and {En{\ss}lin}, T.~A. and {Eriksen}, H.~K. and {Falgarone}, E. and {Fergusson}, J. and {Finelli}, F. and {Forni}, O. and {Frailis}, M. and {Fraisse}, A.~A. and {Franceschi}, E. and {Frejsel}, A. and {Galeotta}, S. and {Galli}, S. and {Ganga}, K. and {Giard}, M. and {Giraud-H{\'e}raud}, Y. and {Gjerl{\o}w}, E. and {Gonz{\'a}lez-Nuevo}, J. and {G{\'o}rski}, K.~M. and {Gratton}, S. and {Gregorio}, A. and {Gruppuso}, A. and {Gudmundsson}, J.~E. and {Hansen}, F.~K. and {Hanson}, D. and {Harrison}, D.~L. and {Henrot-Versill{\'e}}, S. and {Hern{\'a}ndez-Monteagudo}, C. and {Herranz}, D. and {Hildebrandt}, S.~R. and {Hivon}, E. and {Hobson}, M. and {Holmes}, W.~A. and {Hornstrup}, A. and {Hovest}, W. and {Huffenberger}, K.~M. and {Hurier}, G. and {Jaffe}, A.~H. and {Jaffe}, T.~R. and {Jones}, W.~C. and {Juvela}, M. and {Keih{\"a}nen}, E. and {Keskitalo}, R. and {Kisner}, T.~S. and {Kneissl}, R. and {Knoche}, J. and {Kunz}, M. and {Kurki-Suonio}, H. and {Lagache}, G. and {L{\"a}hteenm{\"a}ki}, A. and {Lamarre}, J. -M. and {Lasenby}, A. and {Lattanzi}, M. and {Lawrence}, C.~R. and {Leonardi}, R. and {Lesgourgues}, J. and {Levrier}, F. and {Liguori}, M. and {Lilje}, P.~B. and {Linden-V{\o}rnle}, M. and {L{\'o}pez-Caniego}, M. and {Lubin}, P.~M. and {Mac{\'\i}as-P{\'e}rez}, J.~F. and {Maggio}, G. and {Maino}, D. and {Mandolesi}, N. and {Mangilli}, A. and {Maris}, M. and {Martin}, P.~G. and {Mart{\'\i}nez-Gonz{\'a}lez}, E. and {Masi}, S. and {Matarrese}, S. and {McGehee}, P. and {Meinhold}, P.~R. and {Melchiorri}, A. and {Melin}, J. -B. and {Mendes}, L. and {Mennella}, A. and {Migliaccio}, M. and {Mitra}, S. and {Miville-Desch{\^e}nes}, M. -A. and {Moneti}, A. and {Montier}, L. and {Morgante}, G. and {Mortlock}, D. and {Moss}, A. and {Munshi}, D. and {Murphy}, J.~A. and {Naselsky}, P. and {Nati}, F. and {Natoli}, P. and {Netterfield}, C.~B. and {N{\o}rgaard-Nielsen}, H.~U. and {Noviello}, F. and {Novikov}, D. and {Novikov}, I. and {Oxborrow}, C.~A. and {Paci}, F. and {Pagano}, L. and {Pajot}, F. and {Paoletti}, D. and {Partridge}, B. and {Pasian}, F. and {Patanchon}, G. and {Pearson}, T.~J. and {Perdereau}, O. and {Perotto}, L. and {Perrotta}, F. and {Pettorino}, V. and {Piacentini}, F. and {Piat}, M. and {Pierpaoli}, E. and {Pietrobon}, D. and {Plaszczynski}, S. and {Pointecouteau}, E. and {Polenta}, G. and {Popa}, L. and {Pratt}, G.~W. and {Pr{\'e}zeau}, G. and {Prunet}, S. and {Puget}, J. -L. and {Rachen}, J.~P. and {Rebolo}, R. and {Reinecke}, M. and {Remazeilles}, M. and {Renault}, C. and {Renzi}, A. and {Ristorcelli}, I. and {Rocha}, G. and {Roman}, M. and {Rosset}, C. and {Rossetti}, M. and {Roudier}, G. and {Rubi{\~n}o-Mart{\'\i}n}, J.~A. and {Rusholme}, B. and {Sandri}, M. and {Santos}, D. and {Savelainen}, M. and {Savini}, G. and {Scott}, D. and {Seiffert}, M.~D. and {Shellard}, E.~P.~S. and {Spencer}, L.~D. and {Stolyarov}, V. and {Stompor}, R. and {Sudiwala}, R. and {Sunyaev}, R. and {Sutton}, D. and {Suur-Uski}, A. -S. and {Sygnet}, J. -F. and {Tauber}, J.~A. and {Terenzi}, L. and {Toffolatti}, L. and {Tomasi}, M. and {Tristram}, M. and {Tucci}, M. and {Tuovinen}, J. and {T{\"u}rler}, M. and {Umana}, G. and {Valenziano}, L. and {Valiviita}, J. and {Van Tent}, B. and {Vielva}, P. and {Villa}, F. and {Wade}, L.~A. and {Wandelt}, B.~D. and {Wehus}, I.~K. and {Weller}, J. and {White}, S.~D.~M. and {Yvon}, D. and {Zacchei}, A. and {Zonca}, A.},
        title = "{Planck 2015 results. XXIV. Cosmology from Sunyaev-Zeldovich cluster counts}",
      journal = {\aap},
         year = 2016,
        month = sep,
       volume = {594},
          eid = {A24},
        pages = {A24},
          doi = {10.1051/0004-6361/201525833},
archivePrefix = {arXiv},
       eprint = {1502.01597},
 primaryClass = {astro-ph.CO},
       adsurl = {https://ui.adsabs.harvard.edu/abs/2016A&A...594A..24P}
}

@ARTICLE{2014A&A...571A..21P,
       author = {{Planck Collaboration} and {Ade}, P.~A.~R. and {Aghanim}, N. and {Armitage-Caplan}, C. and {Arnaud}, M. and {Ashdown}, M. and {Atrio-Barandela}, F. and {Aumont}, J. and {Baccigalupi}, C. and {Banday}, A.~J. and {Barreiro}, R.~B. and {Bartlett}, J.~G. and {Battaner}, E. and {Benabed}, K. and {Beno{\^\i}t}, A. and {Benoit-L{\'e}vy}, A. and {Bernard}, J. -P. and {Bersanelli}, M. and {Bielewicz}, P. and {Bobin}, J. and {Bock}, J.~J. and {Bonaldi}, A. and {Bond}, J.~R. and {Borrill}, J. and {Bouchet}, F.~R. and {Bridges}, M. and {Bucher}, M. and {Burigana}, C. and {Butler}, R.~C. and {Cardoso}, J. -F. and {Carvalho}, P. and {Catalano}, A. and {Challinor}, A. and {Chamballu}, A. and {Chiang}, H.~C. and {Chiang}, L. -Y. and {Christensen}, P.~R. and {Church}, S. and {Clements}, D.~L. and {Colombi}, S. and {Colombo}, L.~P.~L. and {Comis}, B. and {Couchot}, F. and {Coulais}, A. and {Crill}, B.~P. and {Curto}, A. and {Cuttaia}, F. and {Da Silva}, A. and {Danese}, L. and {Davies}, R.~D. and {Davis}, R.~J. and {de Bernardis}, P. and {de Rosa}, A. and {de Zotti}, G. and {Delabrouille}, J. and {Delouis}, J. -M. and {D{\'e}sert}, F. -X. and {Dickinson}, C. and {Diego}, J.~M. and {Dolag}, K. and {Dole}, H. and {Donzelli}, S. and {Dor{\'e}}, O. and {Douspis}, M. and {Dupac}, X. and {Efstathiou}, G. and {En{\ss}lin}, T.~A. and {Eriksen}, H.~K. and {Finelli}, F. and {Flores-Cacho}, I. and {Forni}, O. and {Frailis}, M. and {Franceschi}, E. and {Galeotta}, S. and {Ganga}, K. and {G{\'e}nova-Santos}, R.~T. and {Giard}, M. and {Giardino}, G. and {Giraud-H{\'e}raud}, Y. and {Gonz{\'a}lez-Nuevo}, J. and {G{\'o}rski}, K.~M. and {Gratton}, S. and {Gregorio}, A. and {Gruppuso}, A. and {Hansen}, F.~K. and {Hanson}, D. and {Harrison}, D. and {Henrot-Versill{\'e}}, S. and {Hern{\'a}ndez-Monteagudo}, C. and {Herranz}, D. and {Hildebrandt}, S.~R. and {Hivon}, E. and {Hobson}, M. and {Holmes}, W.~A. and {Hornstrup}, A. and {Hovest}, W. and {Huffenberger}, K.~M. and {Hurier}, G. and {Jaffe}, A.~H. and {Jaffe}, T.~R. and {Jones}, W.~C. and {Juvela}, M. and {Keih{\"a}nen}, E. and {Keskitalo}, R. and {Kisner}, T.~S. and {Kneissl}, R. and {Knoche}, J. and {Knox}, L. and {Kunz}, M. and {Kurki-Suonio}, H. and {Lacasa}, F. and {Lagache}, G. and {L{\"a}hteenm{\"a}ki}, A. and {Lamarre}, J. -M. and {Lasenby}, A. and {Laureijs}, R.~J. and {Lawrence}, C.~R. and {Leahy}, J.~P. and {Leonardi}, R. and {Le{\'o}n-Tavares}, J. and {Lesgourgues}, J. and {Liguori}, M. and {Lilje}, P.~B. and {Linden-V{\o}rnle}, M. and {L{\'o}pez-Caniego}, M. and {Lubin}, P.~M. and {Mac{\'\i}as-P{\'e}rez}, J.~F. and {Maffei}, B. and {Maino}, D. and {Mandolesi}, N. and {Marcos-Caballero}, A. and {Maris}, M. and {Marshall}, D.~J. and {Martin}, P.~G. and {Mart{\'\i}nez-Gonz{\'a}lez}, E. and {Masi}, S. and {Massardi}, M. and {Matarrese}, S. and {Matthai}, F. and {Mazzotta}, P. and {Melchiorri}, A. and {Melin}, J. -B. and {Mendes}, L. and {Mennella}, A. and {Migliaccio}, M. and {Mitra}, S. and {Miville-Desch{\^e}nes}, M. -A. and {Moneti}, A. and {Montier}, L. and {Morgante}, G. and {Mortlock}, D. and {Moss}, A. and {Munshi}, D. and {Naselsky}, P. and {Nati}, F. and {Natoli}, P. and {Netterfield}, C.~B. and {N{\o}rgaard-Nielsen}, H.~U. and {Noviello}, F. and {Novikov}, D. and {Novikov}, I. and {Osborne}, S. and {Oxborrow}, C.~A. and {Paci}, F. and {Pagano}, L. and {Pajot}, F. and {Paoletti}, D. and {Partridge}, B. and {Pasian}, F. and {Patanchon}, G. and {Perdereau}, O. and {Perotto}, L. and {Perrotta}, F. and {Piacentini}, F. and {Piat}, M. and {Pierpaoli}, E. and {Pietrobon}, D. and {Plaszczynski}, S. and {Pointecouteau}, E. and {Polenta}, G. and {Ponthieu}, N. and {Popa}, L. and {Poutanen}, T. and {Pratt}, G.~W. and {Pr{\'e}zeau}, G. and {Prunet}, S. and {Puget}, J. -L. and {Rachen}, J.~P. and {Rebolo}, R. and {Reinecke}, M. and {Remazeilles}, M. and {Renault}, C. and {Ricciardi}, S. and {Riller}, T. and {Ristorcelli}, I. and {Rocha}, G. and {Rosset}, C. and {Rossetti}, M. and {Roudier}, G. and {Rubi{\~n}o-Mart{\'\i}n}, J.~A. and {Rusholme}, B. and {Sandri}, M. and {Santos}, D. and {Savini}, G. and {Scott}, D. and {Seiffert}, M.~D. and {Shellard}, E.~P.~S. and {Spencer}, L.~D. and {Starck}, J. -L. and {Stolyarov}, V. and {Stompor}, R. and {Sudiwala}, R. and {Sunyaev}, R. and {Sureau}, F. and {Sutton}, D. and {Suur-Uski}, A. -S. and {Sygnet}, J. -F. and {Tauber}, J.~A. and {Tavagnacco}, D. and {Terenzi}, L. and {Toffolatti}, L. and {Tomasi}, M. and {Tristram}, M. and {Tucci}, M. and {Tuovinen}, J. and {Umana}, G. and {Valenziano}, L. and {Valiviita}, J. and {Van Tent}, B. and {Varis}, J. and {Vielva}, P. and {Villa}, F. and {Vittorio}, N. and {Wade}, L.~A. and {Wandelt}, B.~D. and {White}, S.~D.~M. and {Yvon}, D. and {Zacchei}, A. and {Zonca}, A.},
        title = "{Planck 2013 results. XXI. Power spectrum and high-order statistics of the Planck all-sky Compton parameter map}",
      journal = {\aap},
         year = 2014,
        month = nov,
       volume = {571},
          eid = {A21},
        pages = {A21},
          doi = {10.1051/0004-6361/201321522},
archivePrefix = {arXiv},
       eprint = {1303.5081},
 primaryClass = {astro-ph.CO},
       adsurl = {https://ui.adsabs.harvard.edu/abs/2014A&A...571A..21P}
}

@ARTICLE{2016A&A...594A..27P,
       author = {{Planck Collaboration} and {Ade}, P.~A.~R. and {Aghanim}, N. and {Arnaud}, M. and {Ashdown}, M. and {Aumont}, J. and {Baccigalupi}, C. and {Banday}, A.~J. and {Barreiro}, R.~B. and {Barrena}, R. and {Bartlett}, J.~G. and {Bartolo}, N. and {Battaner}, E. and {Battye}, R. and {Benabed}, K. and {Beno{\^\i}t}, A. and {Benoit-L{\'e}vy}, A. and {Bernard}, J. -P. and {Bersanelli}, M. and {Bielewicz}, P. and {Bikmaev}, I. and {B{\"o}hringer}, H. and {Bonaldi}, A. and {Bonavera}, L. and {Bond}, J.~R. and {Borrill}, J. and {Bouchet}, F.~R. and {Bucher}, M. and {Burenin}, R. and {Burigana}, C. and {Butler}, R.~C. and {Calabrese}, E. and {Cardoso}, J. -F. and {Carvalho}, P. and {Catalano}, A. and {Challinor}, A. and {Chamballu}, A. and {Chary}, R. -R. and {Chiang}, H.~C. and {Chon}, G. and {Christensen}, P.~R. and {Clements}, D.~L. and {Colombi}, S. and {Colombo}, L.~P.~L. and {Combet}, C. and {Comis}, B. and {Couchot}, F. and {Coulais}, A. and {Crill}, B.~P. and {Curto}, A. and {Cuttaia}, F. and {Dahle}, H. and {Danese}, L. and {Davies}, R.~D. and {Davis}, R.~J. and {de Bernardis}, P. and {de Rosa}, A. and {de Zotti}, G. and {Delabrouille}, J. and {D{\'e}sert}, F. -X. and {Dickinson}, C. and {Diego}, J.~M. and {Dolag}, K. and {Dole}, H. and {Donzelli}, S. and {Dor{\'e}}, O. and {Douspis}, M. and {Ducout}, A. and {Dupac}, X. and {Efstathiou}, G. and {Eisenhardt}, P.~R.~M. and {Elsner}, F. and {En{\ss}lin}, T.~A. and {Eriksen}, H.~K. and {Falgarone}, E. and {Fergusson}, J. and {Feroz}, F. and {Ferragamo}, A. and {Finelli}, F. and {Forni}, O. and {Frailis}, M. and {Fraisse}, A.~A. and {Franceschi}, E. and {Frejsel}, A. and {Galeotta}, S. and {Galli}, S. and {Ganga}, K. and {G{\'e}nova-Santos}, R.~T. and {Giard}, M. and {Giraud-H{\'e}raud}, Y. and {Gjerl{\o}w}, E. and {Gonz{\'a}lez-Nuevo}, J. and {G{\'o}rski}, K.~M. and {Grainge}, K.~J.~B. and {Gratton}, S. and {Gregorio}, A. and {Gruppuso}, A. and {Gudmundsson}, J.~E. and {Hansen}, F.~K. and {Hanson}, D. and {Harrison}, D.~L. and {Hempel}, A. and {Henrot-Versill{\'e}}, S. and {Hern{\'a}ndez-Monteagudo}, C. and {Herranz}, D. and {Hildebrandt}, S.~R. and {Hivon}, E. and {Hobson}, M. and {Holmes}, W.~A. and {Hornstrup}, A. and {Hovest}, W. and {Huffenberger}, K.~M. and {Hurier}, G. and {Jaffe}, A.~H. and {Jaffe}, T.~R. and {Jin}, T. and {Jones}, W.~C. and {Juvela}, M. and {Keih{\"a}nen}, E. and {Keskitalo}, R. and {Khamitov}, I. and {Kisner}, T.~S. and {Kneissl}, R. and {Knoche}, J. and {Kunz}, M. and {Kurki-Suonio}, H. and {Lagache}, G. and {Lamarre}, J. -M. and {Lasenby}, A. and {Lattanzi}, M. and {Lawrence}, C.~R. and {Leonardi}, R. and {Lesgourgues}, J. and {Levrier}, F. and {Liguori}, M. and {Lilje}, P.~B. and {Linden-V{\o}rnle}, M. and {L{\'o}pez-Caniego}, M. and {Lubin}, P.~M. and {Mac{\'\i}as-P{\'e}rez}, J.~F. and {Maggio}, G. and {Maino}, D. and {Mak}, D.~S.~Y. and {Mandolesi}, N. and {Mangilli}, A. and {Martin}, P.~G. and {Mart{\'\i}nez-Gonz{\'a}lez}, E. and {Masi}, S. and {Matarrese}, S. and {Mazzotta}, P. and {McGehee}, P. and {Mei}, S. and {Melchiorri}, A. and {Melin}, J. -B. and {Mendes}, L. and {Mennella}, A. and {Migliaccio}, M. and {Mitra}, S. and {Miville-Desch{\^e}nes}, M. -A. and {Moneti}, A. and {Montier}, L. and {Morgante}, G. and {Mortlock}, D. and {Moss}, A. and {Munshi}, D. and {Murphy}, J.~A. and {Naselsky}, P. and {Nastasi}, A. and {Nati}, F. and {Natoli}, P. and {Netterfield}, C.~B. and {N{\o}rgaard-Nielsen}, H.~U. and {Noviello}, F. and {Novikov}, D. and {Novikov}, I. and {Olamaie}, M. and {Oxborrow}, C.~A. and {Paci}, F. and {Pagano}, L. and {Pajot}, F. and {Paoletti}, D. and {Pasian}, F. and {Patanchon}, G. and {Pearson}, T.~J. and {Perdereau}, O. and {Perotto}, L. and {Perrott}, Y.~C. and {Perrotta}, F. and {Pettorino}, V. and {Piacentini}, F. and {Piat}, M. and {Pierpaoli}, E. and {Pietrobon}, D. and {Plaszczynski}, S. and {Pointecouteau}, E. and {Polenta}, G. and {Pratt}, G.~W. and {Pr{\'e}zeau}, G. and {Prunet}, S. and {Puget}, J. -L. and {Rachen}, J.~P. and {Reach}, W.~T. and {Rebolo}, R. and {Reinecke}, M. and {Remazeilles}, M. and {Renault}, C. and {Renzi}, A. and {Ristorcelli}, I. and {Rocha}, G. and {Rosset}, C. and {Rossetti}, M. and {Roudier}, G. and {Rozo}, E. and {Rubi{\~n}o-Mart{\'\i}n}, J.~A. and {Rumsey}, C. and {Rusholme}, B. and {Rykoff}, E.~S. and {Sandri}, M. and {Santos}, D. and {Saunders}, R.~D.~E. and {Savelainen}, M. and {Savini}, G. and {Schammel}, M.~P. and {Scott}, D. and {Seiffert}, M.~D. and {Shellard}, E.~P.~S. and {Shimwell}, T.~W. and {Spencer}, L.~D. and {Stanford}, S.~A. and {Stern}, D. and {Stolyarov}, V. and {Stompor}, R. and {Streblyanska}, A. and {Sudiwala}, R. and {Sunyaev}, R. and {Sutton}, D. and {Suur-Uski}, A. -S. and {Sygnet}, J. -F. and {Tauber}, J.~A. and {Terenzi}, L. and {Toffolatti}, L. and {Tomasi}, M. and {Tramonte}, D. and {Tristram}, M. and {Tucci}, M. and {Tuovinen}, J. and {Umana}, G. and {Valenziano}, L. and {Valiviita}, J. and {Van Tent}, B. and {Vielva}, P. and {Villa}, F. and {Wade}, L.~A. and {Wandelt}, B.~D. and {Wehus}, I.~K. and {White}, S.~D.~M. and {Wright}, E.~L. and {Yvon}, D. and {Zacchei}, A. and {Zonca}, A.},
        title = "{Planck 2015 results. XXVII. The second Planck catalogue of Sunyaev-Zeldovich sources}",
      journal = {\aap},
         year = 2016,
        month = sep,
       volume = {594},
          eid = {A27},
        pages = {A27},
          doi = {10.1051/0004-6361/201525823},
archivePrefix = {arXiv},
       eprint = {1502.01598},
 primaryClass = {astro-ph.CO},
       adsurl = {https://ui.adsabs.harvard.edu/abs/2016A&A...594A..27P}
}

@ARTICLE{2015ApJS..216...27B,
       author = {{Bleem}, L.~E. and {Stalder}, B. and {de Haan}, T. and {Aird}, K.~A. and {Allen}, S.~W. and {Applegate}, D.~E. and {Ashby}, M.~L.~N. and {Bautz}, M. and {Bayliss}, M. and {Benson}, B.~A. and {Bocquet}, S. and {Brodwin}, M. and {Carlstrom}, J.~E. and {Chang}, C.~L. and {Chiu}, I. and {Cho}, H.~M. and {Clocchiatti}, A. and {Crawford}, T.~M. and {Crites}, A.~T. and {Desai}, S. and {Dietrich}, J.~P. and {Dobbs}, M.~A. and {Foley}, R.~J. and {Forman}, W.~R. and {George}, E.~M. and {Gladders}, M.~D. and {Gonzalez}, A.~H. and {Halverson}, N.~W. and {Hennig}, C. and {Hoekstra}, H. and {Holder}, G.~P. and {Holzapfel}, W.~L. and {Hrubes}, J.~D. and {Jones}, C. and {Keisler}, R. and {Knox}, L. and {Lee}, A.~T. and {Leitch}, E.~M. and {Liu}, J. and {Lueker}, M. and {Luong-Van}, D. and {Mantz}, A. and {Marrone}, D.~P. and {McDonald}, M. and {McMahon}, J.~J. and {Meyer}, S.~S. and {Mocanu}, L. and {Mohr}, J.~J. and {Murray}, S.~S. and {Padin}, S. and {Pryke}, C. and {Reichardt}, C.~L. and {Rest}, A. and {Ruel}, J. and {Ruhl}, J.~E. and {Saliwanchik}, B.~R. and {Saro}, A. and {Sayre}, J.~T. and {Schaffer}, K.~K. and {Schrabback}, T. and {Shirokoff}, E. and {Song}, J. and {Spieler}, H.~G. and {Stanford}, S.~A. and {Staniszewski}, Z. and {Stark}, A.~A. and {Story}, K.~T. and {Stubbs}, C.~W. and {Vanderlinde}, K. and {Vieira}, J.~D. and {Vikhlinin}, A. and {Williamson}, R. and {Zahn}, O. and {Zenteno}, A.},
        title = "{Galaxy Clusters Discovered via the Sunyaev-Zel'dovich Effect in the 2500-Square-Degree SPT-SZ Survey}",
      journal = {\apjs},
         year = 2015,
        month = feb,
       volume = {216},
       number = {2},
          eid = {27},
        pages = {27},
          doi = {10.1088/0067-0049/216/2/27},
archivePrefix = {arXiv},
       eprint = {1409.0850},
 primaryClass = {astro-ph.CO},
       adsurl = {https://ui.adsabs.harvard.edu/abs/2015ApJS..216...27B}
}

@ARTICLE{2013JCAP...07..008H,
       author = {{Hasselfield}, Matthew and {Hilton}, Matt and {Marriage}, Tobias A. and {Addison}, Graeme E. and {Barrientos}, L. Felipe and {Battaglia}, Nicholas and {Battistelli}, Elia S. and {Bond}, J. Richard and {Crichton}, Devin and {Das}, Sudeep and {Devlin}, Mark J. and {Dicker}, Simon R. and {Dunkley}, Joanna and {D{\"u}nner}, Rolando and {Fowler}, Joseph W. and {Gralla}, Megan B. and {Hajian}, Amir and {Halpern}, Mark and {Hincks}, Adam D. and {Hlozek}, Ren{\'e}e and {Hughes}, John P. and {Infante}, Leopoldo and {Irwin}, Kent D. and {Kosowsky}, Arthur and {Marsden}, Danica and {Menanteau}, Felipe and {Moodley}, Kavilan and {Niemack}, Michael D. and {Nolta}, Michael R. and {Page}, Lyman A. and {Partridge}, Bruce and {Reese}, Erik D. and {Schmitt}, Benjamin L. and {Sehgal}, Neelima and {Sherwin}, Blake D. and {Sievers}, Jon and {Sif{\'o}n}, Crist{\'o}bal and {Spergel}, David N. and {Staggs}, Suzanne T. and {Swetz}, Daniel S. and {Switzer}, Eric R. and {Thornton}, Robert and {Trac}, Hy and {Wollack}, Edward J.},
        title = "{The Atacama Cosmology Telescope: Sunyaev-Zel'dovich selected galaxy clusters at 148 GHz from three seasons of data}",
      journal = {\jcap},
         year = 2013,
        month = jul,
       volume = {2013},
       number = {7},
          eid = {008},
        pages = {008},
          doi = {10.1088/1475-7516/2013/07/008},
archivePrefix = {arXiv},
       eprint = {1301.0816},
 primaryClass = {astro-ph.CO},
       adsurl = {https://ui.adsabs.harvard.edu/abs/2013JCAP...07..008H}
}

@ARTICLE{2020MNRAS.497.1332B,
       author = {{Bolliet}, Boris and {Brinckmann}, Thejs and {Chluba}, Jens and {Lesgourgues}, Julien},
        title = "{Including massive neutrinos in thermal Sunyaev Zeldovich power spectrum and cluster counts analyses}",
      journal = {\mnras},
         year = 2020,
        month = sep,
       volume = {497},
       number = {2},
        pages = {1332-1347},
          doi = {10.1093/mnras/staa1835},
archivePrefix = {arXiv},
       eprint = {1906.10359},
 primaryClass = {astro-ph.CO},
       adsurl = {https://ui.adsabs.harvard.edu/abs/2020MNRAS.497.1332B}
}

@ARTICLE{2018A&A...614A..13S,
       author = {{Salvati}, Laura and {Douspis}, Marian and {Aghanim}, Nabila},
        title = "{Constraints from thermal Sunyaev-Zel'dovich cluster counts and power spectrum combined with CMB}",
      journal = {\aap},
         year = 2018,
        month = jun,
       volume = {614},
          eid = {A13},
        pages = {A13},
          doi = {10.1051/0004-6361/201731990},
archivePrefix = {arXiv},
       eprint = {1708.00697},
 primaryClass = {astro-ph.CO},
       adsurl = {https://ui.adsabs.harvard.edu/abs/2018A&A...614A..13S}
}

@ARTICLE{2002ARA&A..40..643C,
       author = {{Carlstrom}, John E. and {Holder}, Gilbert P. and {Reese}, Erik D.},
        title = "{Cosmology with the Sunyaev-Zel'dovich Effect}",
      journal = {\araa},
         year = 2002,
        month = jan,
       volume = {40},
        pages = {643-680},
          doi = {10.1146/annurev.astro.40.060401.093803},
archivePrefix = {arXiv},
       eprint = {astro-ph/0208192},
 primaryClass = {astro-ph},
       adsurl = {https://ui.adsabs.harvard.edu/abs/2002ARA&A..40..643C}
}

@ARTICLE{1999PhR...310...97B,
       author = {{Birkinshaw}, M.},
        title = "{The Sunyaev-Zel'dovich effect}",
      journal = {\physrep},
         year = 1999,
        month = mar,
       volume = {310},
       number = {2-3},
        pages = {97-195},
          doi = {10.1016/S0370-1573(98)00080-5},
archivePrefix = {arXiv},
       eprint = {astro-ph/9808050},
 primaryClass = {astro-ph},
       adsurl = {https://ui.adsabs.harvard.edu/abs/1999PhR...310...97B}
}

@ARTICLE{2024PhRvD.109b3528M,
       author = {{McCarthy}, Fiona and {Hill}, J. Colin},
        title = "{Component-separated, CIB-cleaned thermal Sunyaev-Zel'dovich maps from Planck PR4 data with a flexible public needlet ILC pipeline}",
      journal = {\prd},
         year = 2024,
        month = jan,
       volume = {109},
       number = {2},
          eid = {023528},
        pages = {023528},
          doi = {10.1103/PhysRevD.109.023528},
archivePrefix = {arXiv},
       eprint = {2307.01043},
 primaryClass = {astro-ph.CO},
       adsurl = {https://ui.adsabs.harvard.edu/abs/2024PhRvD.109b3528M}
}

@ARTICLE{2014MNRAS.440.3645M,
       author = {{McCarthy}, I.~G. and {Le Brun}, A.~M.~C. and {Schaye}, J. and {Holder}, G.~P.},
        title = "{The thermal Sunyaev-Zel'dovich effect power spectrum in light of Planck}",
      journal = {\mnras},
         year = 2014,
        month = jun,
       volume = {440},
       number = {4},
        pages = {3645-3657},
          doi = {10.1093/mnras/stu543},
archivePrefix = {arXiv},
       eprint = {1312.5341},
 primaryClass = {astro-ph.CO},
       adsurl = {https://ui.adsabs.harvard.edu/abs/2014MNRAS.440.3645M}
}

@ARTICLE{1999ApJ...526L...1K,
       author = {{Komatsu}, Eiichiro and {Kitayama}, Tetsu},
        title = "{Sunyaev-Zeldovich Fluctuations from Spatial Correlations between Clusters of Galaxies}",
      journal = {\apjl},
         year = 1999,
        month = nov,
       volume = {526},
       number = {1},
        pages = {L1-L4},
          doi = {10.1086/312364},
archivePrefix = {arXiv},
       eprint = {astro-ph/9908087},
 primaryClass = {astro-ph},
       adsurl = {https://ui.adsabs.harvard.edu/abs/1999ApJ...526L...1K}
}

@ARTICLE{2021ApJ...908..199R,
       author = {{Reichardt}, C.~L. and {Patil}, S. and {Ade}, P.~A.~R. and {Anderson}, A.~J. and {Austermann}, J.~E. and {Avva}, J.~S. and {Baxter}, E. and {Beall}, J.~A. and {Bender}, A.~N. and {Benson}, B.~A. and {Bianchini}, F. and {Bleem}, L.~E. and {Carlstrom}, J.~E. and {Chang}, C.~L. and {Chaubal}, P. and {Chiang}, H.~C. and {Chou}, T.~L. and {Citron}, R. and {Moran}, C. Corbett and {Crawford}, T.~M. and {Crites}, A.~T. and {de Haan}, T. and {Dobbs}, M.~A. and {Everett}, W. and {Gallicchio}, J. and {George}, E.~M. and {Gilbert}, A. and {Gupta}, N. and {Halverson}, N.~W. and {Harrington}, N. and {Henning}, J.~W. and {Hilton}, G.~C. and {Holder}, G.~P. and {Holzapfel}, W.~L. and {Hrubes}, J.~D. and {Huang}, N. and {Hubmayr}, J. and {Irwin}, K.~D. and {Knox}, L. and {Lee}, A.~T. and {Li}, D. and {Lowitz}, A. and {Luong-Van}, D. and {McMahon}, J.~J. and {Mehl}, J. and {Meyer}, S.~S. and {Millea}, M. and {Mocanu}, L.~M. and {Mohr}, J.~J. and {Montgomery}, J. and {Nadolski}, A. and {Natoli}, T. and {Nibarger}, J.~P. and {Noble}, G. and {Novosad}, V. and {Omori}, Y. and {Padin}, S. and {Pryke}, C. and {Ruhl}, J.~E. and {Saliwanchik}, B.~R. and {Sayre}, J.~T. and {Schaffer}, K.~K. and {Shirokoff}, E. and {Sievers}, C. and {Smecher}, G. and {Spieler}, H.~G. and {Staniszewski}, Z. and {Stark}, A.~A. and {Tucker}, C. and {Vanderlinde}, K. and {Veach}, T. and {Vieira}, J.~D. and {Wang}, G. and {Whitehorn}, N. and {Williamson}, R. and {Wu}, W.~L.~K. and {Yefremenko}, V.},
        title = "{An Improved Measurement of the Secondary Cosmic Microwave Background Anisotropies from the SPT-SZ + SPTpol Surveys}",
      journal = {\apj},
         year = 2021,
        month = feb,
       volume = {908},
       number = {2},
          eid = {199},
        pages = {199},
          doi = {10.3847/1538-4357/abd407},
archivePrefix = {arXiv},
       eprint = {2002.06197},
 primaryClass = {astro-ph.CO},
       adsurl = {https://ui.adsabs.harvard.edu/abs/2021ApJ...908..199R}
}

@ARTICLE{2002MNRAS.336.1256K,
       author = {{Komatsu}, E. and {Seljak}, U.},
        title = "{The Sunyaev-Zel'dovich angular power spectrum as a probe of cosmological parameters}",
      journal = {\mnras},
         year = 2002,
        month = nov,
       volume = {336},
       number = {4},
        pages = {1256-1270},
          doi = {10.1046/j.1365-8711.2002.05889.x},
archivePrefix = {arXiv},
       eprint = {astro-ph/0205468},
 primaryClass = {astro-ph},
       adsurl = {https://ui.adsabs.harvard.edu/abs/2002MNRAS.336.1256K}
}

@ARTICLE{1972CoASP...4..173S,
       author = {{Sunyaev}, R.~A. and {Zeldovich}, Ya. B.},
        title = "{The Observations of Relic Radiation as a Test of the Nature of X-Ray Radiation from the Clusters of Galaxies}",
      journal = {Comments on Astrophysics and Space Physics},
         year = 1972,
        month = nov,
       volume = {4},
        pages = {173},
       adsurl = {https://ui.adsabs.harvard.edu/abs/1972CoASP...4..173S}
}

@ARTICLE{2016A&A...594A..22P,
       author = {{Planck Collaboration} and {Aghanim}, N. and {Arnaud}, M. and {Ashdown}, M. and {Aumont}, J. and {Baccigalupi}, C. and {Banday}, A.~J. and {Barreiro}, R.~B. and {Bartlett}, J.~G. and {Bartolo}, N. and et al.},
        title = "{Planck 2015 results. XXII. A map of the thermal Sunyaev-Zeldovich effect}",
      journal = {\aap},
         year = 2016,
        month = sep,
       volume = {594},
          eid = {A22},
        pages = {A22},
          doi = {10.1051/0004-6361/201525826},
archivePrefix = {arXiv},
       eprint = {1502.01596},
 primaryClass = {astro-ph.CO},
       adsurl = {https://ui.adsabs.harvard.edu/abs/2016A&A...594A..22P}
}

@ARTICLE{2014A&A...571A...9P,
       author = {{Planck Collaboration} and {Ade}, P.~A.~R. and {Aghanim}, N. and {Armitage-Caplan}, C. and {Arnaud}, M. and {Ashdown}, M. and {Atrio-Barandela}, F. and {Aumont}, J. and {Baccigalupi}, C. and {Banday}, A.~J. and et al.},
        title = "{Planck 2013 results. IX. HFI spectral response}",
      journal = {\aap},
         year = 2014,
        month = nov,
       volume = {571},
          eid = {A9},
        pages = {A9},
          doi = {10.1051/0004-6361/201321531},
archivePrefix = {arXiv},
       eprint = {1303.5070},
 primaryClass = {astro-ph.IM},
       adsurl = {https://ui.adsabs.harvard.edu/abs/2014A&A...571A...9P}
}

@ARTICLE{2018ApJS..239...36Y,
       author = {{Yao}, Jian and {Zhang}, Le and {Zhao}, Yuxi and {Zhang}, Pengjie and {Santos}, Larissa and {Zhang}, Jun},
        title = "{Testing the ABS Method with the Simulated Planck Temperature Maps}",
      journal = {\apjs},
         year = 2018,
        month = dec,
       volume = {239},
       number = {2},
          eid = {36},
        pages = {36},
          doi = {10.3847/1538-4365/aaef7a},
archivePrefix = {arXiv},
       eprint = {1807.07016},
 primaryClass = {astro-ph.CO},
       adsurl = {https://ui.adsabs.harvard.edu/abs/2018ApJS..239...36Y}
}

@ARTICLE{2018MNRAS.477.4957B,
       author = {{Bolliet}, Boris and {Comis}, Barbara and {Komatsu}, Eiichiro and {Mac{\'\i}as-P{\'e}rez}, Juan Francisco},
        title = "{Dark energy constraints from the thermal Sunyaev-Zeldovich power spectrum}",
      journal = {\mnras},
         year = 2018,
        month = jul,
       volume = {477},
       number = {4},
        pages = {4957-4967},
          doi = {10.1093/mnras/sty823},
archivePrefix = {arXiv},
       eprint = {1712.00788},
 primaryClass = {astro-ph.CO},
       adsurl = {https://ui.adsabs.harvard.edu/abs/2018MNRAS.477.4957B}
}

@ARTICLE{2003PhRvD..68h3506B,
       author = {{Battye}, Richard A. and {Weller}, Jochen},
        title = "{Constraining cosmological parameters using Sunyaev-Zel'dovich cluster surveys}",
      journal = {\prd},
         year = 2003,
        month = oct,
       volume = {68},
       number = {8},
          eid = {083506},
        pages = {083506},
          doi = {10.1103/PhysRevD.68.083506},
archivePrefix = {arXiv},
       eprint = {astro-ph/0305568},
 primaryClass = {astro-ph},
       adsurl = {https://ui.adsabs.harvard.edu/abs/2003PhRvD..68h3506B}
}
\bibliographystyle{JHEP}

\end{document}